\pdfoutput=1
\documentclass[english,mt,mech]{maddoc}

\usepackage[latin1]{inputenc}
\usepackage{graphicx}
\usepackage[labelfont=bf]{caption}
\usepackage{textcomp} 
\usepackage[nohyperlinks, printonlyused]{acronym}
\usepackage{url}
\usepackage{multicol}
\usepackage{datetime}
\usepackage[section]{placeins}
\usepackage{hyperref}

\hypersetup{linktoc=all}

\begin{document}
\clearpage
  \begin{deckblatt}
    \Titel{An open-source sensor platform for analysis of group dynamics}
    \Name{Hopfeng"artner}
    \Vorname{Michael}
    \Geburtsort{Pegnitz}
    \Geburtsdatum{February 11, 1995}
    \Betreuer{Oren Lederman (MIT), Robert Richer, Prof. Dr. Alex "'Sandy'' Pentland (MIT), Prof. Dr. Bjoern Eskofier}
    \Start{May 30, 2018}
    \Ende{November 30, 2018}
    %\ZweitInstitut{test}
  \end{deckblatt}

\cleardoublepage

Ich versichere, dass ich die Arbeit ohne fremde Hilfe und ohne Benutzung
anderer als der angegebenen Quellen angefertigt habe und dass die Arbeit
in gleicher oder "ahnlicher Form noch keiner anderen Pr"ufungsbeh"orde
vorgelegen hat und von dieser als Teil einer Pr"ufungsleistung
angenommen wurde. Alle Ausf"uhrungen, die w"ortlich oder sinngem"a"s
"ubernommen wurden, sind als solche gekennzeichnet.
\\

Die Richtlinien des Lehrstuhls f"ur Bachelor- und Masterarbeiten
habe ich gelesen und anerkannt, insbesondere die Regelung des
Nutzungsrechts. \\[15mm]

\newdateformat{myformat}{\THEDAY{.} \monthname[\THEMONTH] \THEYEAR}
Erlangen, den \myformat\today \hspace{6.0cm} \\[10mm]

\cleardoublepage

\begin{center}
\bfseries
"Ubersicht
\normalfont
\end{center}

Die Zusammenarbeit mehrerer Personen in Gruppen gewinnt in der heutigen Zeit immer mehr an Bedeutung. Gruppenarbeit wird h"aufig bei Entscheidungsfindungsprozessen und f"ur die L"osung von komplexen Problemen eingesetzt.
Die Forschung in diesem Bereich besch"aftigt sich mit der Quantifizierung und Analyse des Verhaltens innerhalb und zwischen Gruppen.
Durch den Einsatz von tragbaren elektronischen Ger"aten, wie beispielsweise sogenannten Badges, kann diese Quantifizierung automatisiert und mit einer hohen Genauigkeit durchgef"uhrt werden.
Ziel dieser Arbeit sind das Design und die Implementierung einer neuen Firmware f"ur die Badges des Rhythm Projektes - ein Open-Source Projekt der Human Dynamics Group des MIT Media Labs.
%Die Firmware ist durch eine modulare und erweiterbare Architektur gekennzeichnet und kombiniert unterschiedliche entwickelte Techniken.
Die Firmware ist durch eine modulare und erweiterbare Architektur gekennzeichnet und kombiniert unterschiedliche Techniken.
Dazu z"ahlen ein Filesystem, welches auf der Basis einer virtuellen Speicherabstraktion sequentiell generierte Daten effizient speichert, eine Serialisierungsbibliothek, die einen plattformunabh"angigen Austausch von strukturierten Daten erm"oglicht, als auch eine spezielle Technik zur Zeitsynchronisation, die Abweichungen der Oszillatorfrequenz kompensiert.
Au"serdem wurde f"ur die Verifikation einzelner funktionaler Softwarekomponenten der Applikation eine automatisierte Testumgebung entworfen.
Als Ergebnis detaillierter Analysen und spezieller Ma"snahmen konnte der Energieverbrauch im Vergleich zur vorherigen Implementierung signifikant reduziert werden.
Aufgrund des hierarchischen und modularen Aufbaus sowie des hohen Grades an Abstraktion k"onnen die entwickelten Techniken auch in anderen Projekten und Plattformen integriert werden.

\vspace{5.0cm}
\newpage
\begin{center}
\bfseries
Abstract
\normalfont
\end{center}

The collaboration of several people in groups is becoming more and more important nowadays.
Teamwork is often used for decision-making processes and for solving complex problems.
Research in this area focuses on the quantification and analysis of behavior within and between groups.
By using wearable electronic devices, such as badges, this quantification can be performed automatically and with high accuracy.
The goal of this work is the design and implementation of a new firmware for the badges of the Rhythm project - an open-source project of the Human Dynamics Group of the MIT Media Lab.
The firmware is characterized by a modular and extensible architecture and combines different techniques.
These include a filesystem, which efficiently stores sequentially generated data based on a virtual memory abstraction, a serialization library, which enables a platform-independent exchange of structured data, and a time synchronization technique, which compensates frequency deviations of the oscillator.
In addition, an automated test environment was designed for the verification of individual functional software components of the application.
As a result of detailed analysis and special measures, the power consumption could be significantly reduced compared to the previous implementation.
Due to the hierarchical and modular structure and the high degree of abstraction, the developed techniques can also be integrated into other projects and platforms.

\cleardoublepage

\tableofcontents

\cleardoublepage \pagenumbering{arabic}

\chapter{Introduction}

The cooperation of several people within a group is becoming more and more important nowadays. 
A strong trend can be observed from individual-based work to teamwork. This is especially relevant for problem solving or decision-making tasks \cite{wuchty07:TID}.
Also in companies, teamwork is often necessary to efficiently handle issues \cite{Drew96:TTT}.
According to a survey, executives spend on average up to 40-50\% of their working time in meetings \cite{Doyle93:HTM}.
Teams are frequently deployed to work on problems or to make decisions which, due to their complexity, exceed the abilities or knowledge of a single person or to quickly find a solution to a certain problem. In addition, the interactions within the group provide individuals with new perspectives and lead them to new solution strategies \cite{Levi15:GDF}.
Futhermore, it is assumed that the results of teamwork are more creative and of a higher quality than results of individual-based work. One reason for this is the continuous mutual quality control that takes place during the collaboration \cite{Nahavandi94:RTF}.
Although the potential of teamwork is immense, it is often not completely exhausted. Reasons for this are for example conflicts within the group \cite{Jehn01:TDN}, individuals who do not contribute to the group due to their insufficient self-confidence \cite{Tannen95:TPO} or generally the consideration of too few different points of view \cite{Whyte91:DFW}.
This leads to the question which methods can be applied to increase the productivity and performance of groups.
In order to answer this question, the behavior within and between the groups has to be understood and analyzed.
In social and organizational science, the social interaction and behavior of individuals within the group as well as the behavior between different groups are examined \cite{Carley02:COS} \cite{Greenberg03:BIO}.
Research in this area is broadly diversified: from developing methods to improve team performance \cite{Samrose18:CCC} \cite{Calacci16:BAO} \cite{Dimicco04:IGP} \cite{Leshed09:VRT}, to the optimization of workspaces to enhance communication between employees \cite{Brown14:TAO} \cite{Stryker12:FFT}, to the prediction of the behavior of individuals and groups \cite{Pentland10:DYR}.
In order to investigate group behavior, techniques that accurately quantify group dynamics must be applied.

Group dynamics describe the behavioral and social processes within a group and between groups, such as face-to-face communication of members \cite{Forsyth18:GD}.
Pentland found that there exists a relationship between group dynamics and group performance, especially the communication patterns are a strong indicator for the success of a team \cite{Dong10:QGP} \cite{Pentland12:TNS}. 
In addition, it was shown that the group performance can be influenced by giving feedback to the group \cite{Nadler79:TEO}.
There are several approaches to quantify group dynamics and to influence group performance through real-time feedback.
In \cite{Leshed09:VRT} a virtual group meeting system is presented. It is based on the analysis of web-chat messages between the participants and gives real-time feedback on the engagement and word choice of the users to influence their behavior. 
Calacci et al. \cite{Calacci16:BAO} developed an online communication platform called Breakout that analyzes visual and audial data in real-time, such as turn-taking and speech overlap. The Meeting Mediator visualization from Kim et al. \cite{Kim08:MME} is used to provide real-time feedback to the users. This tool displays the participants as nodes and a ball in the center that moves towards the conversation dominating user.
Another automated online collaboration system with integrated feedback assistent developed by Samrose et al. is called Collaboration Coach \cite{Samrose18:CCC}. In addition to acoustic parameters, for instance turn-taking and speech overlap, visual parameters, such as emotional valence and shared smiles of the participants, are analyzed.
DiMicco et al. showed that the behavior of people in a discussion can be influenced by displaying the participation in the discussion based on recorded audio data. Especially the dominating members reduced their participation because of the displayed feedback \cite{Dimicco04:IGP}.
In a patent from Chappel et al. (Appendix \ref{sec:US7216088}) a statistical approach to analyze the interdependencies of team members is described. These quantified interdependence relationships can be used by the team manager to optimize the collaboration between team members. 
Brown et al. \cite{Brown14:TAO} deployed active \ac{RFID} devices from Cattuto et al. \cite{Cattuto10:DOP} to measure the proximity between employees during the work. These data are used to optimize the architecture of workplaces to enhance the communication between employees.
To measure the team performance in call centers and to correlate it with the face-to-face interactions between employees, Watanabe et al. \cite{Watanabe14:ERB} used electronic wearable devices from Wakisaka et al. \cite{Wakisaka09:BSS}. These devices record the proximity to other employees with \ac{IR} transceivers and the movement of the employees with an accelerometer. The same devices were used to measure the influence of interventions to enhance social networks of older people living in the same community \cite{Masumoto17:MAV}.
Herman et al. own a patent (Appendix \ref{sec:WO2010099488}) for a system that uses an electronic portable system
to track the proximity between patients, visitors, nurses and doctors in hospitals. The history of proximity data can be used to analyze and minimize the spread of diseases.
Another system that tracks human interactions, patented by Olguin et al. (Appendix \ref{sec:US10049336}), includes a wearable electronic badge that is equipped with different environmental sensors to measure social interactions.

The Human Dynamics group of the MIT Media Lab has developed different frameworks for the analysis and feedback of social interactions and behavior as well as a number of different electronic wearable devices, called badges \cite{Choudhury02:TSA} \cite{Olguin06:WCB} \cite{Olguin10:SBO} \cite{Lederman16:OB} \cite{Lederman18:RAU}. 
Through the deployment of wearable devices, data for the analysis of social interactions can be automatically recorded from many probands simultaneously and with high accuracy.
A further advantage of this method compared to conventional methods for measuring social behavior, such as surveys, is that the behavior of the probands is hardly influenced by the badges \cite{Olguin09:SOT}. 
Normally the conscious perception of the badges disappears within an hour \cite{Pentland12:TNS}. The electronic wearable badges are optimized to collect social interaction data, such as the absolute location, physical proximity to other people and vocal activity. In a second step, the recorded data are analyzed to obtain communication patterns of the participants during meetings, workshops, discussions or events.
It should be noted that comparable electronic devices on the market are not suitable for this application: The open-source platform RuuviTag \cite{Ruuvi18:WIR} does not have an integrated microphone for recording voice activity, which is necessary for the analysis of communication patterns. Another device from Nordic Semiconductor is called Thingy:52 \cite{Nordic17:NT5}. It includes several sensors, such as an accelerometer, a gyroscope, a humidity- and temperature-sensor as well as a microphone. The disadvantage of this device is the small \ac{NVM} in which the recorded data must be stored for later transmission.

The latest badge project of the Human Dynamics group is called Rhythm \cite{Lederman18:RAU}. It is an open-source project that includes the source code of all system components and tools, as well as the schematic and layout of the electronic badges.
To facilitate the setup of the system and the development environment, Docker containers are used \cite{Lederman16:OBP}. Docker containers are isolated, lightweight operating system virtualizations that contain all dependencies of an application, allowing easy porting of applications between different platforms \cite{Merkel14:DLL}.
One important point is that the badges have to meet requirements such as low energy consumption and efficient data storage in order to maximize the runtime of the system. Furthermore, especially by using many badges in a small area, fast data transfer to the central data receivers is necessary, as otherwise badges that cannot transmit their data in time will have to overwrite them.

The purpose of this work is the reimplementation and extension of the existing sensor platform firmware. Furthermore, a testing framework is implemented to facilate the embedded software development and to achieve software with higher quality.
The focus is on the modularity, testability and maintainability of the new firmware in contrast to the previous one, which has mutual dependencies and no clearly defined interfaces between different functional units.
Therefore, a hierarchical modular approch for the architecture of the firmware, which is characterized by a high degree of abstraction, is applied. Due to this high degree of generic abstraction, the developed techniques might be applied in other projects and on different hardware platforms as well.
The developed techniques include a filesystem that enables simple and efficient storage of data in \acp{NVM} and a serialization library that converts structured data into an efficient byte sequence representation. This serialization library is used, for example, to provide a flexible communication protocol.
Additionally, several methods to minimize the power-consumption were applied. Furthermore, a technique for accurate time synchronization was developed and movement detection based on an accelerometer was incorporated.
The goal of this work is not only to optimize the previous firmware, but also to provide a structured basis for future software developments. Since this project is an open-source project available on GitHub \cite{Lederman16:OBP}, it can also serve as a reference project for modular embedded software development that allows simple testing and verification of the functionality of individual units.

The work is organized as follows: In chapter \ref{cha:Methods} the different system components, the testing framework, the architecture of the firmware as well as the developed methods for data recording, data serialization, data storing, data streaming and data exchange are explained. 
%These include a filesystem for storing data, an efficient serialization technique and transmission protocol, and a time synchronization method.
Chapter \ref{cha:Results} covers the evaluation and results of the applied methods. In chapter \ref{cha:Discussion} the results and methods are discussed and compared to the previous firmware. Finally, the most important results are summarized in Chapter \ref{cha:Conclusion} and an outlook for the future development of the system is given.
   % Einfuehrung (\chapter{Einf"uhrung})
\cleardoublepage
\chapter{Methods} \label{cha:Methods}
The purpose of the thesis is the optimization and extension of the existing sensor platform with respect to modularity and maintainability. This chapter provides a detailed description of the data-recording system and introduces different methods and tools used to achieve the objective. 
First, the setup of the system is presented, afterwards the architecture of the new firmware is introduced, followed by the explanation of the testing framework.
Finally, the developed techniques for data recording, serialization, storage and exchange are described in more detail.

\section{System overview}

The open-source Rhythm project \cite{Lederman18:RAU} encompasses electronic wearable badges and online tools to analyze social interaction between groups and group members.

\vfill

\begin{minipage}{\textwidth}
  \begin{minipage}[b]{0.45\textwidth}
%  \begin{figure}[h]
    \centering
   	\includegraphics[page=1, trim=10cm 7.8cm 10cm 4.1cm, clip, height=5.5cm]{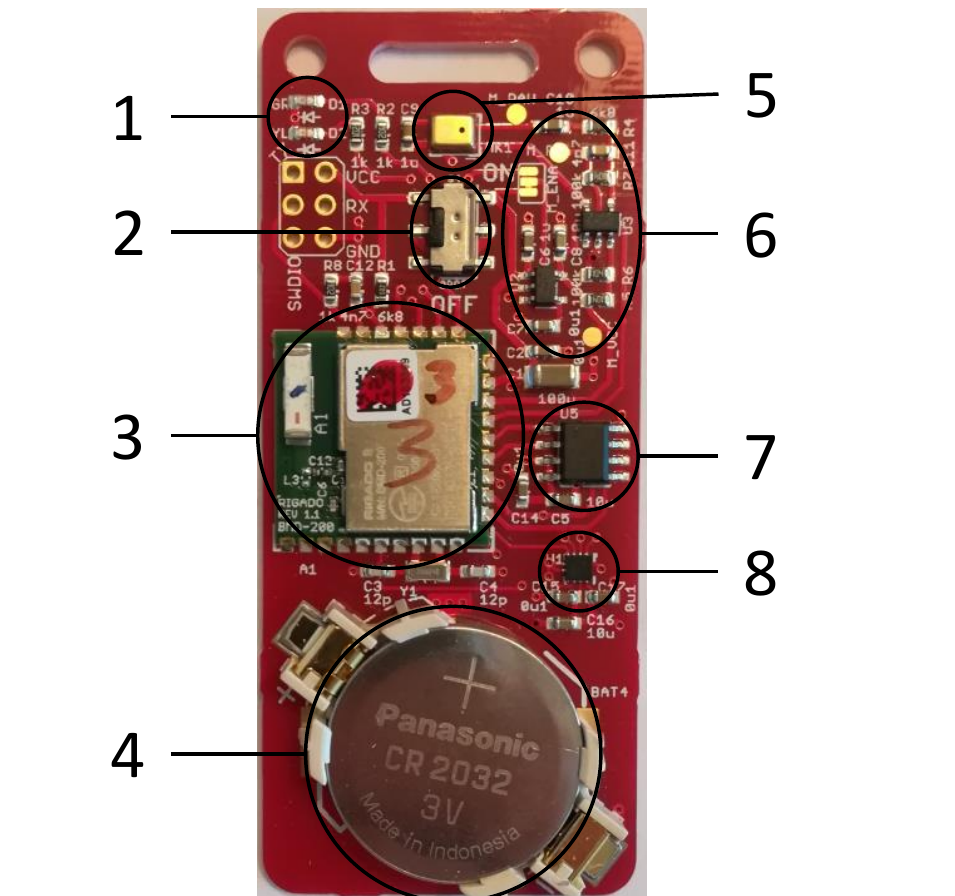}
    \captionof{figure}[Rhythm badge]{\textbf{Rhythm badge}. The custom hardware platform that collects data from different environmental sensors.}%
 %    \captionof{figure}[Rhythm badge]{\textbf{Rhythm badge}. The figure shows the custom hardware platform that collects data from different environmental sensors.}%
\label{fig:rhythm_badge}
%\end{figure}
  \end{minipage}
  \hfil
  \begin{minipage}[b]{0.45\textwidth}
 % 	\begin{table}[h]
    \centering
    \scalebox{0.93}{%
    \begin{tabular}{c|p{6cm}}
      \textbf{No.} & \textbf{Description} \\ \hline
        1 & 2x Status LEDs \\ \hline	
        2 & Power switch \\ \hline
        3 & BMD-200-B BLE-module (Rigado) \\ \hline
%        3 & BMD-200-B module \newline (nRF51822 BLE-chip + antenna) \\ \hline
        4 & 3V coin cell battery \\ \hline	
        5 & Analog microphone (Knowles) \\ \hline	
        6 & Analog filter + amplification circuit \\ \hline	
        7 & M95M02 external EEPROM \\ \hline	
        8 & LIS2DH12 3-axis accelerometer
      \end{tabular}}
      \captionof{table}[Badge components]{\textbf{Badge components}. The table lists the most important components of the badge.}
      \label{tab:badge_components}
%    \end{table}
    \end{minipage}
 \end{minipage}

%\begin{figure}[h]
%\begin{floatrow}
%\ffigbox{%
%	\includegraphics[page=1, trim=10cm 7.8cm 10cm 4.1cm, clip, height=5cm]{img/rhythm_badge_labeld.pdf}
	%\fbox{\includegraphics[page=1, trim=10cm 7.8cm 10cm 4cm, height=5.5cm]{img/rhythm_badge_labeld.pdf}}
%}{%
%\caption[Rhythm badge]{\textbf{Rhythm badge}. The figure shows the custom hardware platform that collects data from different environmental sensors.}%
%\label{fig:rhythm_badge}
%}
%\capbtabbox{%
%  \begin{tabular}{c|c|c} 
%	\textbf{No.} & \textbf{Description} \\ \hline
%	1 & a \\ \hline	
%  \end{tabular}
%}{%
%  \caption{A table}%
%}
%\end{floatrow}
%\end{figure}

%\centering
%%\fbox{\includegraphics[page=1, trim=4.6cm 7.8cm 3.6cm 4cm, height=5.5cm]{img/rhythm_badge_labeld.pdf}}
%\includegraphics[page=1, trim=4.6cm 7.8cm 3.6cm 4cm, clip, height=5.5cm]{img/rhythm_badge_labeld.pdf}
%\caption[Rhythm badge]{\textbf{Rhythm badge}. The figure shows the custom hardware platform that collects data from different environmental sensors.}% which are used to analyze social interactions.}
%\label{fig:rhythm_badge}

\newpage

\begin{figure}[h]
\centering
%\fbox{\includegraphics[page=3, trim=4.5cm 19cm 4.2cm 1.8cm, width=1.0\linewidth]{img/Rhythm.pdf}}
\includegraphics[page=3, trim=4.5cm 19cm 4.2cm 1.8cm, clip, width=0.95\linewidth]{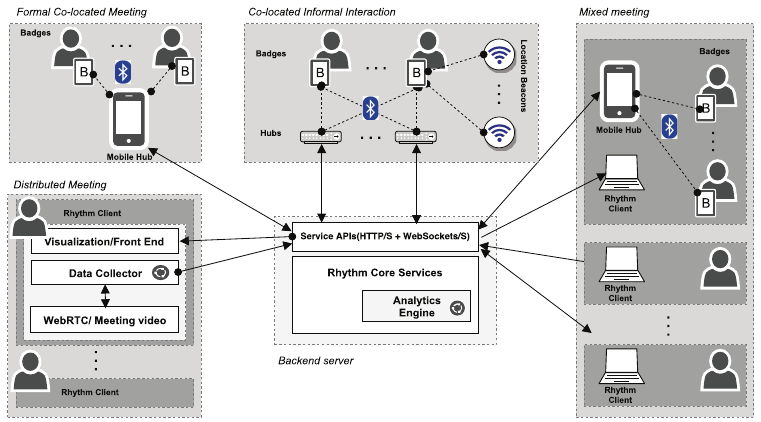}
\caption[Rhythm framework]{\textbf{Rhythm framework} \cite{Lederman18:RAU}. The Rhythm framework consists of different components: The badges worn by the participants record the required data for the subsequent analysis.
Location beacons are used to provide location information.
The recorded data of the badges are retrieved by hubs and/or mobile hubs and sent to a backend server. 
Finally, the server aggregates and analyzes the data.
Additionally, a visual frontend for real-time feedback can be used.}
%The framework consists of badges that collect the data, location beacons that provide location information, hubs and/or mobile hubs that retrieves the data from the badges and a backend server that aggregates and analyzes the data. Furthermore, a visual frontend for real-time feedback can be used.}
\label{fig:framework_overview}
\end{figure}

The badge shown in Figure \ref{fig:rhythm_badge} fits in a standard plastic name tag holder and is worn in front of the participant's chest. It is used to record different types of data in order to analyze the social interaction of the participant. In addition to vocal activity, proximity to other badges, movement and location can be measured.
The most important components of the badge are listed in Table \ref{tab:badge_components}.
%Table \ref{tab:badge_components} lists the most important components of the badge.
The Rhythm framework (Figure \ref{fig:framework_overview}) combines badges with other devices and tools to examine different types of social interactions and meetings.
%There are different types of social interactions and meetings that can be examined by the Rhythm Framework shown in Figure \ref{fig:framework_overview}.

\subsection{Bluetooth Low Energy}
The communication between badges and hubs is realized with \ac{BLE}. \ac{BLE} is a short-range, low-power wireless communication technology developed by the Bluetooth Special Interest Group and specified in the Bluetooth 4.0 standard \cite{Raza15:BS}. The 2.4 GHz \ac{ISM} band is divided into 40 channels with a channel spacing of 2 MHz.
A \ac{BLE} device uses three of these channels for periodical broadcast of advertisment packets. A \ac{BLE} device that only receives data through the three advertisment channels is called scanner. In Bluetooth 4.2 the number of bytes for an advertisment packet is limited to 31 \cite{Raza15:BS}. Additionally, a bi-directional connection for data exchange can be established between two \ac{BLE} devices: the master/central and the slave/peripheral.  The advertisment, discovering and connection processes are managed by the \ac{GAP}. The master or initiator listens for advertisment packets of the slave. These advertisment packets inform the master whether the slave is a connectable device or not. When the master receives such a packet, it transmits a connection request message to a connectable slave to establish a point-to-point connection. A \ac{BLE} slave can only be connected to one master device at a time. For the data exchange during a connection, an adaptive frequency hopping mechanism selects one of the 37 available data channels for a specific time interval \cite{Gomez12:OAE}. The \ac{ATT} defines server and client roles for the communication between two connected devices: The peripheral device acts as server and the central as client.
The \ac{GATT} defines a framework that uses the \ac{ATT} for discovering services and to access their characteristics \cite{Chang14:BAV}.
Services are a collection of information to provide a specific functionality, for example a Heart Rate Service. A service contains one or more characteristics that represent a single data point, such as the current heart rate or the temperature \cite{Gomez12:OAE}. 
The service used for the communication between the badges and the hubs is the \ac{NUS} that has a RX-characteristic to receive bytes and a TX-characteristic to send bytes. The actual interpretation of the bytes is done by a customized protocol that is described in section \ref{sec:Data exchange}.

\subsection{Badges} \label{sec:Badges}
Each participant of a meeting, a workshop, a discussion or an event obtains a badge that is worn in front of the chest. 
The badge is made of a \ac{PCB} assemblied with various components. The core element of the badge is the BMD-200-B \ac{SoM} from Rigado. It combines the embedded 2.4GHz transceiver of the nRF51822 \ac{SoC} from Nordic Semiconductor with an on-module chip antenna and essential peripherals, such as an \ac{ADC}, a \ac{SPI}, an \ac{UART}-interface, an \ac{I2C}-interface, \ac{I/O}-pins, a \ac{RTC} and different timers \cite{Rigado:BMD200}. 
The nRF51822 enables Bluetooth 4.2 LE connectivity and includes an ARM\textsuperscript{\textregistered} Cortex\texttrademark{} M0 32-bit CPU with 256 kB embedded flash program memory and 32 kB \ac{RAM}.

A common setup of the system consists of many badges that record the data but only a few hubs that pull the data from the badges.  Therefore, the badges need to buffer their data until they are transmitted. The integrated \ac{RAM} of the microcontroller is not able to buffer the generated amount of data. Consequently the recorded data have to be stored in a \ac{NVM} such as the embedded flash-memory. To minimze the loss of data when the badge records data for a long time without contact to a hub, the \ac{NVM} is increased by using an external \ac{EEPROM}. The M95M02-DR from STMicroelectronics is an \ac{SMD} with 256 kB of \ac{EEPROM} and is controlled by an \ac{SPI} bus master \cite{EEPROM:M95M02}.
Flash-memory and \ac{EEPROM} differ in the way data are erased and stored, in their speed as well as in the maximum number of store/erase cycles. In flash-memory a whole page has to be erased before words can be stored to this page. The built-in flash-memory of the nRF51822 provides 256 pages with a page size of 1024 bytes. Due to the word size of 32 bits, the minimum number of bytes that can be stored is four. The time required to erase a page is 22.3 ms and 46.3 \textmu s to store a word. The number of store/erase cycles is limited to 20,000 cycles \cite{NRF51822:PS}.
On the other hand, the incorporated \ac{EEPROM} allows to erase and store single bytes. When a store operation is started, the affected bytes are erased automatically.
The \ac{EEPROM} chip provides 1024 pages with a page size of 256 bytes and requires 10 ms for storing one byte or a whole page. In contrast to flash-memory, the number of 4 million store/erase cycles is significantly higher for \ac{EEPROM} \cite{EEPROM:M95M02}.
Due to the limited resources, the different types of data have to be stored efficiently in the \acp{NVM}. Therefore, a flexible filesystem that is decribed in section \ref{sec:Data storage} was developed. This filesystem requires an uniform storage interface that allows to store and read single bytes. In order to meet this requirement an abstraction layer was implemented that enables the application to address the \ac{EEPROM} as well as the flash-memory on byte level.

During the activation of the badge a two bytes \ac{ID} and a one byte group number is assigned to distinguish between different badges and their membership to different groups/projects.
In addition to the \ac{ID} and the group number, the supply voltage, the \ac{MAC} address and status flags are advertised by each badge.
The proximity to other badges can be determined through scanning for surrounding Bluetooth devices. When a Bluetooth device is discoverd, the \ac{RSSI} value is reported. The \ac{RSSI} value is proportional to the received signal strength \cite{Favalli07:WC}. Based on this value, the distance to the discovered device can be approximated with a radio propagation model \cite{Xu10:DMM}\cite{Jianyong14:RBB}\cite{Jianwu09:ROD}. After the received advertising packet is decoded, the group membership is checked to ignore unrelated badges. To save storage space, the \ac{ID} of the discovered device is stored instead of the 6-byte \ac{MAC} address.

To record vocal activity, the analog \ac{MEMS} microphone SPU0414HR5H-SB from Knowles is incorporated. It is a high-performance, low-power acoustic sensor with an analog output pin \cite{SPU0414HR5H-SB:PDS}. The output signal of the microphone is amplified and low-pass filtered by an external operational amplifier. The resulting signal is routed to an \ac{ADC} pin of the BMD-200-B. An external voltage regulator (LP2985) supplies the microphone and amplification circuit with 1.8 Volts.

Moreover, the ultra-low-power, three-axis accelerometer LIS2DH12 from STMicroelectronics is integrated to detect and record movement. The accelerometer is configured via the \ac{SPI} bus and supports various operating modes, datarates and the capability to generate interrupts when preconfigured events, such as free-fall or motion, are detected. Additionally, the accelerometer has a built-in 32-level \ac{FIFO} buffer for the acceleration data. This feature can be used to reduce the reading frequency of the \ac{SPI} bus master because up to 32 data points can be retrieved with only one \ac{SPI} transfer \cite{LIS2DH12:DS}.

The power supply of the badge is provided by a coin cell and can be turned on/off by an external power switch. For status indication and signaling a green and a red \ac{LED} are integrated.

\subsection{Location beacons}
An important parameter for the analysis of social interaction data is the absolute location of the people wearing a badge. To measure the absolute location in a room or a building, different approaches can be applied on basis of the \ac{RSSI} value received from location beacons. Location beacons are stationary mounted devices with known position and periodically transmit a uniquely assignable data packet. The badges described in section \ref{sec:Badges} or iBeacons can be configured to act as location beacons by assigning an \ac{ID} greater or equal to 16,000.

Several localization methods can be used. The naive method is to assign a person to the closest location beacon \cite{Lederman18:RAU}. Another potential localization technique used by Yang et al. \cite{Yang12:LIF} is \ac{RSSI} fingerprinting. This approach requires a map of the building and is divided into two steps. In the first step, \ac{RSSI} values are recorded at different positions of the map and stored together with the position. In a second step, where the actual position should be determined, the currently recorded \ac{RSSI} values are matched with the \ac{RSSI} values of the first step to obtain the most likely position.
A further feasible method for localization based on \ac{RSSI} values is trilateration \cite{Shchekotov14:ILM}. On the basis of a radio propagation model, the distance to the location beacons can be estimated. When the distance to at least three of them is known, the absolute position in a two-dimensional map can be computed with trilateration.

Due to the limited \ac{NVM}, the different devices discovered in a Bluetooth scan have to be filtered according to their relevance. Location beacons and badges with a high \ac{RSSI} value are more important than badges with a low \ac{RSSI} value. 
%To obtain reliable location information location beacons are prioritized. 
This prioritization is accomplished by a sorting mechanism that is described in section \ref{sec:Data recording}.

\subsection{Hubs}
\begin{figure}[h]
\centering
\setlength{\fboxsep}{0pt} 
\setlength{\fboxrule}{2pt} 
\fbox{\includegraphics[height=7cm]{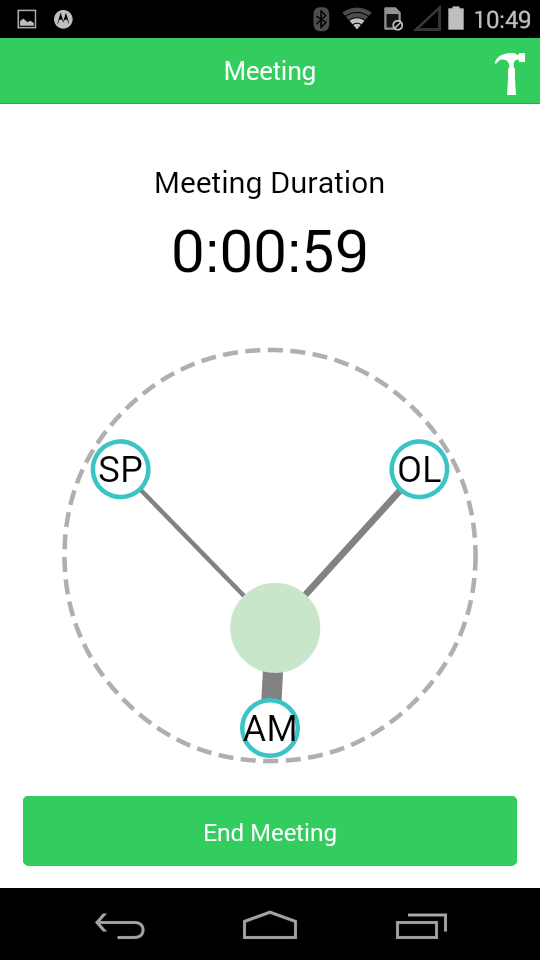}}
\caption[Mobile hub]{\textbf{Mobile hub}. \cite{Lederman18:RAU}. The mobile application that can act as hub and as real-time visualization tool.}
\label{fig:mobile_hub}
\end{figure}

Hubs are used to control, activate and retrieve the recorded data from the badges. There are two different types of hubs used in the project. A Python based implementation and a mobile hub application. In addition to the actual badge control and data request, the mobile hub can be configured as a real-time feedback visualization tool that is shown in Figure \ref{fig:mobile_hub}. The so called Round Table-visualization displays the participants of a meeting around a round table and a green filled circle which is connected to the individual participants by a line. The circle and the lines represent the partiticipation of each user \cite{Lederman18:RAU}. The implementation of the mobile application can also be found on GitHub.

The Python based implementation of the hub software is deployed on a Raspberry Pi with a \ac{BLE} interface. The code is wrapped in a Docker container to simplify the installation and setup. There are two operating modes of the Python based hub: Standalone and server mode. In standalone mode only one hub is active. It retrieves a list of available badges from a local file. After the activation of the available badges, the received data are also stored in local files. In server mode, one or more hubs communicate with a server that manages a list of available badges. In this setup, the received data from the badges are forwarded to the server where the data are processed in a second step. In a setup with multiple hubs, it is essential to synchronize the local clock of the hubs by using the \ac{NTP} \cite{Lederman16:OHP}.

\subsection{Server}
After the recorded data of the platform are stored on the server, a server-side application processes the data in real-time. This analytics engine analyzes the different data and generates statistics about the communication behavior and pattern of the participants. The real-time computed data can be retrieved via \acp{API}.
In addition, the server provides monitoring information about the functionality of system components such as badges, location beacons and hubs.

\section{Firmware architecture}
The main topic of this work is the reimplementation of the badges' firmware. The firmware should be optimized for modularity, extensibility and maintainability. 
As programming language C is used. C is a high-level programming language that is often applied for the programming of embedded systems with limited resources. It allows the developer to interact with the hardware in an easy way and the code generated by the C compilers is in general very efficient \cite{Barr06:PES}. 
This section describes the requirements of the firmware as well as its architecture and the interaction of the different modules.

\subsection{Requirements}
The badge offers the possibility to record the following data: the audio signal from the microphone, the \ac{RSSI} values of surrounding Bluetooth devices, the battery voltage, the raw acceleration data and interrupt events generated by the accelerometer when a predefined acceleration threshold is exceeded.
Due to the diversity of these data sources an appropriate and efficient way to store the datasets in the \acp{NVM} is needed. Therefore, a filesystem with separated partitions for each data source was implemented.
To simplify the implementation of the filesystem, a single interface to interact with the storage is required. The flash-memory only supports the storing of words (4 bytes), not single bytes, and the associated flash pages have to be erased before the words can be stored to the flash cells. In contrast to this, the \ac{EEPROM} offers the possibility to store data at byte level and prevents the application from erasing the \ac{EEPROM} cells before storing to them. As a result, a uniform, easy to use storage interface was developed that combines the flash-memory and the \ac{EEPROM} to one virtual memory with byte level access.

Data handling consists of two steps: sampling, which is the recording of data, and processing, which includes, for example, sorting and the storage of the recorded data.
The processing step requires closed data units. Therefore, the sampled data is splitted into data chunks with a certain size. Each data chunk consists of a header and the actual data points. The header contains information about the chunk, such as a timestamp and how many data points are included in this chunk. 
An efficient mechanism to decouple the data sampling from the data processing is needed because the data processing, such as storing, could be a time-consuming procedure and should not influence or delay the data sampling. Therefore, a so called chunk-\ac{FIFO} was developed to allow an efficient data sampling and to decouple data sampling from data processing.

Before any type of structured data, for instance a chunk of audio data, can be stored to the \acp{NVM} via the filesystem, the structured data must be serialized or encoded. The result of serialization is a sequence of bytes representing the structured data. In C a sequence of bytes can be expressed through a byte-array.
This serialization process is crucial for the efficiency of data storage. The less bytes are required to represent the structured data, the more data can be stored in the \acp{NVM}. To reconstruct structured data from a serialized byte sequence, deserialization or decoding has to be applied. 
The process of de-/serialization has to be generic to facilitate the declaration and use of new structured data types. Otherwise the developer has to implement a de-/serialization procedure for every type of data. 

Another application of the serialization and deserialization procedure is the data exchange between the badges and the hubs. As described in the previous section, the Nordic UART Service is used for the data exchange between the devices. This \ac{BLE} service allows the transmission of raw bytes that have to be interpreted by another instance such as a data exchange protocol. Ultimately, a protocol is nothing else than a description of how to interpret bytes. The data representation for storing and transmission does not have to be equal. One reason for this is that different protocol versions can have different data representations. Therefore, the representation of data that are stored on the badge is independent from the data representation of the protocol to prevent compatibility problems.

There are many different techniques to represent data structures or objects and to serialize them for transfer or storage. The \ac{XML} defines a language for data representation and serialization in a human- and machine-readable format. The representation that is produced during the serialization has usually a larger size compared to other serialization techniques such as the \ac{JSON} due to redundant tags in \ac{XML}. \ac{JSON} represents data in the human- and machine-readable object notation of JavaScript. It is a popular alternative to \ac{XML} because of its better readability and smaller serialized representation. Another serialization technique developed by Google is Protocol Buffers. On the basis of a predefined schema, classes and functions to serialize and deserialize structured data are generated by the Protocol Buffers compiler. It is characterized by a small binary representation of the data and a fast serialization. The mentioned techniques support various programming languages, such as C++, Java and Python \cite{Maeda12:PEO}.
As part of this work, a serialization library called Tinybuf was implemented. It is orientated at Google's Protocol Buffers library.
Tinybuf was developed to overcome some drawbacks of Protocol Buffers: High \ac{RAM} allocation for the serialized message, large overhead produced by additional field identifiers and incompatibility with the previous protocol implementation.
Tinybuf uses a parser to analyze a schema file with the definition of data structures and generates source code for the programming languages C and Python. It is optimized for efficient data serialization to enable an effective data storage and transmission. The Tinybuf library is described in section \ref{sec:Data serialization}.

Testing plays an important rule during the development of software. Implementation errors can lead to unpredictable behaviors and complete system failures which are unacceptable in applications that are deployed for instance in airplanes or cars \cite{Desikan06:STP}.
Although the firmware for the badges is not as critical as applications for airplanes, it is necessary to verify the correct behaviour of the implemented software. 
There are two main concepts to verify the correctness of software components: Static analysis and dynamic testing. Static analysis can be performed by humans or by automated static analysis tools. Humans can apply techniques such as desk checking and code walkthrough to detect implementation bugs \cite{Desikan06:STP}. In general, static analysis tools are applied directly to the source code to be checked, not to the compiled or executable binary. 
These tools are established in the development of applications with high security requirements \cite{Chess04:SAF} or in safety-critical systems for instance in the automotive and aerospace industry \cite{Blanchet03:ASA}.
On the other hand, dynamic testing methods actually run the code against predefined test cases to verify its functionality: The code has to produce the results that are expected by the test cases. 
For instance, unit, regression and integration tests are types of dynamic testing. Unit tests consist of a set of independend test cases that verify the functionality of small functional units, such as functions or modules \cite{Hamill04:UTF}. Regression tests ensure that an already existing functionality is not influenced when the code is changed due to bug fixes or code extension. Integration tests focus on the interaction between multiple components of the system to verify that these components work together as specified \cite{Desikan06:STP}. 

In this work, a dynamic testing framework based on Google Test \cite{Google08:GT} was developed to verify the functionality of the implemented modules. The testing framework combines Google Test with a code coverage analysis tool and the capability to simulate interrupts to test the asynchronous parts of the code. The framework is described in section \ref{sec:Testing framework}.

\subsection{Modules overview}
This section explains the general module interaction and the main implementation concepts that are used.
The functionality of the application can be divided into the following individual functional units: Recording, storage, streaming and exchange of data. Each of these units consists of one ore more interacting modules. Beside these main functional units, additional modules are required for the correct operation of the application and to facilitate the development. 
An instance that provides reliable time basis is such an additional module. Accurate timestamps are needed to assign the correct time to a data chunk. This is neccessary to combine and correlate the data recorded from different badges in a second step. Furthermore, a timeout mechanism was implemented that calls a timeout handler function if a predefined time interval elapses. The mechanism can be deferred by calling a reset function for the timeout. One application of the timeout mechanism is, for example, the automatic stop of data recording after a predefined interval if no connection between the badge and a hub has been established within this interval.
Another crucial aspect for the correct behavior of the badge is to ensure the proper operation of the \ac{HW}. Therefore, a selftest module provides the capability to test selected peripheral components, such as the \ac{EEPROM}, the flash-memory, the microphone, the battery measurement and the accelerometer. The incorporated \acp{LED} signal the testing process and indicate an error if necessary.
For development and test purposes, the \ac{UART}-interface can be enabled manually by using the corresponding compile target in the Makefile. Via the \ac{UART}-interface the badge can communicate with a serial monitor such as Cutecom for debug or logging purposes. Additionally, commands can be sent via the \ac{UART}-interface to the badge to perform specific actions, for example restarting the badge.

\begin{figure}[h]
\centering
%\fbox{\includegraphics[page=1, trim=2.5cm 25.5cm 2.9cm 2.6cm]{img/code.pdf}}
\includegraphics[page=1, trim=2.5cm 25.3cm 2.6cm 2.3cm, clip, width=14cm]{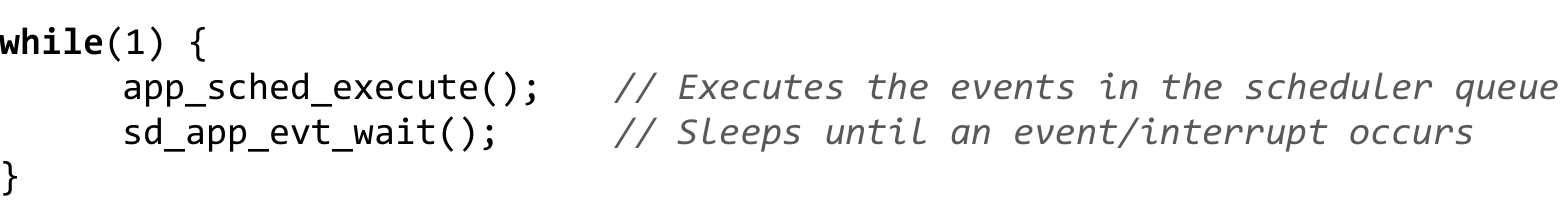}
\caption[Main loop]{\textbf{Main loop}. The main loop with scheduler execution and sleep function.}
\label{fig:main_loop}
\end{figure}

The programming paradigm used for the application is an event-driven approach. Event-driven software architectures are based on the processing of events. When an event, such as a timer event or an \ac{I/O} pin interrupt, occurs, a registered callback function is invoked to handle the event \cite{Dabek02:EDP}. The callback function is executed in a predefined priority context and suspends lower priority contexts such as the main context that has the lowest context priority.
In general, the execution time of callback functions has to be very short to minimize the delay of lower priority contexts or to prevent the loss of new callbacks with the same or lower priority. The number of available priority levels for the application depends on the \ac{HW}.
Nordic Semiconductor provides a \ac{SDK} that abstracts the \ac{HW}-interfaces in drivers, includes useful libraries and simplifies the usage of the Bluetooth connectivity via a so called SoftDevice. SoftDevices are precompiled and linked binary software components with an embedded \ac{BLE} protocol stack. The SoftDevice type S130 is used for the badges. This SoftDevice type enables a \ac{BLE} application to act as central, peripheral, broadcaster or observer device.
The latest \ac{SDK} version that supports the incorporated nRF51822 chip is v12.3.0 \cite{NordicSemiconductor18:INF}.
The \ac{SDK} contains two modules that are crucial for the previous described event-driven programming paradigm: The app-timer and the app-scheduler module.
The app-timer module uses the \ac{RTC} to generate interrupts or events at configurable time intervals. These events are handled by a previously registered callback function. The callback functions are executed in a low priority context to enable the handling of higher priority events in the meantime.
As already mentioned, it is crucial that the time for the execution of callback handlers/functions is short. There are cases, for example when an event triggers a storage operation, where the time required for the callback function would be too long. The app-scheduler module elimates this problem by transferring the execution from the callback context to the main context. The concept of the scheduler is based on an event queue.  Events are inserted into the scheduler \ac{FIFO} queue. The main context processes the events from the queue by deleting them and calling the corresponding handler functions. Another application of the scheduler is the postponement of functions that cannot be executed because the required resources are currently not available.
Due to this event-driven approach, the main function that is invoked at application-start only consists of initialization functions and the loop that is shown in Figure \ref{fig:main_loop}.

The \textit{app\_{}sched\_{}execute()} function reads out available events from the scheduler queue and calls the corresponding handler functions. The \textit{sd\_{}app\_{}evt\_{}wait()} function is an SoftDevice \ac{API} function that enters sleep mode until an interrupt occurs. This is important to reduce power consumption of the chip.
When the \ac{CPU} of the nRF51822 chip is running, a typical value for the current consumption is about 4.2 mA whereas the sleep mode consumes only about 3.8\textmu{}A \cite{NRF51822:PS}.

\begin{figure}[p]
\centering
%\fbox{\includegraphics[page=1, trim=1.8cm 6.3cm 5.3cm 2cm]{img/modules_overview.pdf}}
\includegraphics[page=1, trim=1.6cm 6.3cm 5.5cm 2cm, clip, height=20cm]{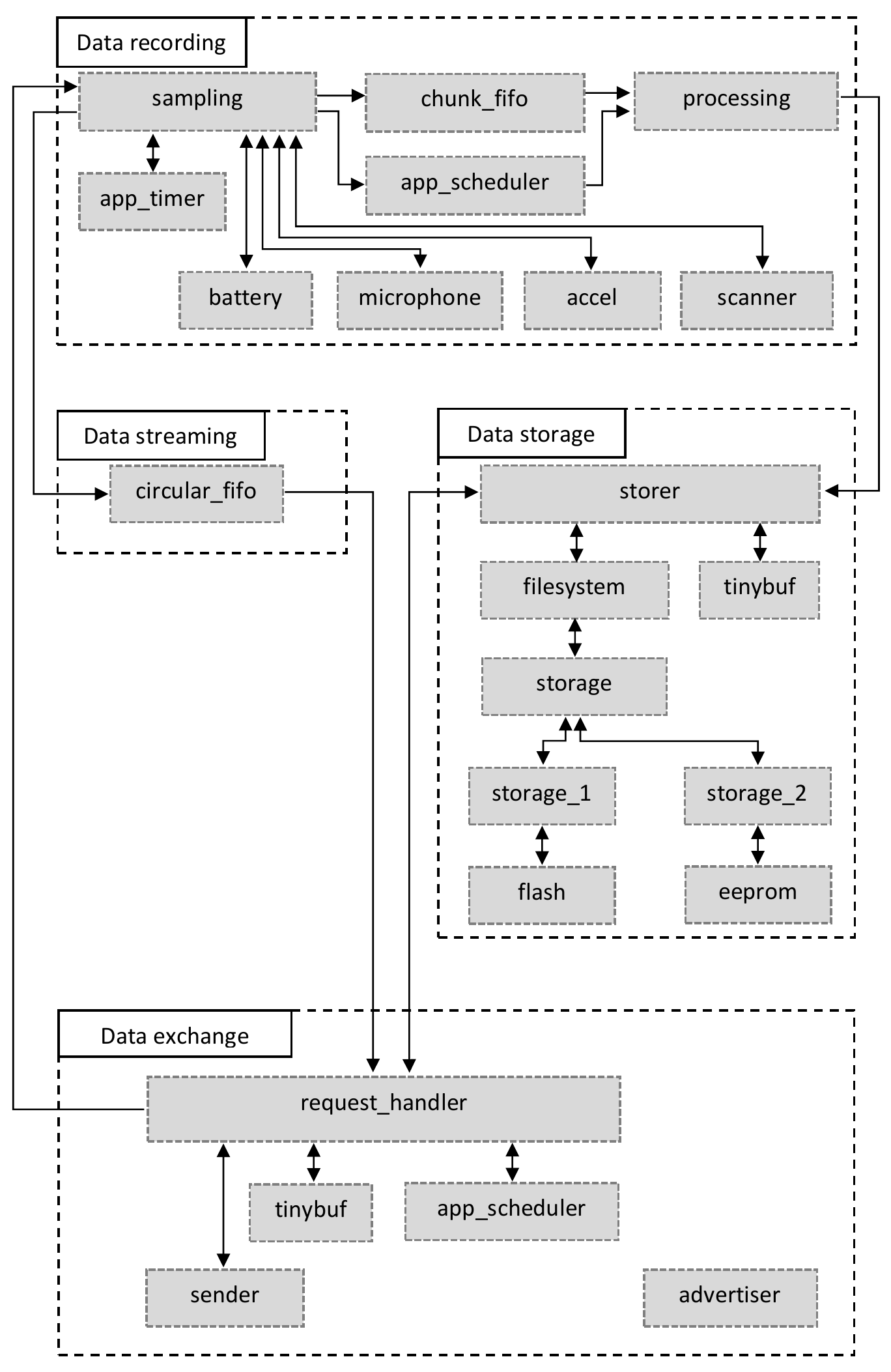}
\caption[Modules overview]{\textbf{Modules overview}. The main functional units data recording, data streaming, data storage and data exchange consist of one or more modules to provide the functionality. The arrows illustrate the dataflow and/or the controlflow between the components. In this overview only the modules are depicted that are necessary to understand the general behavior of the firmware. }
\label{fig:modules_overview}
\end{figure}

Figure \ref{fig:modules_overview} gives an overview about the interaction between the functional units: data recording, data streaming, data storage and data exchange. Furthermore, the different modules of each functional unit are displayed. In the following a brief explanation of their functionality is presented.

The data recording unit includes the sampling, app-timer, app-scheduler, chunk-\ac{FIFO} and processing module. Additionally, it uses the \ac{HW} abstraction modules for the data sources battery, microphone, accelerometer and scanner. The sampling module controls the data recording process. It collects data from the available data sources at predefined time intervals through the app-timer module or at occurness of specific events such as a motion-interrupt from the accelerometer. 
As already mentioned, the sampled data is splitted into data chunks for each data source. For an efficient and more importantly synchronized exchange of the data chunks between the sampling and the processing module, the chunk-\ac{FIFO} module is applied. The processing module reads the available data chunks from the chunk-\ac{FIFO} and processes the chunks, for instance sorting the data or storing to the \acp{NVM}. In general, this processing step is time consuming and should not be handled in a high priority context. Therefore, the execution is transferred to the main context by the app-scheduler module.
The \ac{HW} abstraction modules represent the functionality of the corresponding \ac{HW} components. The battery module provides a function to read the supply voltage of the battery while the microphone module reads the audio signal from the analog microphone. Both modules use the \ac{ADC} to generate the data. To synchronize the access on the \ac{ADC} an abstraction layer based on the \ac{SDK} \ac{ADC}-driver was implemented. All required \ac{SDK} drivers have to be enabled in the sdk-config header file of the project.
The incorporated accelerometer is controlled by the accelerometer (accel) module. This module enables the access to the recent acceleration data and calls a configurable callback function when the accelerometer generates an interrupt event. To interact with the accelerometer chip, the \ac{SPI}-driver of the \ac{SDK} is used. Since the \ac{SPI}-driver is also required by the eeprom module, an additional abstraction layer was implemented to synchronize multiple \ac{SPI}-driver access. Furthermore, the abstraction layer includes an automatic control of the \ac{SS}-pin to choose the correct \ac{HW} device for \ac{SPI} communication.
The last data type that can be recorded is the proximity to other badges.
The proximity to surrounding Bluetooth devices can be reconstructed using the RSSI values from a Bluetooth scan. The \ac{BLE} scan functionality is provided by the SoftDevice. To facilitate the interaction with the SoftDevice, a \ac{BLE} module was implemented that abstracts only the needed \ac{BLE} functionality, such as scanning, advertising and bi-directional communication. The scanner module, that is associated to the data recording unit, is based on this \ac{BLE} abstraction and provides the functionality to scan for surrounding Bluetooth devices, such as location beacons or other badges.
A detailed description of the sampling and processing strategy of each data source is presented in section \ref{sec:Data recording}.

After the data from the different data sources have been recorded, they can be directly transmitted to a central \ac{BLE} device and/or stored to the \acp{NVM} for later transmission. 
The first case, in which the data arrive continuously, is called data streaming \cite{Babcock02:MAI}.
This data streaming process is handled by the data streaming unit in cooperation with the data exchange unit.
The circular-\ac{FIFO} module is the only module in the data streaming unit. In contrast to the \ac{FIFO} implementation of the \ac{SDK}, the circular-\ac{FIFO} allows synchronized overwriting of unread data if the \ac{FIFO} is full.
This has the advantage that the data in the \ac{FIFO} is always up to date.
Each data source has its own circular-\ac{FIFO}, which contains single data points (e.g. the current microphone values) and no complete data chunks. The circular-\ac{FIFO} as well as the streaming messages are presented in section \ref{sec:Data streaming}.

In the second case, the recorded data are stored in the \acp{NVM} to be transmitted at a later timepoint. The data storage unit is responsible for the storing process of the generated data chunks. The storer module controls the storage to and the reading from the \acp{NVM}. Therefore, the tinybuf module is used to efficiently serialize the data chunks to a byte sequence that can be stored to the \acp{NVM} via the filesystem module.
The filesystem consists of multiple seperated partitions for each class of data, such as the microphone data or the accelerometer data etc.
The filesystem module is based on the storage module, which combines multiple memories to one virtual memory. Between the actual \ac{HW} abstraction modules for the memories and the storage module, another abstraction layer is integrated to enable an uniform interface for the different memory types. The flash module is the \ac{HW} abstraction of the integrated flash-memory and enables certain flash operations, such as erasing pages and storing or reading words. The interaction with the flash-memory is done by the fstorage module of the \ac{SDK} and the SoftDevice. To minimize the SoftDevice conflicts between \ac{BLE} activity and flash operation, the fstorage module splits the requested flash operation into smaller units that are successively processed.
The \ac{EEPROM} is controlled by the eeprom module and enables the application to store and read single bytes or byte sequences. As already mentioned, the communication between the nRF51822 and the \ac{EEPROM} chip is based on the \ac{SPI}. To synchronize the \ac{SPI} access between the eeprom and the accelerometer module, the abstraction layer on top of the \ac{SPI}-driver is applied.
To retrieve the data chunks again from the \acp{NVM}, the serialized data chunks are read from the corresponding filesystem partition and deserialized via the tinybuf module. In section \ref{sec:Data storage} the functionality of the filesystem as well as the uniform storage abstraction is explained. 

Ultimately, data has to be exchanged with a central \ac{BLE} device such as a hub. The data exchange unit is responsible for this task (see Figure \ref{fig:modules_overview}). Advertising related functionality, for instance the advertising packet management, is implemented in the advertiser module.  The sender module enables the bi-directional communication during an established \ac{BLE} connection between a central \ac{BLE} device and the badge through the Nordic UART Service. The advertiser module as well as the sender module is based on the \ac{BLE} module that abstracts the interaction with the SoftDevice.
The entire data exchange process is controlled by the request handler module. This includes the processing of received messages, the reading of stored data chunks and the transmission of stream or data chunk messages.
The general procedure is the following: The sender module receives a message or notification from the remote \ac{BLE} device and generates a callback for the request handler module. The request handler module processes this message by deserializing it with the tinybuf module to an interpretable request structure. This request is analyzed in a second step and if necessary a response for the remote device is generated and transmitted. In general, the process of analyzing the request and generating the response is time consuming and is therefore transferred to the main execution context through the app-scheduler module. Additionally, the app-scheduler module is used to suspend an operation that is currently unavailable and to retry it at a later timepoint. 
After the reponse is generated, the tinybuf module is used to serialize the response to a byte sequence that can be transmitted through the sender module. Section \ref{sec:Data exchange} presents the data exchange unit and the corresponding modules.

\section{Testing framework} \label{sec:Testing framework}
As part of this work, a testing framework was developed to apply dynamic testing methods to the implemented software.
A testing framework is a software tool that provides an environment to easily create and execute tests and reports the results of these tests.
Usually, unit tests are created simultaneously with the development of the application software.
Unit tests are based on the application modules to be tested, but exist only in the test framework and not in the application itself. A unit test checks a specific behavior of the application software. If the test is succesful, this specific behavior of the application is verfied. In general, the most basic functionality of the application should be tested first, followed by more complex tests that may combine different functional units of the application. One advantage of a well-maintained testing framework is the ability to immediately verify code changes, which enables faster application development \cite{Hamill04:UTF}. 
Altough the implementation of unit tests is an additional expenditure during the implementation, a case study in \cite{Osherove15:TAO} showed that the overall time to provide a bug free application can be reduced when unit testing is applied during the development process.

The basis for the testing framework is Google Test. Google Test is a cross platform testing framework written in C++. It is compiled and linked together with the application modules to be tested. The framework enables an easy declaration of tests and provides a lot of useful features. In Google Test, one single test is called test and the set of various tests associated to one component is called test case.
The basic concept is the following: The user creates a C++ file to setup a test case for a specific component of the application. Within this file one or more single tests are defined to verify the component's functionality.
To verify the functionality of a component or a function,  assertions are made about its behavior, which have to be fulfilled.
Google Test displays a failure message when the assertion fails. This failure message contains the line number where the test failed, additional information about the cause of failure and optionally a custom failure message.
Assertions are divided into two groups: Fatal and nonfatal assertions. When a fatal assertion fails, the current test is aborted, whereas a nonfatal assertion would not abort the current test. An overview of common used assertions is presented in Table \ref{tab:assertions}.

\begin{table}[h]
\renewcommand{\arraystretch}{1.5}
\centering

\begin{tabular}{c|c|c} 
	\textbf{Fatal assertion}	& \textbf{Nonfatal assertion} & \textbf{Verifies} \\ \hline
	\textit{ASSERT\_{}TRUE(condition)} &	\textit{EXPECT\_{}TRUE(condition)} &  \textit{condition} is true \\ \hline
	\textit{ASSERT\_{}FALSE(condition)} &	\textit{EXPECT\_{}FALSE(condition)} &  \textit{condition} is false \\ \hline
	\textit{ASSERT\_{}EQ(val1, val2)} &	\textit{EXPECT\_{}EQ(val1, val2)} &  \textit{val1} == \textit{val2} \\ \hline
	\textit{ASSERT\_{}NE(val1, val2)} &	\textit{EXPECT\_{}NE(val1, val2)} &  \textit{val1} != \textit{val2} \\ \hline
	\textit{ASSERT\_{}LT(val1, val2)} &	\textit{EXPECT\_{}LT(val1, val2)} &  \textit{val1} \textless{} \textit{val2} \\ \hline
	\textit{ASSERT\_{}LE(val1, val2)} &	\textit{EXPECT\_{}LE(val1, val2)} &  \textit{val1} \textless{}= \textit{val2} \\ \hline
	\textit{ASSERT\_{}GT(val1, val2)} &	\textit{EXPECT\_{}GT(val1, val2)} &  \textit{val1} \textgreater{} \textit{val2} \\ \hline
	\textit{ASSERT\_{}GE(val1, val2)} &	\textit{EXPECT\_{}GE(val1, val2)} &  \textit{val1} \textgreater{}= \textit{val2}

\end{tabular}

\caption[Assertions]{\textbf{Assertions}. Each fatal or nonfatal assertion performs a specific type of verification.}
\label{tab:assertions}

\end{table}

\begin{figure}[h]
 \begin{minipage}[c][4.5cm]{0.4\textwidth}
 \centering
 \includegraphics[page=2, trim=2.5cm 22cm 10.3cm 2.5cm, width=\textwidth, clip]{img/code.pdf}\\
 \vfill 
 (a)
 \end{minipage}
 \begin{minipage}[c][4.5cm]{0.6\textwidth}
 \centering
 \includegraphics[width=\textwidth]{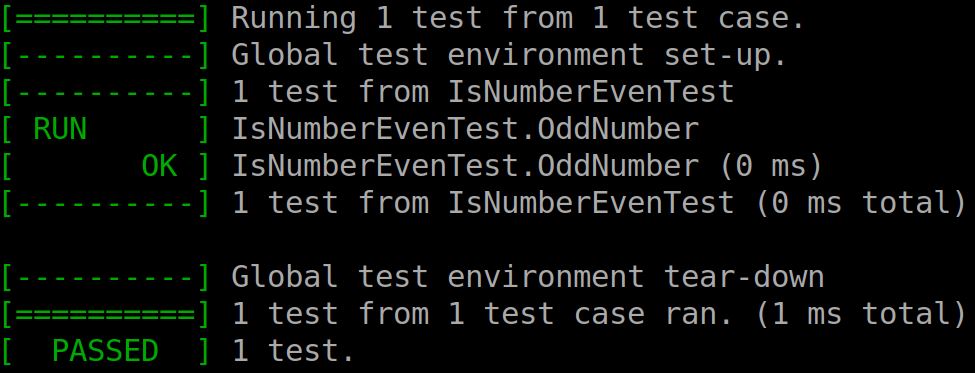}\\
 \vfill 
 (b)
 \end{minipage}
 \caption[Successful unit test]{\textbf{Successful unit test}. The correct implemented function that checks whether a number is even or not and a single test is shown in (a). The test report generated by the Google Test framework after running the test is depicted in (b).}
\label{fig:successful_unit_test}
\end{figure}

\begin{figure}[h!]
 \begin{minipage}[c][7.3cm]{0.4\textwidth}
 \centering
 \includegraphics[page=3, trim=2.5cm 22cm 10.3cm 2.5cm, width=\textwidth, clip]{img/code.pdf}\\
 \vfill 
 (a)
 \end{minipage}
 \begin{minipage}[c][7.3cm]{0.6\textwidth}
 \centering
 \includegraphics[width=\textwidth]{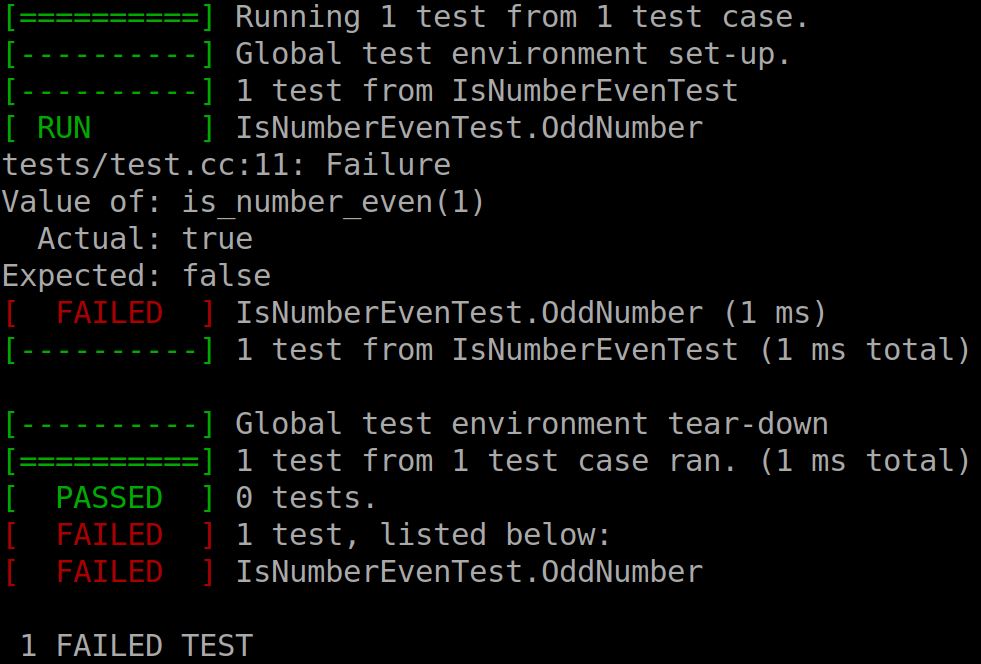}\\
 \vfill 
 (b)
 \end{minipage}
 \caption[Failed unit test]{\textbf{Failed unit test}. An incorrect implementation of the function that checks whether a number is even or not is shown in (a). After execution of the test, the Google Test framework reports the failure of the test (b).}
\label{fig:failed_unit_test}
\end{figure}

Each test uses one or more of these assertions to verify the functionality of a certain component. Figure \ref{fig:successful_unit_test} (a) shows a simple function to be tested and an associated test. The task of the function is to check whether an input number is even or odd by using the modulo operator. If the number is even, the function should return true, otherwise false.
The corresponding test is defined by the TEST(TestCaseName, TestName) macro and uses one of the mentioned assertions to verify a specific behavior of the function. In this case, the test verifies that the function returns false if the input number is 1.
To run the test case, the created C++ test case file, the corresponding modules to test and the Google Test framework modules have to be compiled and linked together. The framework invokes the created tests sequentially. The resulting test report is displayed in Figure \ref{fig:successful_unit_test} (b). The report shows the success of the test and thus also that the function fulfils the behavior asserted in the test.
In contrast, Figure \ref{fig:failed_unit_test} (a) shows the same function, but in this case it is implemented incorrectly: The condition in the \textit{if}-statement is exactly the opposite of the previous implementation.
If the same test is executed for the incorrect implementation of the function, the test fails. The report produced by the Google Test framework shows this failure (Figure \ref{fig:failed_unit_test} (b)).
It provides the line number where the test failed and information about the cause of the failure.

\vfil

\noindent With this information, the implementation bug can be fixed quickly.
Several tests within a test case should be independent of each other, otherwise it might be difficult to find the cause of a failed test. Especially, test of modules or functions with an internal state are affected by this problem. To ensure the same configuration or state for several different tests, the framework provides so called test fixtures: TEST\_{}F(TestCaseName, TestName). Before each test fixture is invoked by the framework, a \textit{SetUp}-function is called to reset the internal state or to create a consistent configuration \cite{Google08:GT}.

\newpage

Another related dynamic testing method is code coverage testing. This type of testing is characterized by specifically designed tests that cover a high percentage of the code. Code coverage can be refined into different types: statement or line coverage, path coverage, condition or branch coverage and function coverage. 
For line coverage, the tests are designed to execute each line of the code to be tested. One assumption is that the more lines covered, the better the functionality is tested.
Path coverage considers the code as individual logical paths that can be executed and not as single lines. This type of coverage is a stronger criterion than line coverage. 
Branch coverage is related to path coverage but provides a stronger criterion than path coverage. This is due to the fact that a different combination of conditions can lead to the same path being executed.
The function coverage represents the number of program functions that are covered by the tests. Furthermore, the number of function calls is monitored, enabling the developer to optimize functions that are called frequently \cite{Desikan06:STP}.

The technique applied for code coverage analysis is called instrumentation. In this technique, additional code or instructions are incorporated into the original source code. These instructions generate information about the instrumented code during runtime. The result of the analysis is the coverage in percent. For example, the line coverage is computed in the following way:
\begin{equation}
\text{Line coverage}=\frac{\text{Number of lines executed}}{\text{Total number of lines}}*100\%
\end{equation}
    
When the tests execute the instrumented code and it turns out that the code coverage is insufficient, either the existing tests have to be adjusted or new tests have to be added to increase the coverage. 
This process is repeated until the code coverage is sufficient. A set of effective tests is the result of this repeated process.

\begin{figure}[h]
 \begin{minipage}[c][3.5cm]{0.62\textwidth}
 \centering
% \fbox{\includegraphics[page=1, trim=3cm 13.3cm 15.8cm 4.7cm]{img/coverage/LCOV_1.pdf}}

 \includegraphics[page=1, trim=3cm 13.1cm 15.8cm 4.7cm, width=\textwidth, clip]{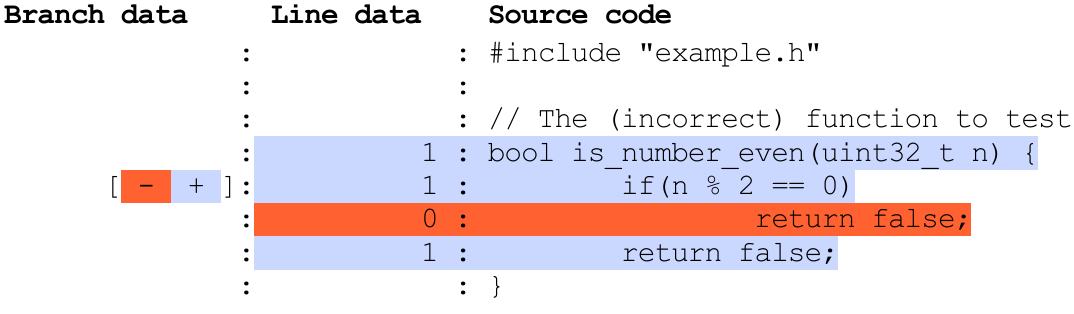} \\
 \vfill 
 (a)
 \end{minipage}
\begin{minipage}[c][3.5cm]{0.36\textwidth}
 \centering
 \includegraphics[page=1, trim=20.8cm 17cm 1.2cm 2cm, width=\textwidth, clip]{img/coverage/LCOV_1.pdf} \\
 \vfill 
 (b)
 \end{minipage}
 \caption[Example coverage analysis]{\textbf{Example coverage analysis}. The detailed source code coverage analysis is depicted in (a). It includes the source code itself, the number how often each line is executed (line data) and information about the branch coverage (branch data). In (b) the statistic about the entire source file coverage is displayed.}
\label{fig:example_lcov}
\end{figure}

In this work the coverage analysing tool gcov is used. This utility is included in the \ac{GCC}. To incorporate the instrumentation into the code, special compiler options have to be added: \textit{-fprofile-arcs -ftest-coverage}. 
During the compilation process for each source file, a .gcno file is generated that contains information that are necessary for gcov such as a flow graph of the program. 
While executing the compiled object files, a .gcda file is created for each of the source files. These .gcda files contain profiling information about the executed object files. 
Both, the .gcno and .gcda files are used by gcov to perform the actual coverage analysis \cite{Gcov05:GAT}. 
Since the generated results are difficult to read, lcov in combination with genhtml is applied to provide a more convenient visualization for the developer. Lcov collects the coverage information generated by gcov and writes them into a file. Based on this file, genhtml creates a \ac{HTML} file that presents the coverage information, such as line, function and branch coverage. \cite{Lcov08:LAG}. The \ac{HTML} file can be interpreted by any standard Browser such as Google Chrome or Firefox.
An example for the code coverage analysis and \ac{HTML} visualization is shown in Figure \ref{fig:example_lcov}. 
The function to be tested is the same as for the previous examples. 
This time the function has another implementation error: Instead of returning true when the condition in the \textit{if}-statement is fulfilled, the function returns erroneously false. 
The functionality of the function is verified with the same test as for the previous implementations: The test asserts that the function returns false when it is called with the input number 1. 
The implementation error has no effect on this assertion, because it is not reached during the test. Consequently, the test reports the correct functionality of the function, although there is obviously an error. 
To solve this problem, and ultimately detect the implementation error, one or more tests with different input parameters have to be added.
The code coverage analysis tool helps the developer to easily find such untested code lines or branches. Figure \ref{fig:example_lcov} (a) shows the coverage analysis generated with gcov, lcov and genhtml for the mentioned test. The red highlighted line containing the implementation error is never called during the test. The reason for this is pointed out by the branch data column: The condition of the \textit{if}-statement is never true. In addition to the detailed source code analysis, a statistic of the entire source file is generated. This is shown in Figure \ref{fig:example_lcov} (b) and includes statistic values for the line, function and branch coverage.
Therefore, the combination of the Google Test framework and gcov/lcov provides an useful tool for software development. The test framework verifies the functionality of the implementation and the coverage tool analyzes how much of the source code is actually executed by the tests.
This enables developers to create better tests and eventually more stable applications.

%Combining gcov/lcov and Google Test together in this framework: Software developers also use coverage testing in concert with testsuites, to make sure software is actually good enough for a release. Testsuites can verify that a program works as expected; a coverage program tests to see how much of the program is exercised by the testsuite. Developers can then determine what kinds of test cases need to be added to the testsuites to create both better testing and a better final product.

%LCOV. Nochmal auf vorheriger Bilder zurück referenzieren, um klar zu machen warum es Sinn macht (Nur ein Case getestet). Auch Schwierigkeit klar machen, dass die Tests gut genug sein sollen.

\begin{figure}[h]
 \centering
% \fbox{\includegraphics[page=1, trim=3.7cm 20.1cm 2.5cm 3.5cm]{img/accel_mock.pdf}}
 \includegraphics[page=1, trim=3.7cm 20.1cm 2.5cm 3.5cm, width=0.9\textwidth, clip]{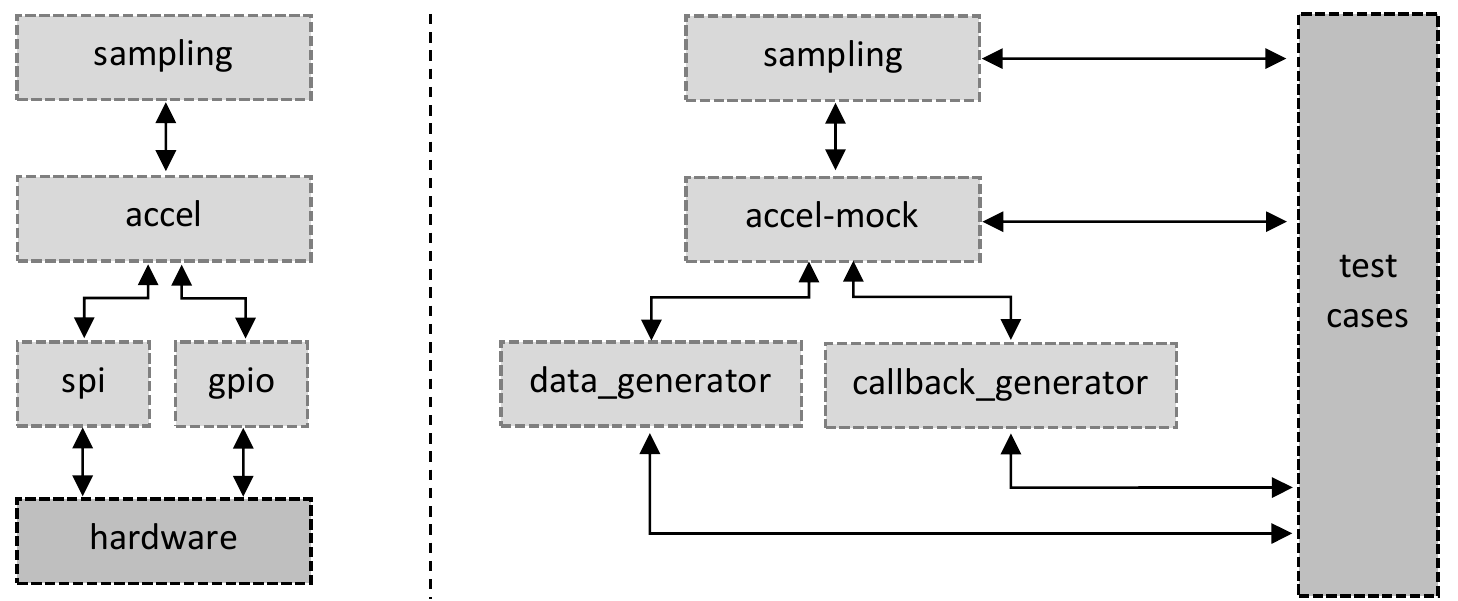} \\
 
 \caption[Accelerometer mock object]{\textbf{Accelerometer mock object}. The left part shows the sampling and accelerometer module when the application is directly running on the \ac{HW}. For the offline testing with the test framework, a \ac{HW}-independent mock object for the accelerometer has to be implemented.}
\label{fig:accel_mock}
\end{figure}

Modules or functions that are based on \ac{HW} or \ac{HW}-drivers cannot be verified with the testing framework because the required functionality is only available on the \ac{HW}. 
Examples for such \ac{HW} components are the \ac{EEPROM}, the flash-memory, the accelerometer, the microphone, the battery and the \ac{BLE} interface.
To verify the correct functionality of these \ac{HW} modules, their tests are integrated into the application running on the \ac{HW}.
The \ac{HW} modules have to be abstracted in so called mock objects, to test other units of the application that interact with these modules.
Since the programming language is C, mock objects are modules with the same interface/header-file as the \ac{HW} modules. Internally, the mock objects no longer access the \ac{HW}, but try to simulate it.
The first aspect of mock object simulation includes the generation of data. The second aspect is the simulation of asynchronous events such as interrupts.
For instance, the mock object for the accelerometer module has to provide both aspects: When the read function of the mock object is called, acceleration data have to be returned. In addition, the mock object has to generate motion interrupts under certain conditions.
The left part of Figure \ref{fig:accel_mock} shows the sampling module that uses the \ac{HW} dependend accelerometer module. To test the implementation of the sampling module, the mock object of the accelerometer module has to be created, which is shown in the right part of the picture. The functionality of the mock object itself is again verified with the test framework. 
The easiest way to generate data, e.g. accelerometer data, is to return predefined default values. In some applications this would already be sufficient, but there are applications that perform complex operations on the returned data, such as sorting or classification.
For these applications, specifically generated data must be returned to test the complex functionality.
To provide this possibility, an abstraction layer for data generation was implemented in this work: If the application calls a read function of the mock object, the function call is forwarded to the data-generator module. In addition to default values, function pointers can be set in this module. These function pointers enable test-dependent functions to be called to generate the data.

\newpage
To simulate asynchronous events, for example a motion interrupt of the accelerometer, the callback-generator module can be used. This module allows to easily register and create asynchronous events at specific configurable time points.
The event generating process is started by calling a trigger-function. The module manages an array of time points that are processed sequentially. These time points are defined during the initialization by the developer. At each time point, a previous registered callback function is called to simulate an asynchronous event. Internally, threads are used to generate this asynchronous behavior. To protect critical sections of the code from asynchronous execution, special synchronization mechanisms must be applied.
In the case of the accelerometer mock object, the callback generator module is used as follows: The test defines the time points at which the motion interrupts should be generated. During the initialization process of the accelerometer mock object, the trigger-function that starts the process is called. At the defined time points an event is generated and the interrupt handler function of the accelerometer mock object is invoked. The interrupt handler is invoked as on the \ac{HW}.

The developed test framework enables the capability for simple creation and execution of tests. In addition, the integrated code coverage analysis tool gives the developer an indication of the quality of the tests.
%To be able to test modules based on the \ac{HW}, mock objects can be used in combination with the data-generator and callback-generator modules.

%Module timer-lib (timing and synchronization mechanism via mutexes..) High priority context could not be interrupted by a lower priority context, these critical areas have to be saved with certain mechanisms.. CRITICAL-REGION-EXIT/ENTER: macro from the app-util-platform module, that 

\section{Data recording} \label{sec:Data recording}
\begin{figure}[h]
\centering
%\fbox{\includegraphics[page=5, trim=2.5cm 24.4cm 8.6cm 2.6cm]{img/code.pdf}}
\includegraphics[page=5, trim=2.5cm 24.4cm 8.6cm 2.5cm, clip]{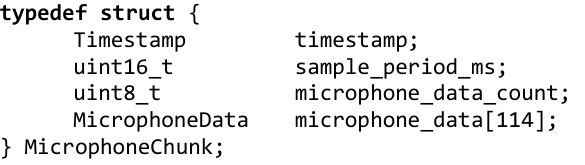}
\caption[Microphone chunk]{\textbf{Microphone chunk}. The figure shows the C structure for the microphone chunk. It consists of a timestamp, the sample period in milliseconds, the number of microphone data points in the microphone data point array and the microphone data point array itself. The MicrophoneData-type consist of one byte representing the audio data.}
\label{fig:microphone_structure}
\end{figure}

This section describes the data recording unit and the recording strategy of the different data sources in more detail. 
At the start of the application, the sampling module initializes the required timers, chunk-\acp{FIFO} and \ac{HW} abstraction modules. The sampling module provides an interface to start and stop data recording for a single data source.
The parameters required for each data source, such as the sampling period for the microphone, are passed via the start function. Depending on the data source, no timer, one timer or two timers are required for reading the data points. These timers are also launched in the start function. Furthermore, the start function sets up the data chunks, where the recorded data points are buffered. In addition to the actual data points, the chunks also contain metadata in which information such as a timestamp and, if applicable, the number of data points are stored. Chunks are internally represented as C structures. Figure \ref{fig:microphone_structure} shows an example of the chunk structure for the microphone data.
After a chunk has been filled with data points, it has to be processed, for example, stored.
The fast data generation of the sampling module must not be delayed by these time-consuming operations. Therefore, an efficient mechanism to exchange finalized chunks is required. Care has to be taken that no copy operations are used, because these operations are rather time-consuming. These requirements are implemented in a specifically designed chunk-\ac{FIFO}.
The functionality of the mechanism is as follows: 
During the initialization of the sampling module, a chunk-\ac{FIFO} for each data-source is created, in which the chunks are inserted sequentially. It allocates memory in \ac{RAM} for a given number of chunks. The chunk-\ac{FIFO} provides interfaces to open or close a chunk for writing or reading.
When opening a chunk for writing, a pointer to the next free position in the chunk-\ac{FIFO} is returned.
This pointer has to be casted in order to access the chunk-structure, for example a microphone-chunk.
This procedure prevents the already mentioned time-consuming copy operations.
If too less memory is allocated during the initialization or the reading of the finalized chunks from the chunk-\ac{FIFO} is performed too late, no writable chunks are available in the chunk-\ac{FIFO}.
In this case the pointer to the last finalized chunk is returned. This prevents overwriting of unread chunks. However, the last written chunk is overwritten by the new chunk.
The reason for this kind of implementation is the following:
If no chunk is currently available for writing, data recording should actually be stopped so that no new data points are generated.
As soon as space is available in the chunk-\ac{FIFO}, data recording should be restarted again. More complex mechanisms would have to be provided to automatically restart data recording. In order to keep the implementation simple, the described method of overwriting the last finalized chunk is used.
In both cases data loss occurs.
Therefore, the number of available chunks in the chunk-\ac{FIFO} must be selected appropriately during initialization.

After a chunk has been closed by the sampling module, a processing function is inserted into the scheduler. Each data-source has its own processing function, which is implemented in the processing module. The scheduler executes this function as soon as possible. 
The processing function reads finalized chunks from the chunk-\ac{FIFO} for the given data-source and processes them sequentially.
Depending on the data source, the processing differs. In the case of microphone chunks, the chunk structure is just serialized and stored by the storer module. In contrast to this, the data points of proximity chunks are sorted before they are serialized and stored.

%The different data-sources and their sampling strategy is described in the following.

\subsection{Proximity data}
The proximity to other people is an important variable to infer communication patterns and face-to-face interactions.
The proximity data can be used to determine which persons have most likely interacted with each other.
The badges retrieve the data by scanning for \ac{BLE} advertising packets of surrounding devices. 
Each badge advertises its packet every 200 ms. The advertising packet is broadcasted on all three advertising channels.
The parameters for the scanning of advertising-packets can be configured remotely.
The selectable parameters for the scan procedure are the interval, the window, the duration and the period. 
The scan interval describes the time interval between two active receiving windows. The active receiving time is represented by the scan window parameter. In each window only one of the three advertising channels is covered.
Therefore, the next window receives on the subsequent advertising channel of the previous window and starts again with the first one when the previous window has covered the last advertising channel \cite{Kindt18:NDL}.
This procedure is repeated until the scan duration is exceeded. Finally, the scan period describes the time interval at which a new scan procedure should be started.
Typical parameters are: an interval of 300 ms, a window of 100 ms, a duration of 3 seconds and a period of 15 seconds.

The proximity and approximate distance is finally calculated based on the \ac{RSSI} values of the received advertising packets.
The advertising packets of each badge contains an \ac{ID} and a group number. Both values are set by the hub during the activation of the badge. In addition to badges, location beacons can also be recorded in a scan. Location beacons are stationary \ac{BLE} devices mounted on a known position that periodically transmit their advertising packet. Both, iBeacons and badges can be deployed as location beacons. Each location beacon is also assigned an \ac{ID} and a group number.
The recording strategy of proximity data is as follows:
A timer is started that periodically invokes a callback function at the specified scan period.
In this callback a new scan-chunk of the scan-chunk-\ac{FIFO} is opened and the actual \ac{BLE} scan is started via the scanner module. Every time an advertising packet is received by the scanner module, a function is called that inserts the received \ac{ID} and \ac{RSSI} value into the chunk. If the group number of the received advertising packet does not match with its own group number, the badge ignores this device and no insertion into the chunk takes place.
Since the available \ac{RAM} is limited, multiple \ac{RSSI} values of the same device have to be aggregated during the scan. Currently two aggregation methods are supported: The average and the maximum of the \ac{RSSI} values. The scan-chunk has the capacity to store up to 255 devices during the sampling procedure. Beside the \ac{ID} and the aggregated \ac{RSSI} value, the number of received packets for the corresponding device is stored in the scan-chunk.
After the scan duration has been expired, a function is called that closes the current scan-chunk of the chunk-\ac{FIFO} and schedules the processing via the app-scheduler module.
The processing function, implemented in the processing module, reads out the available scan-chunk(s) from the chunk-\ac{FIFO}. 
In worst case, a scan-chunk consists of 255 discovered devices. To store a chunk like this into the \acp{NVM}, 1027 bytes are required: 6 bytes for the timestamp, 1 byte for the number of discovered devices, and for each of the 255 devices 2 bytes for the \ac{ID}, 1 byte for the \ac{RSSI} value and 1 byte for the number of received packets. Since the \acp{NVM} are also limited, 1027 bytes for one scan-chunk is not acceptable.
In general, not all of the discovered devices are important for the data analysis. Especially badges with a low \ac{RSSI} value are less important. A low \ac{RSSI} value usually indicates a greater distance and implies that there is probably no face-to-face interaction between the corresponding participants.
In contrast, location beacons are often characterized by a lower \ac{RSSI} value, because in a normal setup there are only few location beacons that are distributed over the entire operational area.
To calculate the absolute position of a person, the distance to at least three, better four location beacons is necessary. 
In order to prioritize a certain number of location beacons and to filter out badges with a low \ac{RSSI} value, a special sorting mechanism is applied: First of all, the location beacons, characterized by an ID greater or equal 16,000, are sorted to the beginning of the scan-chunk. After that, the location beacons are sorted by their aggregated \ac{RSSI} values to obtain the nearest location beacons at the beginning of the chunk. 
Finally, all remaining devices are sorted by their aggregated \ac{RSSI} values.
The result of this multistage sorting mechansim is a chunk that prioritizes location beacons and is sorted according to the proximity to other badges. 
Before serializing and storing to the \acp{NVM} using the storer module, the sorted scan-chunk is reduced by only taking the first 29 devices. The number of 29 devices has been adopted from the previous firmware implementation, but can be changed at any time if required.

\subsection{Audio data}
In addition to the proximity between persons, acoustic information is essential to analyze the communication patterns, such as turn-taking and speech overlap.
To record audio data, the analog microphone of the badge is used.
The microphone transforms the audio signal into an analog electrical signal on its output pin. This signal is amplified and low-pass filtered by the operational amplifier circuit of the badge. The cutoff-frequency of the low-pass filter is approximately 340 Hz.
According to the Nyquist-Shannon sampling theorem, the sampling frequency must be higher than twice the highest frequency occurring in the signal. Otherwise, aliasing effects may appear \cite{Marks12:ITS}. Therefore a sampling frequency of 700 Hz is used.
The pre-processed signal is digitized by the \ac{ADC}. The reference voltage for the \ac{ADC} is the supply voltage of the microphone and amplifier circuit, which is regulated to 1.8 Volts by a regulator.
The resolution of the digital representation is 8 bit. 
The zero point at the output of the operational amplifier circuit is 0.9 Volts. This voltage corresponds to a digital representation value of 128. 
The actual digital audio signal is the absolute difference between the current \ac{ADC} value and this zero point.

For the analysis of communication patterns, the recording of the actual content of the conversation is not required. Only certain language characteristics are relevant, such as how often, how long and at what volume the participants speak.
Therefore, to decrease the amount of data that has to be stored and finally transmitted, the raw microphone amplitude is averaged over 50 ms \cite{Lederman18:RAU}. This averaging period can be setted remotely.
To control the data-recording of the microphone, the sampling module provides a start- and a stop-function.
In the start-function a microphone-chunk of the microphone-chunk-\ac{FIFO} is opened and two timers are started. The first timer invokes a callback function every 1.42 ms (\(\widehat{=}\) 700 Hz). This callback function reads the raw amplitude via the microphone module. The microphone module uses the underlying \ac{ADC}-driver to generate this amplitude.
The second timer calls a function at the specified averaging period, for example every 50 ms. In this function, the previous read amplitudes are averaged and inserted into the microphone-chunk.
As soon as the microphone-chunk is filled, the chunk is closed and a new one is opened for further data recording. Additionally, the corresponding processing function is inserted into the scheduler.
This function reads out the available chunks from the microphone-chunk-\ac{FIFO}, serializes and finally stores them to the \acp{NVM} via the storer module.

\subsection{Acceleration data}
Another useful information about the behavior of participants are their patterns of physical movement.
This can be used, for example, to determinie their physical activity intensity or the number of steps.
A three-axis accelerometer is used to record movement information.
The incorporated accelerometer (LIS2DH12) of the badge offers a wide range of functionalities. 
The accelerometer supports four operating modes: Power-down, low-power, normal and high-resolution mode. In power-down mode, most of the internal units of the accelerometer are switched off to reduce power consumption. In this mode, no acceleration data are available and the power consumption of the accelerometer is reduced to approximately 0.5 \textmu{}A \cite{LIS2DH12:DS}.
The low-power mode provides a resolution of 8 bits for the acceleration. The resolution refers to the configured full scale value. A full scale value of \(\pm\)2 \ac{g}, \(\pm\)4 \ac{g}, \(\pm\)8 \ac{g} or \(\pm\)16 \ac{g} can be selected. In normal mode the resolution is 10 bits and in high-resolution mode 12 bits. The power consumption in these three modes depends on the datarate at which the acceleration is measured. Table \ref{tab:accelerometer_power_consumption} gives an overview of the power consumption.

\begin{table}[h]
\renewcommand{\arraystretch}{1.3}
\centering
\scalebox{0.9}{%
\begin{tabular}{c|c|c|c} 
\textbf{Datarate} \textbf{(Hz)}	& \textbf{Low-power mode} \textbf{(\textmu{}A)} & \textbf{Normal mode} \textbf{(\textmu{}A)} & \textbf{High-resolution mode} \textbf{(\textmu{}A)} \\ \hline
1 & 2 & 2 & 2 \\ \hline
10 & 3 & 4 & 4 \\ \hline
25 & 4 & 6 & 6 \\ \hline
50 & 6 & 11 & 11 \\ \hline
100 & 10 & 20 & 20 \\ \hline
200 & 18 & 38 & 38 \\ \hline
400 & 36 & 73 & 73 
\end{tabular}}
\caption[Accelerometer power consumption]{\textbf{Accelerometer power consumption}. Excerpt of the accelerometer power consumptions in different operation modes and datarates \cite{LIS2DH12:DS}.}
\label{tab:accelerometer_power_consumption}
\end{table}

The accelerometer has an integrated 32-level \ac{FIFO} for the raw acceleration values. This \ac{FIFO} is used to minimize the number of \ac{SPI}-transfers, since up to 32 data points can be read at once by the \ac{SPI}-master. Furthermore, the accelerometer features an optional high-pass filter to eliminate the acceleration due to gravity.
Beside the acceleration values of the three orthogonal axes, the accelerometer provides the capability to enable interrupts that are generated if preconfigured acceleration events are detected. These events are for example motion events, free-fall events or click-events. The events are signaled through an output pin of the accelerometer, which is connected to an interrupt pin of the microcontroller.
The functionality of the accelerometer is abstracted in the accel module of the firmware. This module configures the accelerometer by writing to its internal registers via the \ac{SPI}-driver.
It provides interfaces to setup the accelerometer, to read the current acceleration along the three axes in \ac{mg} and to activate the event detection.

The sampling module of the firmware provides two data-sources based on the accelerometer: Raw acceleration data and motion events.
The start-function for recording raw acceleration data expects certain parameters for the datarate, the operating mode, the full scale and the period to read out the 32-level \ac{FIFO}.
After configuration of the accelerometer via the accel module, the function opens a chunk of the accelerometer-chunk-\ac{FIFO} and starts a timer that periodically reads the latest data from the 32-level \ac{FIFO}. It is important that the specified timer frequency is at least as high as the selected datarate of the accelerometer.
Since the \ac{SPI}-driver is also used by the eeprom module, it can occur that the \ac{SPI}-driver is not available to read the acceleration data from the \ac{FIFO} at the moment at which the timer callback function is executed.
Therefore, it is advisable to select the timer frequency significantly higher than the datarate of the accelerometer.
%In the timer callback function the available acceleration data points of the 32-level \ac{FIFO} are read.
For the data analysis, the amount of movement, but not the exact acceleration of all three axes, is of particular importance.
To eliminate the acceleration offset due to gravity, the integrated high-pass is applied.
To retrieve a measure for the amount of movement, the absolute acceleration values of the three axes are summed up to one representative value \cite{Bouten97:TAA}. This value is inserted into the accelerometer-chunk. If the chunk is filled, the corresponding processing function is called via the scheduler. This function is responsible for serializing and storing the chunk.

The motion event data-source is based on the interrupts generated by the accelerometer. To enable the motion event detection, three parameters have to be selected when calling the corresponding start-function of the sampling module. The first parameter is the acceleration threshold that has to be exceeded by at least one axis. The second parameter represents the minimal duration for which the acceleration must exceed the specified threshold to generate an interrupt. The last parameter describes the minimal duration between two consecutive events. If the next event occurs within this duration, it will be ignored. 
Every time such an event is detected, an interrupt is generated and a callback function in the sampling module is invoked. In this callback function, the event is inserted into the accelerometer-interrupt-chunk, which is closed afterwards. Finally, the corresponding processing function is scheduled to serialize and store the chunk.

\subsection{Battery data}
To monitor the functionality of the badges during their utilization, it is necessary to know the current supply voltage of the coin cell battery. 
The supply voltage can be used to determine the remaining capacity of the battery. Based on this information either the battery or the badge itself can be exchanged in time to avoid data loss. 
In addition, the supply voltage is important for subsequent data analysis, as it can be used to explain irregular behavior of the badges or corrupted data, especially when the voltage is low.

The current voltage can be retrieved through the battery module.
Internally, the battery module periodically triggers the \ac{ADC}-driver to measure the current voltage and averages consecutive values afterwards. 
To determine the supply voltage, the internal band gap reference voltage of the nRF51822 with 1.2 Volts serves as reference voltage for the \ac{ADC} \cite{NRF51822:PS}. Since the external supply voltage exceeds 1.2 Volts in general, it is reduced to one third of the original value by an internal voltage divider. This value is then converted by the \ac{ADC} with a resolution of 10 bits. 
Finally, the current external supply voltage (\(V_{supply}\)) can be calculated based on the converted value (\(Value_{ADC}\)) with the following formula:
\begin{equation}
	V_{supply} = \frac{Value_{ADC}}{2^{10}} * 3 * 1.2 V
\end{equation}
	
Averaging is applied to eliminate voltage drops during high current consumption activities such as flash-memory interactions or \ac{BLE} activity.
The averaging is implemented as \ac{EWMA}. This type of averaging is characterized by a simple computation and by declining weights for older values \cite{Holt04:FSA}. The recursive computation of the current averaged value \(S_t\) at time point \(t\) based on the previous averaged value \(S_{t-1}\) and the current measured value \(Y_t\) is as follows:

\begin{equation}
S_t = 
\begin{cases}
Y_1, & t=1\\
\alpha * Y_t + (1 - \alpha) * S_{t-1}, & t > 1
\end{cases}
\end{equation}

The \(\alpha\) coefficient in this equation ranges from 0 to 1 and affects the influence of older values. The higher this coefficient, the faster the influence of older values decreases.
The periodical voltage measurement is implemented with a timer. Each time the timer callback updates the averaged voltage, the supply voltage of the advertising packet is also updated. This is accomplished using a function of the advertising module. 
The battery voltage can also be stored in the \acp{NVM} for later data analysis. This functionality is again provided by the sampling module. 
As soon as this data-source is activated, the sampling module sets up another timer that periodically reads the current averaged supply voltage from the battery module. This value is inserted into a battery-chunk of the battery-chunk-\ac{FIFO} and the corresponding processing function of the processing module is scheduled. 
The processing of this data-source is similar to the other data-sources: The chunk is read out from the chunk-\ac{FIFO}, then serialized and finally stored to the \acp{NVM}.

\section{Data serialization} \label{sec:Data serialization}

The efficient serialization and deserialization of structured data, such as chunks, is essential for data storage and transfer. 
Especially for data transfer, the de-/serialization mechanism has to be platform independent, since the data is exchanged between different architectures and programming languages.
This section describes first Google's Protocol Buffer library and afterwards the serialization and deserialization technique that was developed in this work.

\begin{figure}[h]
\centering
%\fbox{\includegraphics[page=1, trim=2.5cm 25.4cm 10cm 2.5cm]{img/serialization.pdf}}
\includegraphics[page=1, trim=2.5cm 25.4cm 10cm 2.5cm, clip]{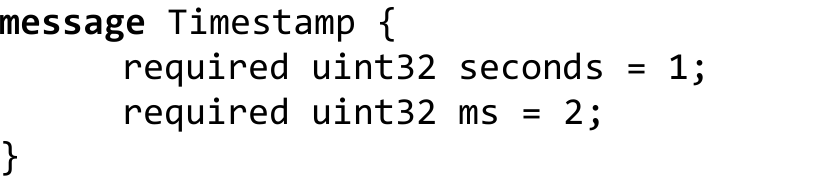}
\caption[Protocol Buffers example]{\textbf{Protocol Buffers example}. The Protocol Buffers message of a timestamp consists of one field representing the seconds and one for the milliseconds. The numbers to the right of the equals sign are the corresponding field identifiers.}
\label{fig:protobuf_example}
\end{figure}

Various types of de-/serialization techniques exist, such as \ac{XML}, \ac{JSON} or Protocol Buffers.
\ac{XML} and \ac{JSON} are not suitable for an efficient serialization due to their high overhead.
In contrast, Google's Protocol Buffers library is characterized by a small overhead due to its efficient binary data representation \cite{Google08:PB}. Protocol Buffers officially supports the programming languages C++, Java, Python and Go, but not C. For C another Protocol Buffers implementation called nanopb exists, which is especially suitable for embedded systems with restricted resources \cite{Aimonen12:NPB}.
Based on a schema-file, the Protocol Buffers compiler generates special source code to serialize strucutured data into a binary representation and to deserialize it back to structured, interpretable data.
In the schema-file, structured data are defined through messages. 
Each message has a name and contains one or more fields that characterizes the message. In Figure \ref{fig:protobuf_example} a simple example message for representing a timestamp is shown.

After the fields of the message have been setted via the automatically generated source code, the message can be serialized for transfer. 
Protocol Buffers support various data types, such as integer, float, double, string, boolean, byte-arrays and even other messages.
Each field is specified with a field rule: required, optional or repeated. Required fields are essential for the message and have to be set/defined before serialization. On the other hand, optional fields need only to be specified if necessary. If an optional field is not set, it will be ignored during the serialization. 
Repeated fields can contain multiple elements of the specified data type, which corresponds to an array-representation.
There are three main reasons why Protocol Buffers has not been used in this work and instead a custom de-/serialization technique was developed.
The first reason is, that Procotol Buffers serialization applies \ac {varint} encoding. The encoded or serialized length of an integer encoded with this mechanism can be higher than the actual byte representation of the integer.
Secondly, each field gets an additional indentifier during the encoding process. Especially for repeated fields with another message as data type, the overhead is extremly high due to the additional identifiers.
Thirdly, Protocol Buffers are not compatible with the previous implementation of the transmission protocol. Support of the previous protocol is particularly necessary to perform integration tests for the new firmware.

\begin{figure}[h]
\centering
%\fbox{\includegraphics[page=2, trim=2.3cm 22.6cm 2.4cm 1cm]{img/serialization.pdf}}
\includegraphics[page=2, trim=2.3cm 22.6cm 2.4cm 1cm, clip, scale=0.88]{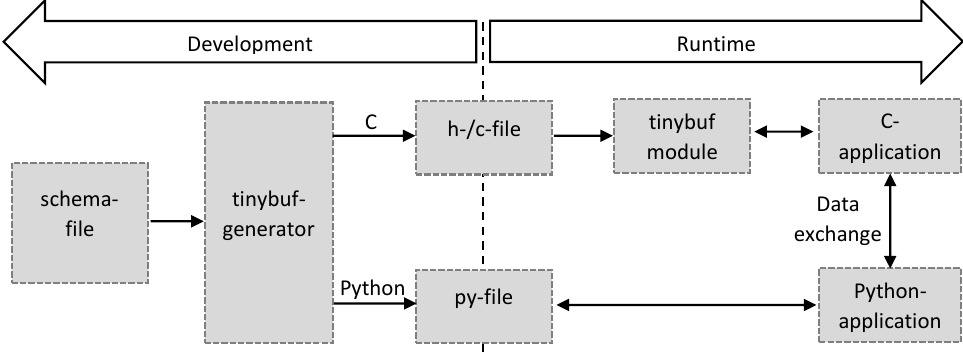}
\caption[Tinybuf workflow]{\textbf{Tinybuf workflow}. The workflow is divided into two parts. The left part takes place during the development process of the system: The schema-file is defined and converted to source code by the tinybuf-generator.
The right part of the figure is executed while the system is running: The generated source code is used to de-/serialize the defined messages, which can be exchanged platform independently between applications with different programming languages.}
\label{fig:tinybuf_workflow}
\end{figure}

To overcome these difficulties a custom de-/serialization techique called Tinybuf has been developed. 
The workflow of Tinybuf is comparable to the workflow of Protocol Buffers. In the beginning, a schema-file has to be created that contains the message declarations.
Afterwards the schema-file is converted to source code. Currently C and Python are supported as programming languages.
Finally, the generated source code can be used to access the messages and to efficiently de-/serialize them. An overview of the workflow is shown in Figure \ref{fig:tinybuf_workflow}.

The tinybuf-generator parses the schema-file in analogy to the Protocol Buffers compiler. This generator is implemented in Python and automatically generates source code for the chosen programming language.
In case of Python, the tinybuf-generator creates one .py-file which represents the messages as Python classes.
At runtime, the application creates objects of these classes and sets the desired message fields. Afterwards, the encode function is called to serialize the message object into a byte sequence. To convert such a byte sequence back to a message object, the decode function of the class has to be called.
If C is selected as programming language, a C header-file and a C source-file is automatically generated. The defined messages are represented as C structures. 
In contrast to the generated python file, the encode and decode functions are not included in the generated C-source file.
These functions are provided by an additional module: the tinybuf module.
The encode and decode functions are generic and can be used for any structure created by the tinybuf-generator.
This approach is more complex than the Python based method, because additional information about the structure is required for encoding or decoding.
Especially when many different messages are defined, this decoupling leads to smaller code size compared to methods where each message has its own encoding and decoding function.
A small code size is essential for embedded systems as the available resources are limited.

\begin{table}[h]
\renewcommand{\arraystretch}{1.1}
\centering
\scalebox{0.88}{%
\begin{tabular}{l|l} 
\textbf{Data type} & \textbf{Comment} \\ \hline
uint8 & 1 byte unsigned integer  \\ \hline
int8 & 1 byte signed integer  \\ \hline
uint16 & 2 byte unsigned integer  \\ \hline
int16 & 2 byte signed integer  \\ \hline
uint32 & 4 byte unsigned integer  \\ \hline
int32 & 4 byte signed integer  \\ \hline
uint64 & 8 byte unsigned integer  \\ \hline
int64 & 8 byte signed integer  \\ \hline
float & 4 byte floating-point number \\ \hline
double & 8 byte floating-point number \\ \hline
\textit{message} & another previously defined message
\end{tabular}}
\caption[Tinybuf data types]{\textbf{Tinybuf data types}. The supported data types are integers, floating-point numbers and other messages.}
\label{tab:tinybuf_data_types}
\end{table}

Tinybuf supports the field data types shown in Table \ref{tab:tinybuf_data_types}. The \textit{message} data type has to be replaced with the actual message name. In contrast to Protocol Buffers, the integers are not encoded with varint, but with their binary representation.
The field rules provided by Tinybuf are the following: required, optional, repeated, fixed-repeated and oneof.
Required, optional and repeated fields have the same characteristics as in Protocol Buffers. Fixed-repeated fields are similar to repeated fields, but have a predefined fixed number of elements. The advantage of fixed-repeated fields is that no additional length information needs to be encoded.
The most complex field rule is the oneof rule. In many applications it is often necessary to exchange not only one message, but different messages. The recipient of the serialized message must know which message has been received in order to deserialize it correctly. Either the recipient knows in advance which message will be received, or additional information must be included in the serialized message.
Oneof fields consists of multiple fields, with only one field set at a time. When the oneof field is serialized, additional information describing which of the fields has been set is automatically included. This is exactly the mechanism that solves the message interpretation problem described above: Each of the different messages is represented as a field in a oneof field.
Another advantage of the oneof field is that in C all the included fields share the same memory using a C union.

The presented de-/serialization technique is used by the data storage unit and the data exchange unit of the firmware. 
The data chunks mentioned in the previous section are represented by Tinybuf messages. Consequential, these chunks can be serialized directly with the tinybuf module, enabling an efficient storage.
Furthermore, the tinybuf module is applied during the data exchange process. The communication protocol is defined by a Tinybuf schema-file which contains all messages required for the communication between the badges and remote \ac{BLE} devices, such as hubs.
By using this automated, generic de-/serialization technique, the developer no longer needs to implement individual de-/serialization functions for each message. This reduces development time and enables the quick integration of new functionalities.
%The remote \ac{BLE} device that communicates with the hub can use the generated python-file to serialize and deserialize the messages.%Static allocation of repeated fields as arrays with given size.%In python the struct.unpack/pack functions are used.%Beispiel Bild mit vielen  %Neben messages werden auch noch defines unterst"utzt. %All das erleichtert extrem das definieren neuer Nachrichten, ohne das speziell f"ur jede neue Nachricht ein eigene serialisierungs Funktion gebaut werden muss.%Protocol Buffers haben den Vorteil, dass Sie backwards kompatibel sind, wenn neue Felder zu einer bestehenden Nachricht hinzugefügt werden. Das kann Tinybuf nicht.

\section{Data storage} \label{sec:Data storage}

The storage of data in the \acp{NVM} is essential for the functionality of the system.
On the one hand, persistent metadata, such as the \ac{ID} and group number of the badge, must survive power losses. On the other hand, badges must store the recorded data chunks locally in order to transmit them when the remote \ac{BLE} device requests it.
To manage the storage of different types of data consistenly and without influencing each other, an efficient filesystem with a simple interface has been developed.
To facilitate the implementation of the filesystem, a uniform storage abstraction of the \acp{NVM} is provided. This storage abstraction combines multiple \acp{NVM} to one large virtual memory.
This section describes the data storage unit of the firmware in more detail. This includes the virtual memory, the filesystem and finally the highest layer that combines the filesystem and the tinybuf module to enable simple storing and reading of structured data.

\subsection{Virtual memory}
The purpose of a uniform storage interface is to simplify the interaction with multiple and different memory types such as the external \ac{EEPROM} and the internal flash-memory.
Flash-memory is organized in pages with a certain size. Each page has to be erased manually before storing data on it \cite{Tal02:TFT}.
The internal flash-memory of the nRF51822 has a page size of 1024 bytes. The smallest storable units are words with a size of 4 bytes or 32 bits \cite{NRF51822:PS} 
The external \ac{EEPROM} (M95M02-DR) allows single bytes to be stored and there is no need to erase them before storing new data \cite{EEPROM:M95M02}.
To combine these two memories into one virtual memory, a similar interface to each of them is required. 

Since the interaction with the flash-memory is rather difficult, an abstraction layer (storage\_1) has been implemented to simplify the access to the flash-memory. This abstraction enables the application to store single bytes to the flash-memory without manually erasing pages in advance and to read single bytes from it.
This is achieved by a mechansim that automatically erases pages  before storing the data. 
This mechanism tracks the address ranges in which data can be stored without erasing the corresponding pages.
If a store operation is within this address range, no erase operation needs to be performed.
The mechanism is optimized for storing data to consecutive addresses.
To store and read single bytes instead of words, a special behavior of flash-memories is used: The erase operation sets all bits of a page to 1. During store operations, bits within one flash word can only be setted from 1 to 0. This can be used to create a word where all bits are 1 except the bits belonging to the byte to be stored.
If the created word is stored to the flash-memory, only the bits with a value of 0 are updated. 
The byte can be extracted from the flash word again during the read operation.
To provide the same interface for the \ac{EEPROM}, a similar abstraction layer (storage\_2) is added. In this case, no automatic erase mechanism is included, because it is already handled by the \ac{EEPROM}.

\begin{figure}[h]
\centering
%\fbox{\includegraphics[page=1, trim=6cm 21.8cm 8.7cm 3.3cm]{img/storage.pdf}}
\includegraphics[page=1, trim=6cm 21.8cm 8.7cm 3.3cm, clip, scale=0.9]{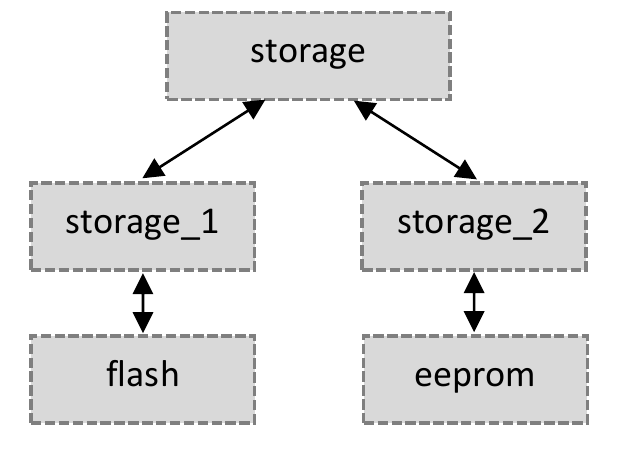}
\caption[Virtual storage]{\textbf{Virtual storage}. The figure shows the modules involved to provide a simple accessible virtual memory.
%The \ac{HW} abstraction modules flash and eeprom represent the functionality of the internal flash-memory and the external \ac{EEPROM}. The storage\_1 and storage\_2 modules provide a uniform interface to access the different memories. Ultimately, the storage module combines storage\_1 and storage\_2 to one virtual memory.
}
\label{fig:storage_interface}
\end{figure}

Finally, both abstraction layers are combined to one virtual memory by the storage module. 
The storage\_1 and storage\_2 modules have their own address spaces, starting with address 0. 
The storage module concatenates both address spaces and generates a new consecutive address space.
%The size of this new address space results from the sum of the sizes of the individual address spaces.
Internally, the addresses of the new address space are mapped to the corresponding addresses of storage\_1 and storage\_2.
The resulting module interaction is shown in Figure \ref{fig:storage_interface}.

\subsection{Filesystem}
In order to simplify the implementation of the filesystem, this virtual storage interface is used as basis.
The filesystem has to meet the following requirements:
Each class of data to be stored must have its own physical partition in the memory to prevent mutual influences. The filesystem has to be optimized for sequential storage and read operations since most of the data to be stored are created and retrieved sequentially. Another advantage of sequential storage operations are an even access of the memory cells, maximizing their lifetime.
The additional overhead produced by the filesystem must be minimal to maximize the available memory for application data.
Another important feature is the ability to detect corrupted data in memory. Furthermore, the filesystem must support the variable length of the elements within a partition to optimize the utilization of the available memory and to support compression of data before storage.
In addition, the filesystem must be able to handle power losses during storage operations.

\begin{figure}[h]
\centering
%\fbox{\includegraphics[page=2, trim=1cm 22.8cm 4cm 2.5cm]{img/storage.pdf}}
\includegraphics[page=2, trim=1cm 22.8cm 4cm 2.5cm, clip,scale=0.9]{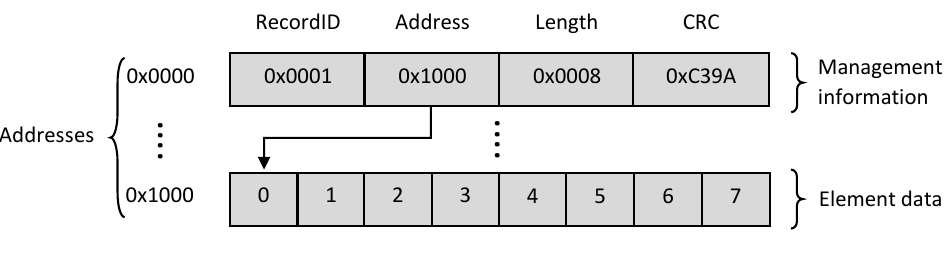}
\caption[Filesystem with management sector]{\textbf{Filesystem with management sector}. The management information for each data element is seperated in a management sector. The management entry references the data element and contains additional information, such as a unique element number, the length of the element and a checksum.}
\label{fig:storage_management}
\end{figure}

In general two different concepts exist for filesystems \cite{Gal05:ATF} \cite{Daberko98:OSI}. The first approach is characterized by a physical seperation of the data elements to be stored and the management information. This additional metadata information is needed to manage the stored elements. 
The management information is stored in a prior defined memory range. Each data element is referenced by an information entry in this management sector. This entry includes, for example, a unique number that identifies the element or record, the address at which the current element is stored in memory, the length of the element, and a checksum to detect corrupted elements. The resulting memory layout is shown in Figure \ref{fig:storage_management}.
This approach is characterized by a relatively simple implementation. However, the approach also suffers from a number of disadvantages. The size of the management area must be defined in advance. This defines the maximum number of elements that can be stored, even if there is enough space for more data elements.
Another drawback of this approach becomes evident when it is implemented with flash-memory as storage. If the entries of the management section are within a flash page and an erase operation is performed on this page, all entries are lost. This data loss can only be prevented if all entries are backed up. However, this backup mechanism heavily demands the memory cells.

The second filesystem concept is characterized by the combination of element data and management information.
An existing filesystem implementation from Nordic Semiconductor works according to this concept and adds an element header with management information to each data element \cite{NordicSemiconductor18:INF}. 
The filesystem of Nordic Semiconductor is not suitable for deployment in this project due to the following reasons.
Firstly, this implementation only supports the internal flash-memory and has no interface to integrate other memories such as the external \ac{EEPROM}.
On the other hand, the additional overhead of 12 bytes is too large and only storing and reading of words is allowed.

The combination of element data and management information is also applied in the filesystem that was developed as part of this work.
During the initialization, the application defines the sizes of the partitions for each class of data and registers the partitions in the filesystem. 
Two types of partitions are available: static partitions and dynamic partitions.
In static partitions, all stored elements must have the same size which is defined during initialization. In contrast, dynamic partitions allow the storage of elements with an arbitrary length.
Each element within a partition is stored along with a header that contains additional management information.
In the case of a static partition, the header contains a record or element number and optionally a \ac{CRC} calculated from the element data.
In dynamic partitions, the header consists of an element number, length information and optionally a \ac{CRC}.
The resulting header size can be retrieved from Table \ref{tab:filesystem_header_length}.

\begin{table}[h]
\renewcommand{\arraystretch}{1.2}
\centering
\begin{tabular}{l|c|c} 
& \textbf{Static partition} & \textbf{Dynamic partition} \\ \hline
\textbf{CRC disabled} & 2 & 4 \\ \hline
\textbf{CRC enabled} & 4 & 6
\end{tabular}
\caption[Element header size]{\textbf{Element header size}.
The table shows the size of the element header in bytes, depending on whether the partition is static or dynamic and whether \ac{CRC} is enabled or not.}
\label{tab:filesystem_header_length}
\end{table}

\begin{figure}[h]
\centering
%\fbox{\includegraphics[page=3, trim=0.8cm 22.9cm 7cm 2.5cm]{img/storage.pdf}}
\includegraphics[page=3, trim=0.8cm 22.9cm 7cm 2.5cm, clip, scale=0.9]{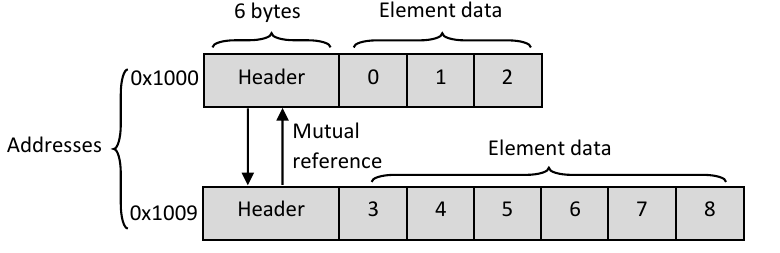}
\caption[Filesystem with doubly linked list]{\textbf{Filesystem with doubly linked list}. Each element in the filesystem has a header that refers to the next and to the previous element. The elements are stored consecutively in the memory.}
\label{fig:storage_linked_list}
\end{figure}

The data structure used in dynamic partitions to manage variable length elements is an \ac{XOR} linked list. 
\ac{XOR} linked lists are special cases of doubly linked lists with reduced space requirements \cite{Berdine05:SMA}.
In doubly linked lists, each element header refers to the next and to the previous element.
Since the elements are stored consecutively in memory, these references can be expressed by the size of the previous element and the size of the current element.
To reduce the header size, the previous element size and the current element size are combined via the \ac{XOR} operator which results in an \ac{XOR} linked list.
The first element header of each partition is a special header. It contains additional metadata (14 Bytes) about the partition, such as the partition \ac{ID}, the partition size, the address of the last element in the partition and the size of the first element. 
With the size of the first element, all other elements can be retrieved from the \ac{XOR} linked list by iterating over the element headers. 
When the partition is full, the storage starts again at the first address of the partition.
Overwriting the header of the first element is particularly problematic in case of a power loss, as all necessary management information is stored in this header and may be lost.
In order to prevent this, a special backup mechanism is applied that stores the header of the first element in a reserved memory section before it is overwritten.
In Figure \ref{fig:storage_linked_list} an extract of the filesystem with a doubly linked list is shown.

To retrieve the elements again from the partitions, an iterator for each partition is provided. 
By default, the iterator refers to the latest stored element. The iterator can be moved to the previous element as well as to the next element.
This allows the application to flexibly read the sequentially stored elements.
In addition, the filesystem offers an interface to clear a partition and to clear the entire memory.
Clearing a partition is performed by deleting the first element header that contains the management information of the partition.
The process of clearing the entire memory is relatively time-consuming compared to deleting the first element header of the partitions, since all bytes in the memory must be reseted to their default values.

\subsection{Filesystem integration}
The storer module of the firmware is another abstraction above the filesystem. 
This module manages the storing and reading of the previous described data chunks in the filesystem.
For each of the data sources of section \ref{sec:Data recording} an own partition with a configurable size is registered in the filesystem.
Furthermore, the \ac{ID} and group number assigned to the badge during the activation are also stored in an own filesystem partition.
The module has an interface to store a data chunk of a certain data source, for example a microphone chunk, in the corresponding partition. 
This interface is used by the processing module of the data recording unit to store the recorded data chunks.
The tinybuf module efficiently serializes the data chunks before they are stored.
Finally, the stored data chunks are retrieved by the data exchange unit for the transmission to a remote \ac{BLE} device.
The current setup of the system requires that data chunks must be retrievable from a certain timestamp in the past until now. 
This functionality is also implemented by the storer module.
First, the iterator of the partition is used to search for the oldest data chunk whose timestamp is greater than or equal to the requested timestamp.
Afterwards, starting from this data chunk, the next one can be retrieved with a function.

\section{Data streaming} \label{sec:Data streaming}
Besides storing the recorded data for later transmission, the data can also be directly streamed. 
In this case, the data are transmitted to the remote \ac{BLE} device immediately after their recording.
The following firmware units are involved in the streaming process: the data recording, the data streaming and the data exchange (see Figure \ref{fig:modules_overview}).

The sampling module of the data recording unit is responsible for collecting the data to be streamed from the individual data sources.
The recorded data points for streaming are not organized in chunks, since the data points have to be transmitted immediately.
A specially developed \ac{FIFO} mechanism is used for sharing the recorded data points between the data recording unit and the data exchange unit.
This \ac{FIFO} mechanism is implemented in the data streaming unit.
The requirement on this \ac{FIFO} in contrast to other \ac{FIFO} implementations is, that unread data points can be overwritten by new data points when the \ac{FIFO} is full.
This is necessary because the remote BLE device prefers to receive the most recent data points.
Conceptually, a circular \ac{FIFO} implementation is used as basis. This circular FIFO is extended to allow a synchronized shifting of the current read and write position of the \ac{FIFO}.

The available data sources for streaming are the same as for storing: i.e. audio, proximity, supply voltage, acceleration data and acceleration events.
Each of the mentioned data sources has its own circular \ac{FIFO}.
In contrast to the storage, the proximity data points and acceleration data points are not aggregated for streaming.
For the proximity data points, the raw RSSI values are used and not the maximum or the average of several RSSI values.
In the case of acceleration data, the raw values along the three axes are inserted into the circular \ac{FIFO}.

The streaming process is controlled by the request handler module of the data exchange unit.
This module is responsible to start the streaming for the various data sources and to retrieve them from the circular \acp{FIFO}.
The data points are read from the \acp{FIFO} and enqueued for the transmission to the remote \ac{BLE} device.
Both, the streaming and the storage of the recorded data can be performed simultaneously. However, streaming is only available when a remote \ac{BLE} device is connected.
The data sources parameters are the same for the streamed and stored data.

\newpage

\section{Data exchange} \label{sec:Data exchange}

In order to access data from the badges for monitoring or analysis, it must be exchanged.
The data exchange unit of the firmware is responsible for the interaction with remote \ac{BLE} devices and for controlling the behavior of the badge.
This section provides an overview of the applied advertising-based and connection-oriented data exchange. In addition, the communication protocol, the processing of external requests and the clock synchronization are described in more detail.

\subsection{Advertising}
The periodic broadcasting of advertising packets by the badges is essential for the functionality of the system. 
On the one hand, the advertising packets are received by other badges during Bluetooth scans. The \ac{RSSI} values are used to determine the proximity between the badges.
On the other hand, useful information can be retrieved to monitor the badges including the current supply voltage and the status of the badge.

The advertising process is managed by the advertising module of the data exchange unit.
With this module the broadcasting of the advertising packet can be started or stopped. 
The device name that appears in the badge's advertising packet is "'HDBDG'', the broadcasting period of the packet is 200 ms.
In addition, the module has interfaces for inserting customized information into the advertising packet. The customized information includes the \ac{ID} and group number, the \ac{MAC} address, the current supply voltage and status information of the badge, e.g. the currently activated data sources and whether its clock is synchronized or not.
To minimize the size of the advertising packet, the status information flags are encoded by the bits of one status byte.
Furthermore, the supply voltage \(V_{supply}\), which has the data type float (4 bytes), is encoded in a single byte \(B\) by the following compression:

\begin{equation} \label{eq:battery}
B = V_{supply}*100-100
\end{equation}

This compression allows a voltage range from 1 V to 3.55 V to be covered in one byte with a resolution of 0.01 V.
The covered range is sufficient for the coin cell batteries. The receiver of the advertising packet can calculate the original voltage by solving equation \ref{eq:battery} for \(V_{supply}\). 

\newpage

\subsection{Connection-based communication}
In addition to unidirectional data exchange based on advertising packets, bi-directional connection-based \ac{BLE} communication can also be selected for data exchange.
The remote \ac{BLE} central device, for example a hub, initiates the connection. After the connection has been established, the Nordic UART Service is used for the bi-directional communication.
This \ac{BLE} service has the RX- and the TX-characteristic.
Using these two characteristics, raw bytes can be sent between the connected \ac{BLE} devices. When data are received, a notification in form of a callback is generated by the SoftDevice to handle the bytes.
The number of bytes that can be transmitted by the application at once is limited to 20 bytes \cite{Gomez12:OAE}.
To enable the transmission and reception of more than 20 bytes, an additional abstraction based on the Nordic UART Service was implemented in the sender module.
Internally, this module uses two \acp{FIFO}, one for transmission and one for reception, to handle packets larger than 20 bytes.
To transmit data, the application inserts the data into the TX-\ac{FIFO}. The sender module reads the first 20 bytes from the TX-\ac{FIFO} and transmits them via the Nordic UART Service. This process is repeated until no more bytes are left in the TX-\ac{FIFO}.
The repetition time of this process is crucial for the resulting transmission speed.
To be able to repeat the process, special mechanisms must be applied in order to avoid blocking the application.
It can be implemented in three different ways.
The first approach registers a callback function in the SoftDevice, which is called as soon as the 20 bytes have been successfully transmitted. 
If the TX-\ac{FIFO} is not empty yet, the callback function reads the next 20 bytes (or less) from the TX-\ac{FIFO} and restarts the transmission.
The second available technique to achieve the repeated execution is the scheduler. The function that reads the bytes from the TX-\ac{FIFO} and starts the transmission is executed again by inserting itself into the scheduler queue.
The third method starts an asynchronous timer that periodically calls a function to read the bytes from the TX-\ac{FIFO} and start the transmission. The timer is stopped as soon as there are no more bytes in the TX-\ac{FIFO}.
The advantage of the last approch is that the timer period is tunable via a parameter so that the transmission speed can be influenced. Additionally, the highest data throughput can be achieved with this method.
Therefore, this timer-based approach is used in the new implementation.
Another method to influence the transmission speed is 
the adjustment of the connection interval settings on the remote \ac{BLE} central device. The central device defines the connection parameters used with the peripheral device. Smaller connection intervals imply higher data throughput.

To handle the received data from the remote \ac{BLE} device, the sender module uses a \ac{FIFO}. As soon as data are received, they are inserted into this RX-\ac{FIFO}. Furthermore, a registerable notification handler is called to inform the application about the reception of data.

An additional feature of the sender module is the timeout mechanism to automatically disconnect if no data exchange occurs within a predefined time interval. This is important in case that the disconnect event of the central device is not received by the badge. The automatic disconnect feature can be disabled during development and testing.

\subsection{Communication protocol}
In addition to the physical data exchange of raw bytes, their correct interpretation by the involved communication partners is essential for the functionality of the entire system.
The interpretation of the raw bytes is enabled by a communication protocol.
This protocol defines rules for reconstruction of the exchanged message.
Since communication takes place between devices developed with different programming languages, the rules of the communication protocol must also be implemented in these languages.
Therefore, the developed serialization library Tinybuf, which supports C and Python, is used for the implementation of the communication protocol.
Tinybuf enables a very simple, flexible and extensible design of messages to be exchanged. Furthermore, it is characterized by an efficient binary representation of the messages after serialization. The smaller the binary representation of the messages, the faster they can be exchanged.
In the implemented communication protocol, messages are divided into two groups: Requests and responses.
All messages transmitted from the remote \ac{BLE} device to the badge are called requests.
Conversely, all messages transmitted from the badge are referred as responses.
All requests as well as all responses are placed in a Tinybuf oneof field.
By using oneof fields, the information which message is encoded can be directly inserted into the binary representation.
This eliminates the need to manually insert an additional header to identify the message.
The sender module described above sequentially transmits all bytes in the TX-\ac{FIFO} in 20 byte packets, without considering message boundaries.
Therefore, additional information must be included in the binary representation so that the recipient can separate the messages.
This additional information is provided in the form of a 2 byte header, which contains the length of the following serialized message.

\begin{figure}[h]
\centering
%\fbox{\includegraphics[page=2, trim=1.8cm 17.8cm 1.8cm 6.3cm, width=\textwidth]{img/exchange.pdf}}
\includegraphics[page=2,  trim=1.7cm 17.8cm 1.7cm 6.3cm, clip, width=\textwidth]{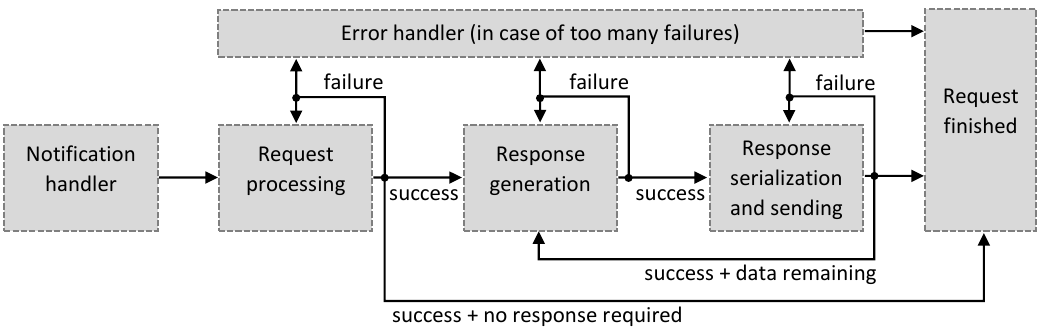}
\caption[Message processing]{\textbf{Message processing}. The flowchart shows the individual steps from receiving a message to generating a response and transmitting it.}
\label{fig:flow_chart_request}
\end{figure}

The entire communication process is managed by the request handler module of the data exchange unit.
This includes processing received requests, controlling the data recording unit, reading recorded data, and generating messages for transmission.
The internal procedure for processing received messages is illustrated in Figure \ref{fig:flow_chart_request}.

When data are received via the \ac{BLE} interface, the notification handler is called to inform the request handler module that new data are available in the RX-\ac{FIFO} of the sender module.
Afterwards, the received data are read out and the request is reconstructed by decoding the binary representation using the tinybuf module.
The resulting request, such as assigning the badge ID and group number, is then processed.
Requests can be divided into two groups: Requests that expect a response and those that do not.
If no response is expected by the remote \ac{BLE} device, the processing of the request is finished.
Otherwise, a response must be generated, for instance from a data chunk read from memory. After the response is generated, it is serialized using the tinybuf module and inserted into the TX-\ac{FIFO} of the sender module.
If further data have to be transmitted, the response generation is triggered again.
If one of the steps is currently not executable because a required resource is in use or not enough space is available in the TX-\ac{FIFO}, the step is executed again by the scheduler at a later time.
In the case that a step fails too often, an error handler is called to reset the internal state and to perform a certain action, such as disconnecting from the remote \ac{BLE} device.

The communication protocol offers a broad spectrum of available requests to control the badge or to retrieve data from the badge.
Each data source can be started by its own request, which also provides the parameters for data recording.
If a start request is received for an already running data source, it will be ignored unless the parameters have changed.
The streaming of data points as well as the storage of data chunks for later transmission can be performed simultaneously.
To stop a specific data source, a stop request has to be sent to the badge.
The stored data chunks of each data source can be retrieved from a particular timestamp in the past until now by a data request.
In addition to the data related requests, further requests exist to manage the badge.
The status request enables the remote \ac{BLE} device to retrieve the current status of the badge and to optionally assign its \ac{ID} and group number.
Furthermore, requests are available to restart the badge, to perform peripheral testing, or to identify the badge by blinking one of its status \acp{LED}.

\subsection{Clock synchronization}
An accurate time basis of the badges is essential for the subsequent data analysis. The timestamps of the data chunks are used to correlate the data of different badges in order to analyze the communication patterns of the participants.
To achieve an accurate time basis, the badges synchronize their internal clocks with an external time source.
After the local time of the badge has been synchronized, the incorporated \ac{RTC} oscillator with a nominal frequency of 32768 Hz \cite{RTC:ABS07} is used to provide the internal time.
At its frequency, the oscillator increments an internal tick counter that starts at 0. Based on these oscillator ticks and the synchronized reference time, the current local time can be calculated with the following equation.

\begin{equation} \label{eq:synchronization}
t_{cur} = \frac{1}{32768 \text{ Hz}}*(ticks_{cur} - ticks^{i}_{sync}) + t^{i}_{sync}
\end{equation}

In this equation, \(t_{cur}\) denotes the calculated current time, \(ticks_{cur}\) is the number of ticks, \(ticks^{i}_{sync}\) represents the number of ticks at the \(i\)-th synchronization timepoint and \(t^{i}_{sync}\) is the \(i\)-th received synchronization time. The \(i\)-th synchronisation point corresponds to the latest synchronisation point. The oscillator frequency is assumed to be constant over time.
In general, this assumption is not correct as the frequency of the oscillators may drift over time.
Although the frequency error is usually quite small, the cumulative error over a long time period results in an observable difference in the calculated time \cite{Hwang04:ESD}.
Several reasons might exist for the frequency deviation of oscillators from their specified frequency.
On the one hand, environmental factors can influence the oscillator frequency. These include variations in temperature or supply voltage as well as physical vibrations or shocks.
On the other hand, the manufacturing process and the aging of the oscillator itself might have an influence on its frequency \cite{Elson02:FGN} \cite{Dian17:ASI}.

To avoid a local clock drift, it is necessary to apply special synchronization techniques.
The simplest approach is a frequent synchronization with an external accurate time source. By ensuring a small time interval between the synchronization points, the time error caused by drift is reduced.
However, in addition to the local clock drift, it is also a challenge to provide an accurate external time source for the synchronization process.
In order to create such an accurate external time base, various influencing factors have to be considered.
First, the accuracy of the external time base itself plays an important role. 
Furthermore, several factors that influence the time for transmission of the synchronization message must be taken into account. These include the time from message creation to transmission, the actual transmission time, and finally the time from message reception to processing \cite{Elson02:FGN}. 
A commonly used synchronization method is \ac{NTP}  \cite{Mills91:ITS}. This technique is often applied for time synchronization of devices in variable-latency networks, such as the internet. It uses a software-based approach to compensate the latency of the synchronization message transfer and to adjust the local oscillator frequency to minimize the local time drift.
%Another approch by Elson et al. \cite{Elson02:FGN} uses linear regression of the received synchronization timestamps to determine the optimal oscillator frequency.
The \ac{TPSN} from Ganeriwal et al. \cite{Ganeriwal03:TSP} is based on the \ac{MAC} layer of the communication interface. It uses the \ac{MAC} layer to exchange the synchronization timestamps between the devices. As a result, fluctuations in packet latency are minimized and can be easily compensated.

In the current setup of the system multiple hubs provide the external time base for the badges. Therefore it is necessary to synchronize the hubs with each other. Since the hubs have access to the internet, the \ac{NTP} is used for this synchronization \cite{Lederman16:OHP}.
Due to its complex protocol and energy inefficiency, \ac{NTP} has not been selected as synchronization technique between the badges and the hubs.
%To synchronize the badges with the hubs \ac{NTP} is not a suitable synchronization technique, because of its complex protocol implementation and energy inefficiency.
The \ac{TPSN} is also not applicable because it requires access to the \ac{MAC} layer of the communication interface, but the used \ac{SDK} allows no direct access to the \ac{BLE} \ac{MAC} layer. 
%Nachteil von linear regression (cite: paper) harder to implement, requires many samples in a short time interval (approximately 10 samples per minute). If time interval too big, ggf schon wieder Oscillator abweichungen.
Due to the drawbacks of the described methods a simple, energy efficient but nevertheless effective time synchronisation technique was developed in this work.
This technique has the advantage that it is compatible with the protocol of the previous firmware and therefore no changes to the existing hub software have to be done.
The time synchronization process of the badge is as follows: The hub connects to the badge and sends a request containing the current timestamp. When the badge receives this timestamp message, it retrieves the current number of oscillator ticks. Finally, the badge takes the number of ticks and the received timestamp to adjust its internal time.

The resulting equations are explained in more detail in the following. 
The current time \(t_{cur}\)  is calculated similarly to equation \ref{eq:synchronization}. The only difference is that the slope of the straight line is no longer constant.
The slope \(\overline{m}_{i}\) is the \ac{EWMA} of the latest optimal slope \(m_{i}\). The latest optimal slope \(m_{i}\) is the slope of the line between the \(i-1\)-th and the \(i\)-th synchronization point.
The optimal slope \(m_{i}\) is limited by plausible values for the oscillator frequency. The limiting parameter is \(f_{dev}\), which describes the maximum deviation from the nominal frequency. This limitation enables robustness against outliers. 

\begin{align} 
t_{cur} & = \overline{m}_{i} * (ticks_{cur} - ticks^{i}_{sync}) + t^{i}_{sync} \label{eq:average_synchronization} \\
\overline{m}_{i} & = \begin{cases}
m_{1}, & i=1\\
\alpha * m_{i} + (1 - \alpha) * \overline{m}_{i-1}, & i > 1
\end{cases} \\
m_{i} & =
\begin{cases}
\frac{1}{32768 \text{ Hz}}, & i=1\\
\min\left\lbrace \max \left\lbrace \frac{t^{i}_{sync} - t^{i-1}_{sync}}{ticks^{i}_{sync} - ticks^{i-1}_{sync}} ; \frac{1}{32768 \text{ Hz} + f_{dev}} \right\rbrace ; \frac{1}{32768 \text{ Hz} - f_{dev}} \right\rbrace, & i > 1
\end{cases}
\end{align}

\begin{figure}[h]
 \begin{minipage}[c]{0.5\textwidth}
 \centering
 \includegraphics[width=\textwidth]{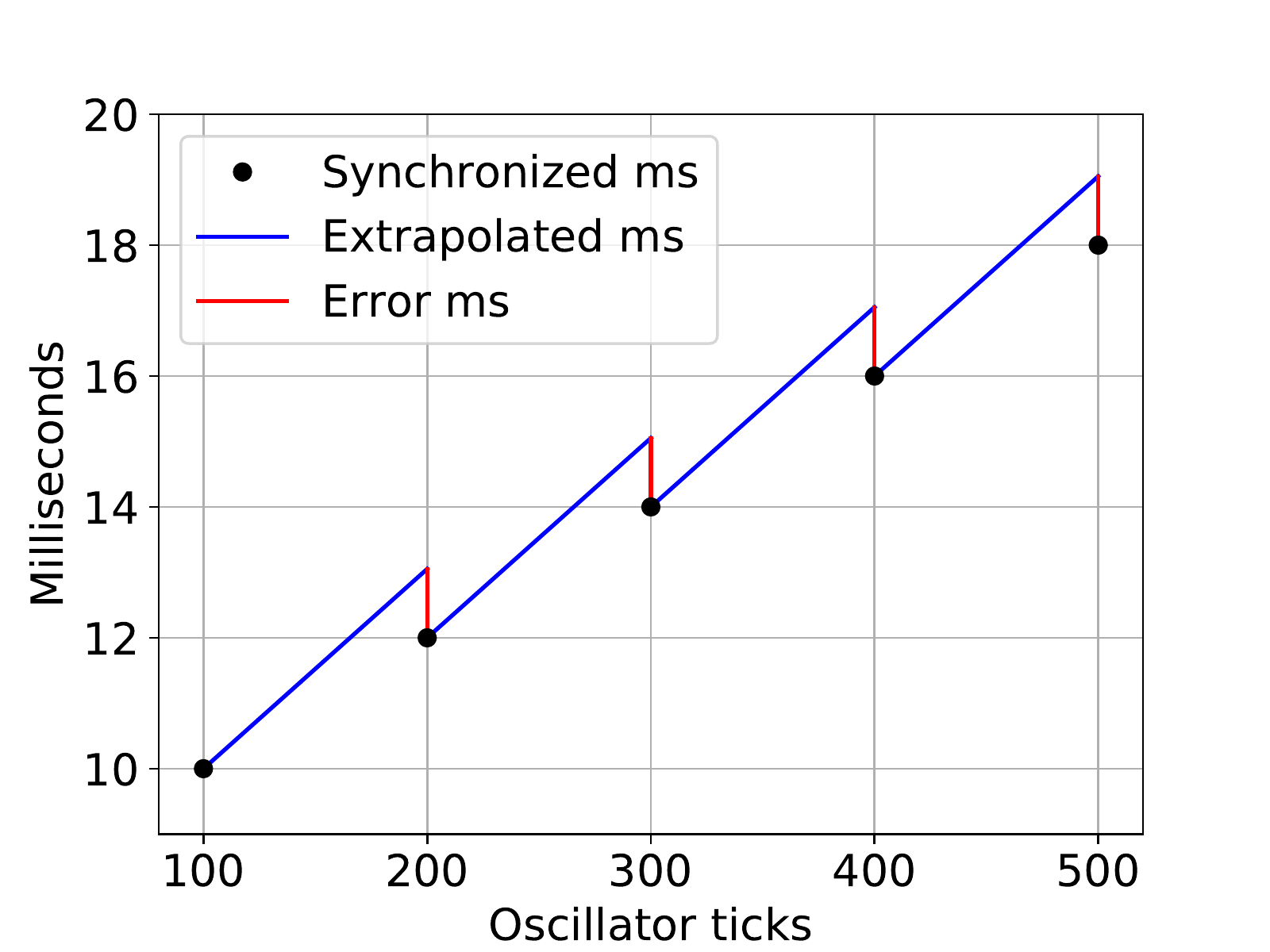}\\
 \vfill 
 (a)
 \end{minipage}
 \begin{minipage}[c]{0.5\textwidth}
 \centering
 \includegraphics[width=\textwidth]{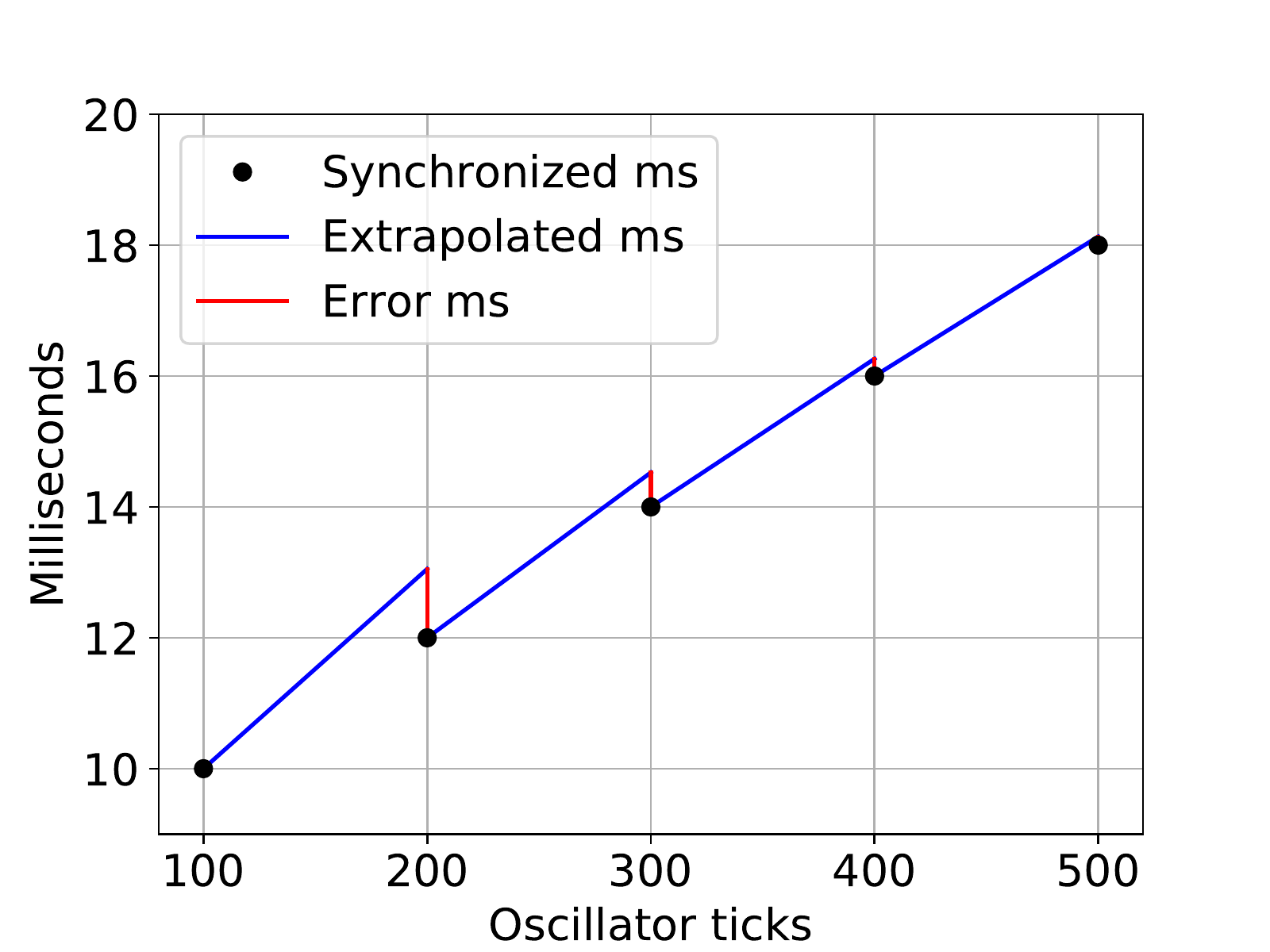}\\
 \vfill 
 (b)
 \end{minipage}
 \caption[Illustration time synchronization]{\textbf{Illustration time synchronization}. Figure (a) shows the constant slope approach. The black circles represents the synchronization points, the blue line corresponds to the line equation \ref{eq:synchronization} with constant slope and the red line illustrate the error between the internal time and the synchronized time. Due to the constant slope of the straight line, the error cannot be reduced. In contrast, Figure (b) illustrates the method with variable slope from equation \ref{eq:average_synchronization}. In this case, the error decreases because the slope of the straight line approaches the true slope.}
\label{fig:illustration_time_sync}
\end{figure}

\newpage
The presented time synchronization technique enables the robust compensation of time drifts caused by oscillator frequency changes and is characterized by a simple iterative implementation with basic arithmetic operations. 
The tunable parameters of the method are the \(\alpha\) coefficient of the \ac{EWMA} and the maximum frequency deviation \(f_{dev}\).
In Figure \ref{fig:illustration_time_sync} the functionality of the method is illustrated and compared to the approch with a constant slope.

   % (\chapter{})
\cleardoublepage
\chapter{Results} \label{cha:Results}

In this chapter, results are presented for the various methods described in detail in the preceding chapter.
First, the developed serialization library Tinybuf is evaluated.
Afterwards, the transmission speed of the different approaches for the exchange of large amounts of data via the bi-directional \ac{BLE} connection is analyzed.
Then the implemented clock synchronization technique is investigated.
Subsequently, the audio data recording strategy is examined.
Finally, the results of the power consumption analysis of the badge in different operating modes is presented.

\section{Serialization performance}
An important factor for both data transmission and data storage is the performance of the de-/serialization technique.
The performance can be divided into two categories.
The first one is the resulting size of the serialized binary representation of a message. The smaller this size, the more data can be transferred or stored.
The second one is the time required to perform the serialization or deserialization of data.
The longer the execution time, the more energy is consumed. Therefore, this time should be as short as possible.

To evaluate the performance of the developed de-/serialization library Tinybuf, Google's Protocol Buffers library is used as reference.
Since most messages of the badge for transmission and storage typically consist of arrays, a simple message with a repeated field is used for evaluation.
The repeated field should contain up to 100 unsigned integer values with a maximum of 2 bytes.
On the one hand, the Tinybuf generator creates a 2 byte unsigned integer array with 100 elements. On the other hand, Protocol Buffers uses an array of 100 \ac{varint} encoded unsigned integer elements.

\begin{figure}[h]
 \centering
 \includegraphics[height=9cm]{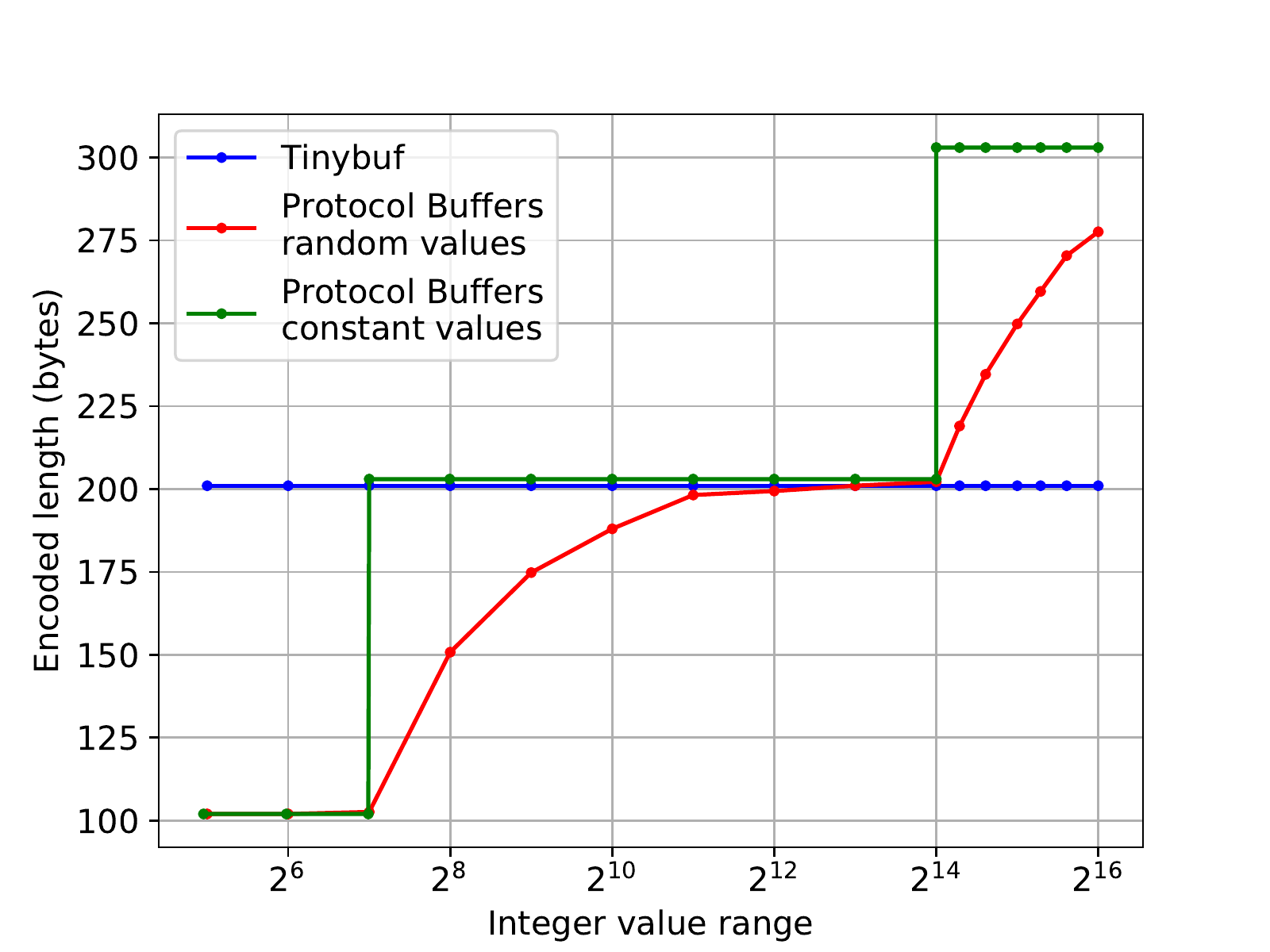} 
 \caption[Encoded length]{\textbf{Encoded length}. The plot shows the comparison between the encoded length for Tinybuf and Protocol Buffers for an array of 100 integer elements.}
\label{fig:serialization_encoded_length}
\end{figure}

Due to the \ac{varint} encoding of the Protocol Buffers library, the resulting encoded length depends on the element values. 
In order to evaluate this influence, two different approaches are used to generate the integer values.
In the first approach all array elements have the same integer value. The integer value range is from \(2^{5}\) to \(2^{16}\).
In the second approach, the integer values are not constant across the array, but are randomly derived from an uniform distribution.
The value range for the uniform distribution starts at 0 and ends with the same values as for the first approach. The resulting encoded length for Tinybuf and Protocol Buffers is shown in Figure \ref{fig:serialization_encoded_length}. The encoded length for Tinybuf is constant for the investigated integer values. For Protocol Buffers, the encoded length depends on the integer values of the array.

\begin{figure}[h]
 \begin{minipage}[c]{0.5\textwidth}
 \centering
 \includegraphics[width=\textwidth, trim=0cm 0cm 1cm 1cm, clip]{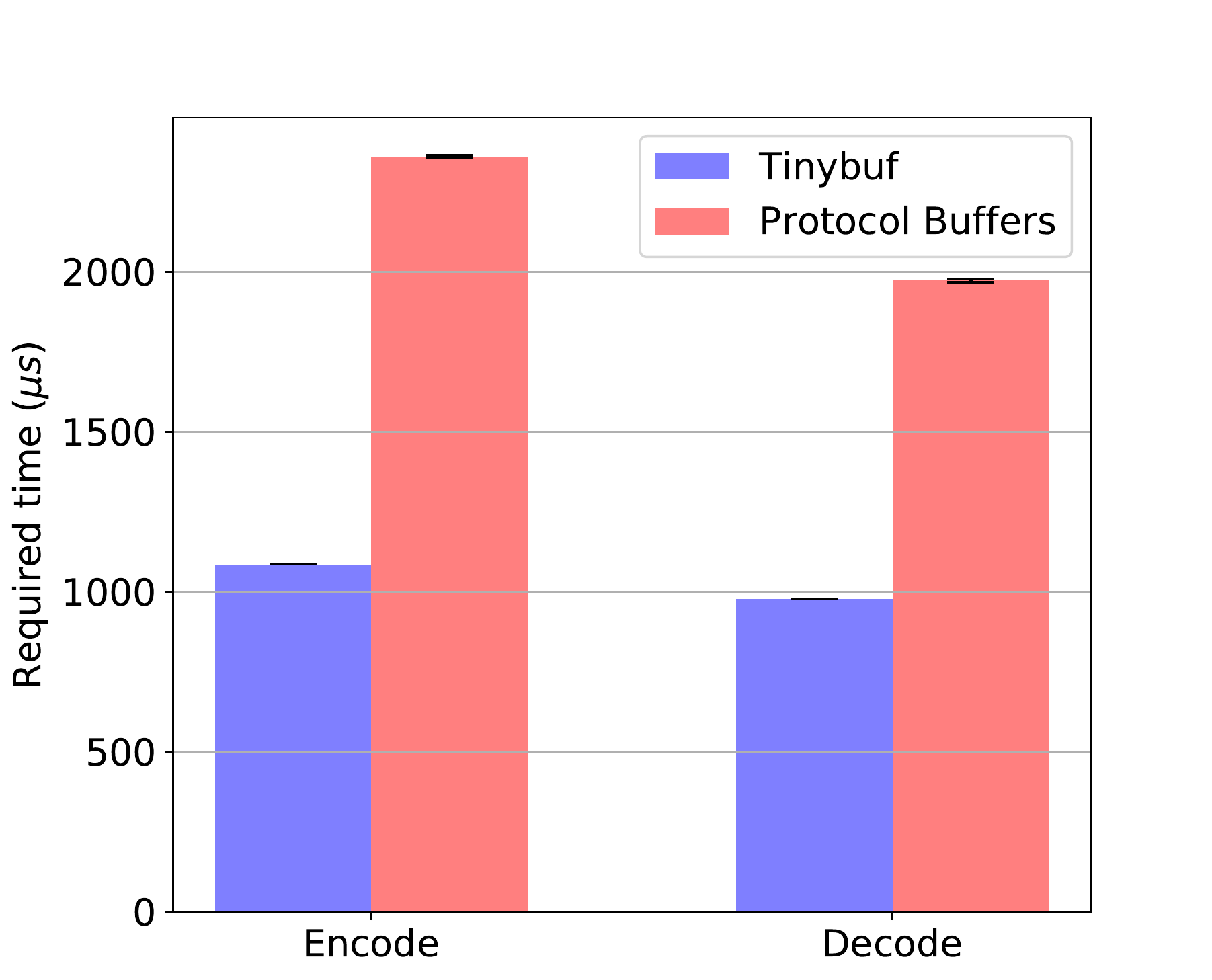}\\
 \vfill 
 (a)
 \end{minipage}
 \begin{minipage}[c]{0.5\textwidth}
 \centering
 \includegraphics[width=\textwidth, trim=0cm 0cm 1cm 1cm, clip]{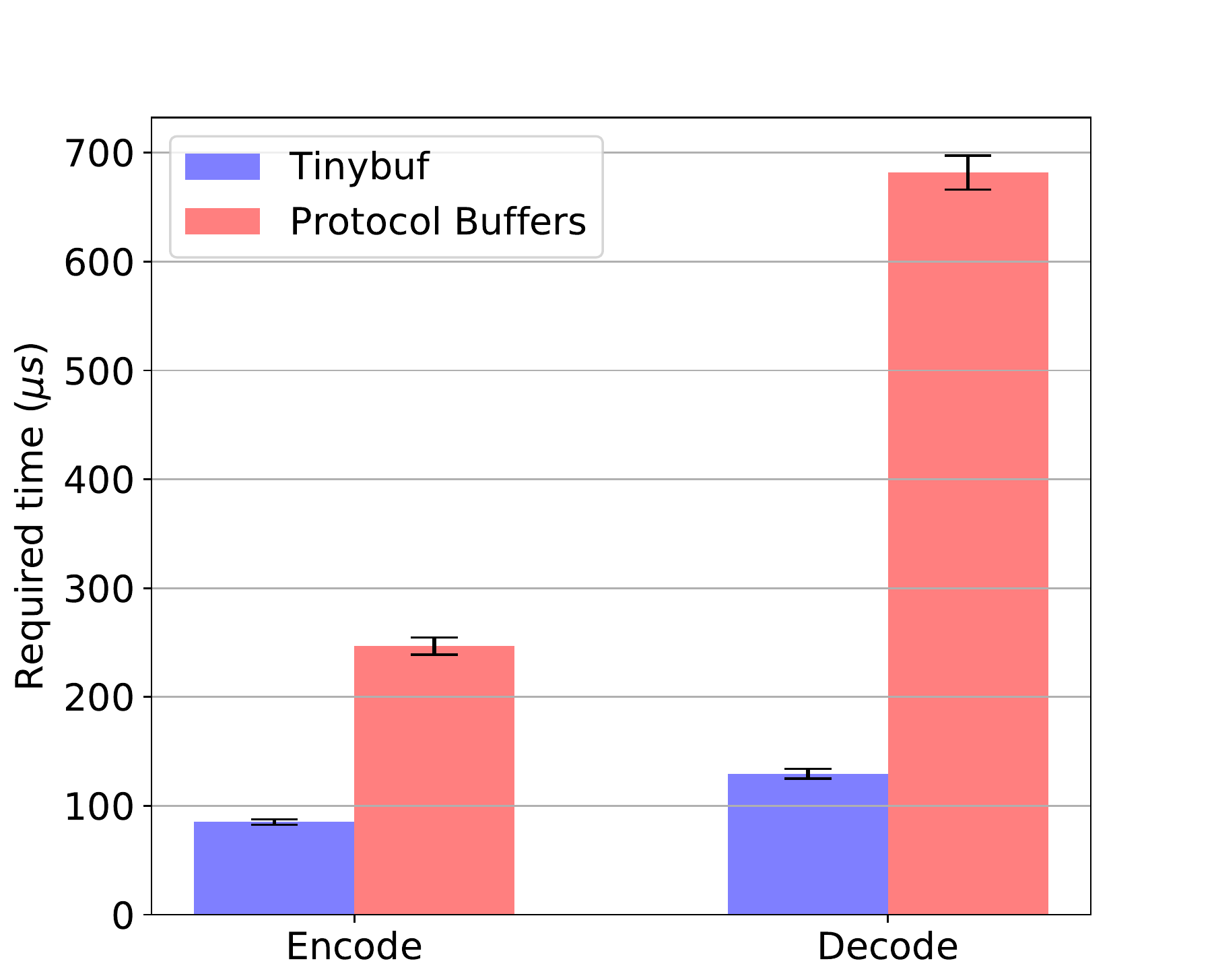}\\
 \vfill 
 (b)
 \end{minipage}
 \caption[Encoding and decoding execution times]{\textbf{Encoding and decoding execution times}. For the analysis of the execution times, integer values have been chosen so that the resulting encoded length for Tinybuf and Protocol Buffers is identical. In (a) the required time of the C implementations executed on the nRF51822 is depicted. The results for the Python implementations are shown in (b).
The execution times have been averaged over multiple measurements.}
%The variation of the measured execution times is represented with the standard deviation.}
\label{fig:serialization_time}
\end{figure}

Both Tinybuf and Protocol Buffers support the programming languages C and Python. For the evaluation of the execution time for encoding and decoding, the described message with a repeated field consisting of 100 integer elements is also used. The evaluation of the C implementation has been performed on the nRF51822. The Python based implementation has been analyzed on a 64-bit Windows 10 machine with i5 2.5 GHz dual core.
The results of the required times of both implementations are shown in Figure \ref{fig:serialization_time}. It can be seen, that for both implementations Tinybuf is faster than Protocol Buffers.

\section{Transmission speed}
The transmission speed or data throughput of the \ac{BLE} interface is an important factor for the functionality of the system.
The transfer of multiple microphone-chunks is used to determine the data throughput of the badge.
The transferred microphone-chunks have a constant size of 127 bytes. Based on the number of chunks received and the time elapsed since the request for the data, the remote \ac{BLE} device can calculate the transmission speed. The used \ac{BLE} connection interval is 50 ms.

The data throughput of the implemented timer-based approach, scheduler-based approach and SoftDevice callback approach are analyzed.
The data throughput of the timer-based approach, which periodically tries to transmit the next 20 bytes via the Nordic UART Service, depends on the selected timer period.
The theoretically possible data throughput as a function of the timer period \(T_{Timer}\) can be computed as follows:

\begin{equation}
\text{Throughput} = \frac{20 \text{ bytes}}{T_{Timer}}
\end{equation}

\begin{figure}[h]
 \centering
\includegraphics[height=8.5cm]{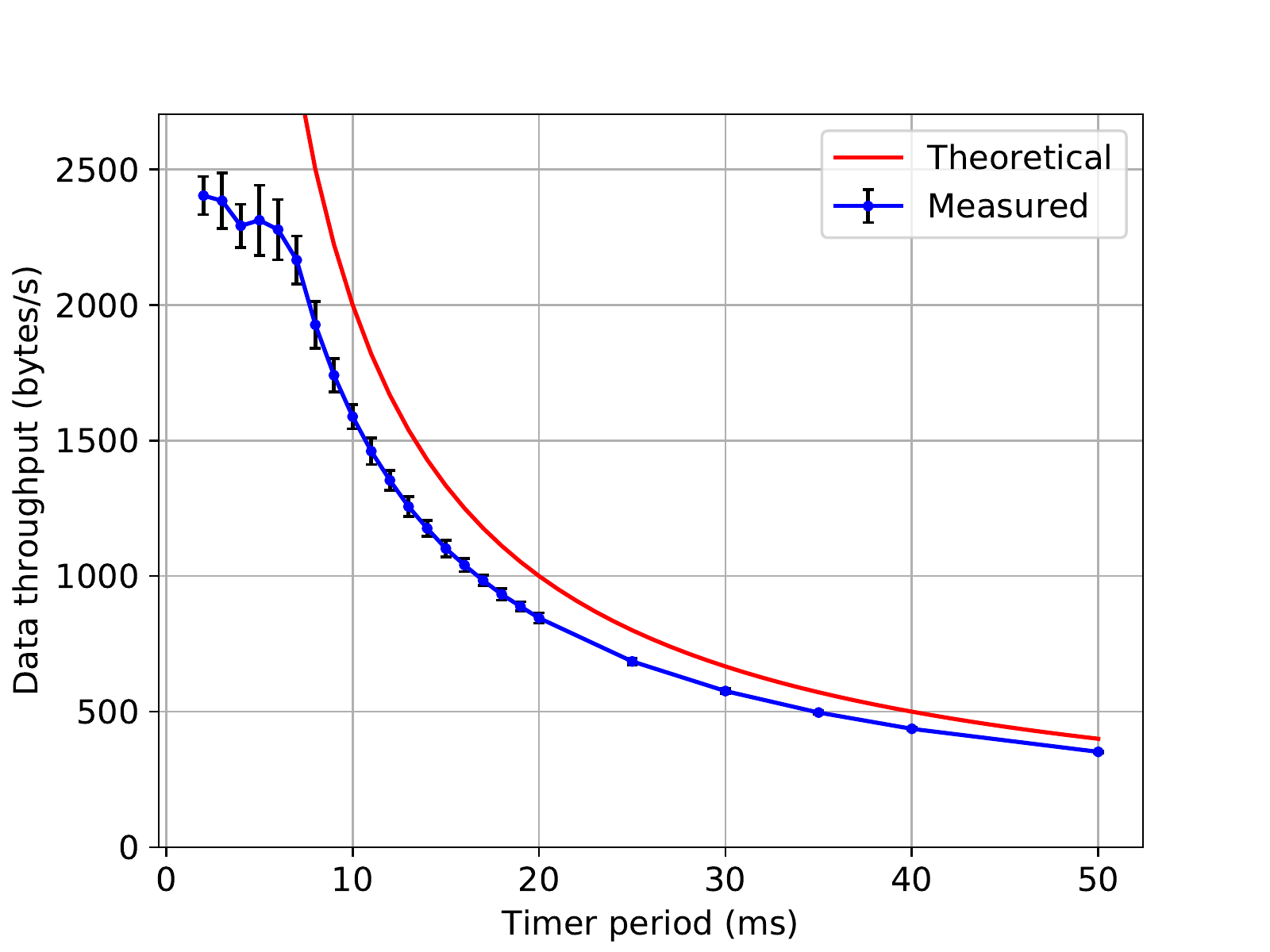}
 \caption[Throughput timer-based approach]{\textbf{Throughput timer-based approach}. The measured data throughput of the timer-based approach as function of the timer period is depicted by the blue line. In contrast, the red line illustrates the theoretical possible data throughput.}
\label{fig:speed_timer_period}
\end{figure}

\begin{figure}[h]
 \centering
\includegraphics[height=8.5cm]{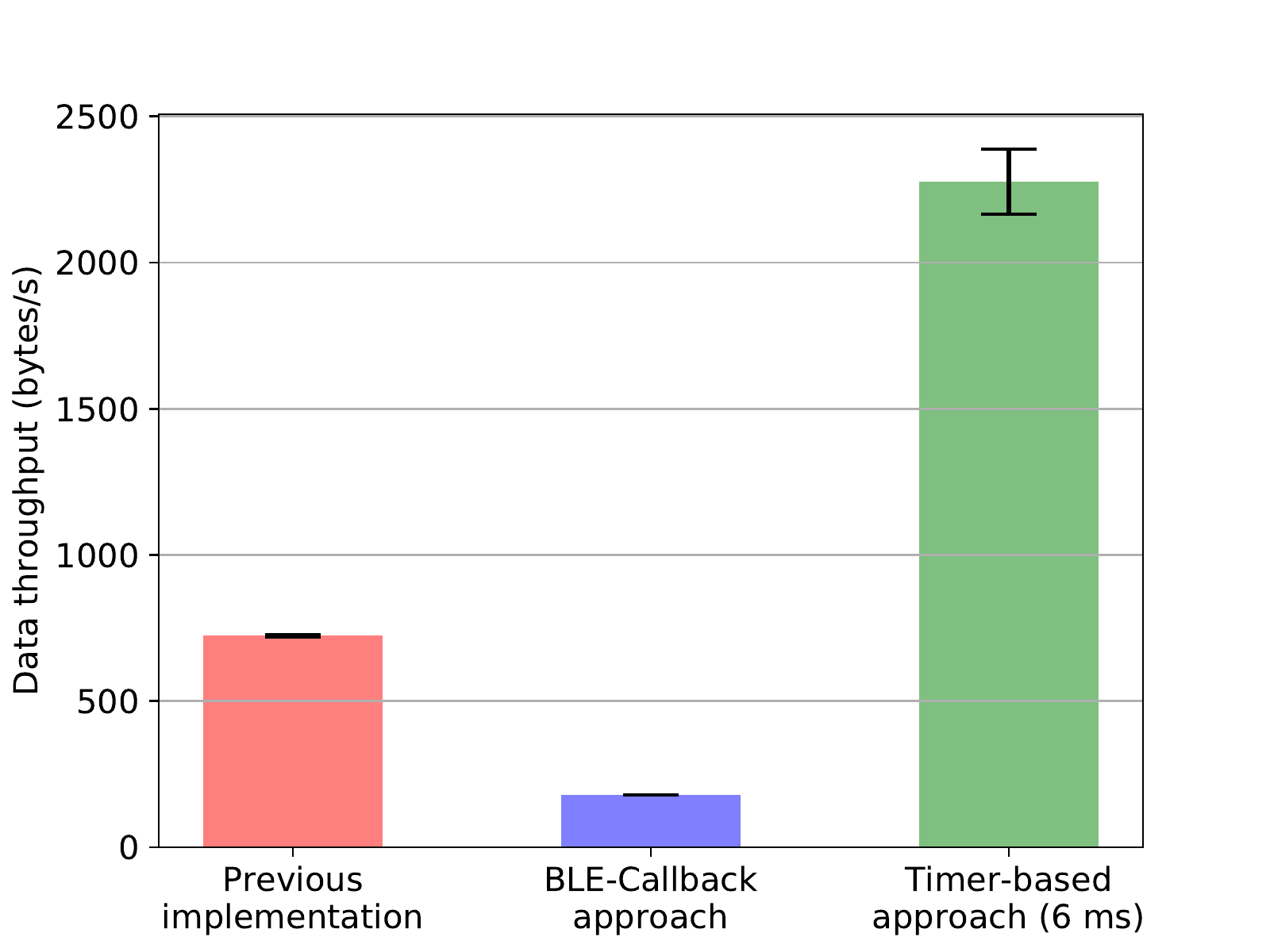}
 \caption[Comparison data throughput]{\textbf{Comparison data throughput}. The previous implementation of the firmware uses the scheduler-based approach for data transmission. The SoftDevice callback approach is the slowest approach. The timer-based approach developed in this work is the fastest for a certain timer period.}
\label{fig:speed_comparision}
\end{figure}

\noindent Figure \ref{fig:speed_timer_period} shows the measured data throughput of the timer-based approach in dependency of the timer period.
%The measured data throughput of the timer-based approach in dependency of the timer period is shown in Figure \ref{fig:speed_timer_period}.
Figure \ref{fig:speed_comparision} presents a comparision between the three different approaches.
%A comparison between the data throughput of the three different approaches is presented in Figure \ref{fig:speed_comparision}.

% Discussion: Break-even point: Durch das Auslesen beeinflusst

\section{Clock synchronization}
In order to determine the accuracy of the implemented clock synchronization technique, synchronisation messages have been sent to the badge in randomized intervals between 0 and 600 seconds, which closely approaches a realistic deployment scenario.
Each time the badge receives a message, it outputs the received timestamp and the oscillator ticks at the reception point via its serial interface.
This serial output is recorded and afterwards the set of tuples is used to evaluate the synchronisation accuracy.

\begin{figure}[h]
 \centering
 \includegraphics[height=9cm]{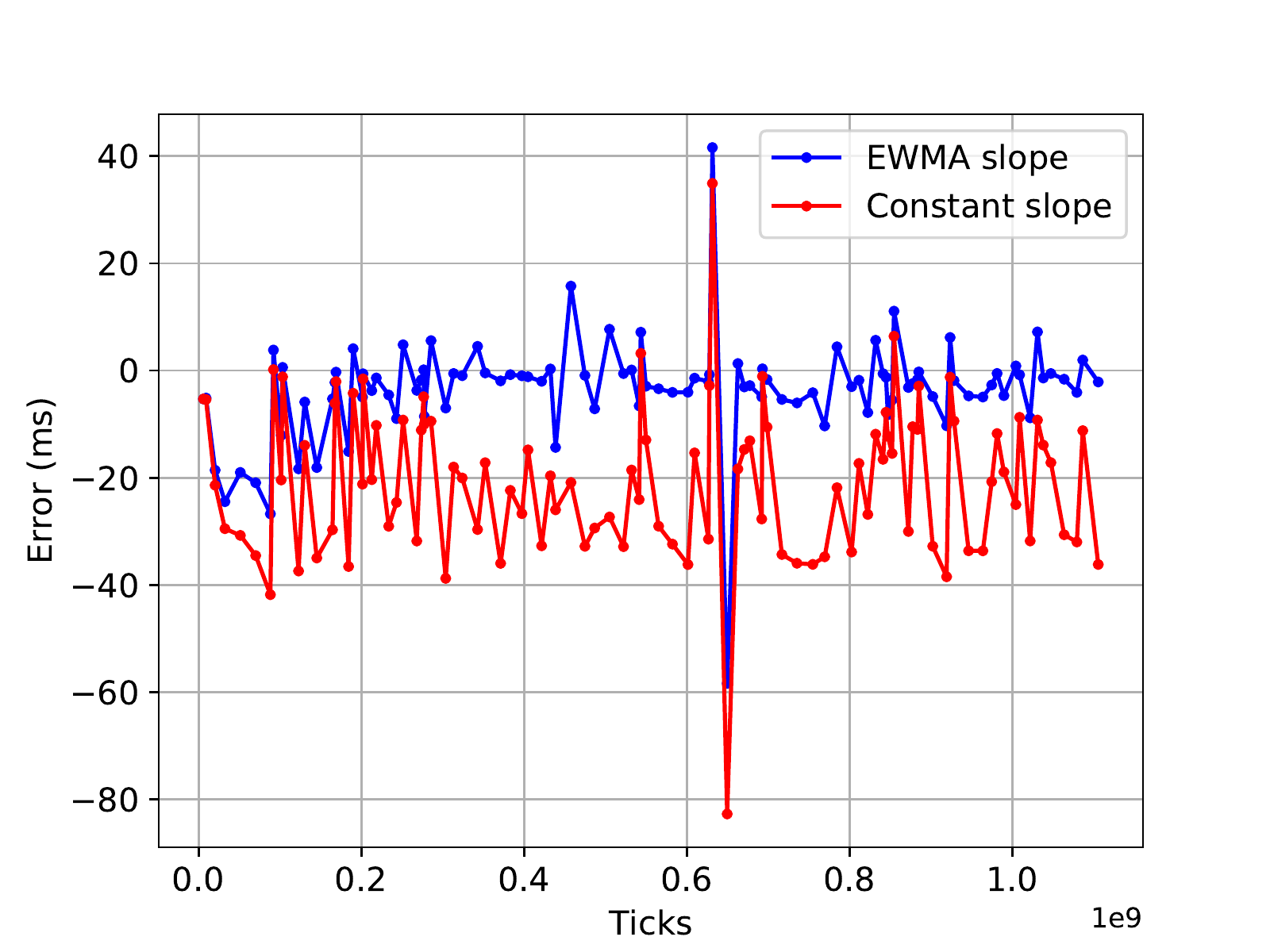} 
 \caption[Typical error graphs]{\textbf{Typical error graphs}. The red graph represents the errors at the synchronization points when a constant oscillator frequency of 32768 Hz is assumed. In contrast, the blue graph shows the errors when the \ac{EWMA} slope technique with the parameters \(\alpha = 0.1\) and \(f_{dev} = 4.0\) is applied. A negative error implies that the internal time of the badge is slower than the external time.}
\label{fig:ticks_millis_error}
\end{figure}

For the evaluation of accuracy, the difference between the received timestamp and the internal time at the reception point is used. The higher this difference (error), the less accurately the internal clock is synchronized.
In Figure \ref{fig:ticks_millis_error} a typical error graph over a period of approximately 9.5 hours is shown.

\begin{figure}[h]
 \centering
 \includegraphics[height=9cm]{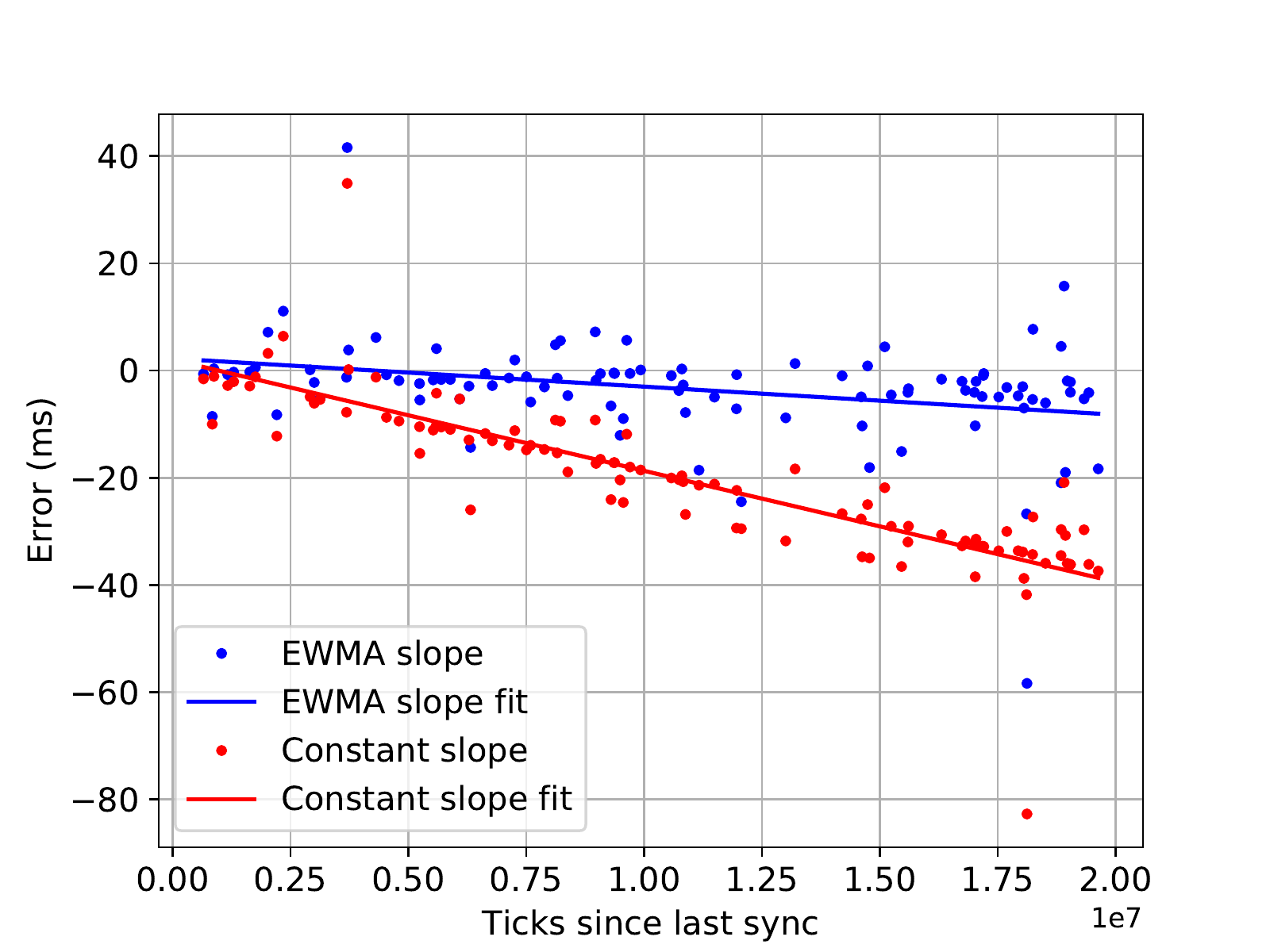} 
 \caption[Frequency offset]{\textbf{Frequency offset}. The x axis represents the time difference to the last synchronization message. The plots show the measured time error and a fitted straight line to estimate the frequency deviation.}
\label{fig:ticks_since_last_sync_millis_error}
\end{figure}

The error of the constant slope method can be explained by a constant frequency deviation of the oscillator frequency. This frequency offset is illustrated in Figure \ref{fig:ticks_since_last_sync_millis_error}. It shows the synchronization error in dependency of the time since the last synchronization. 
%In general, if the time between two synchronization points is longer, the resulting synchronization error increases.
The longer the time since the last synchronization point, the greater the accumulation of the internal time error due to a deviating oscillator frequency. Especially, the constant slope method suffers from this effect. The \ac{EWMA} slope technique can almost completely compensate this oscillator deviation.

\begin{figure}[h]
 \centering
 \includegraphics[height=9cm]{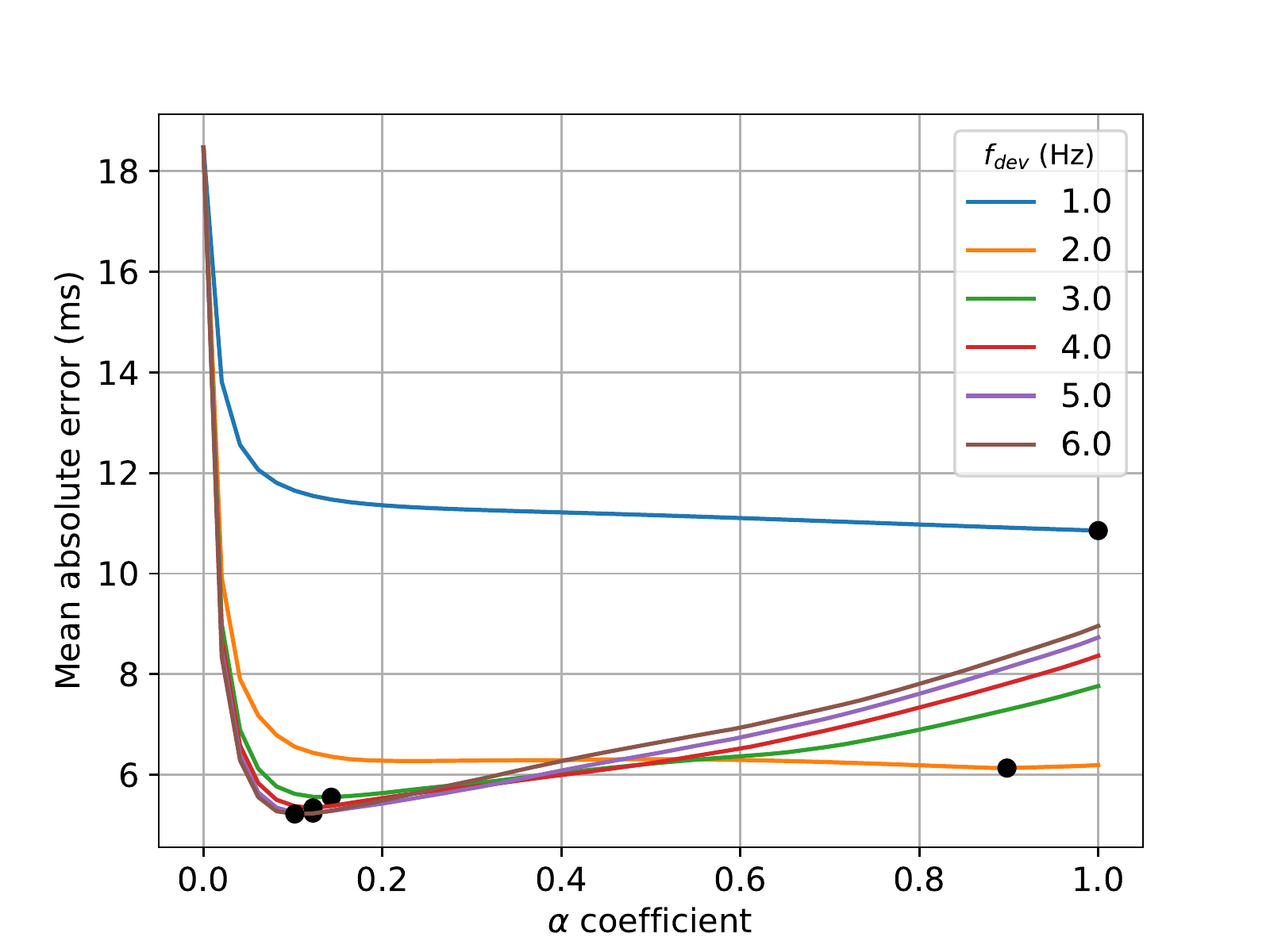} 
 \caption[Clock synchronization optimization]{\textbf{Clock synchronization optimization}. The plot shows the mean absolute error in dependence of the parameters \(\alpha\) and \(f_{dev}\). The optimal point for each parameter combination is marked with a black circle. In this measurement, the optimal parameter combination is at \(\alpha = 0.102\) and \(f_{dev} = 6\).}
\label{fig:parameter_optimization}
\end{figure}

The tunable parameters for the developed \ac{EWMA} slope technique are the \(\alpha\) coefficient and the maximum frequency deviation \(f_{dev}\).
These parameters have to be optimized to achieve a small synchronization error. 
As measure for the goodness of a parameter combination, the mean of the absolute synchronization errors has been taken.
The smaller this value, the better the parameter combination. 
An example for an optimization of the two parameters is shown in Figure \ref{fig:parameter_optimization}. A value of 0 for \(\alpha\) corresponds to the initial case with constant slope and its average absolute error can be regarded as a reference value for the optimization.
The optimal parameter combination found for several measurements is shown in Table \ref{tab:alpha_fdev}.

\begin{table}[h]
\renewcommand{\arraystretch}{1.5}
\centering
\begin{tabular}{c|c|c}
 & \textbf{\(\boldsymbol{\mu}\)} & \textbf{\(\boldsymbol{\sigma}\)} \\ \hline
 \(\boldsymbol{\alpha}\) & 0.11 & 0.089 \\ \hline
 \textbf{\(\boldsymbol{f_{dev}}\)} & 3.833 & 0.898
\end{tabular}

\caption[Optimal parameters clock synchronization]{\textbf{Optimal parameters clock synchronization}. The table shows the mean (\(\mu\)) and the standard deviation (\(\sigma\)) of the distribution of the optimal parameters \(\alpha\) and \(f_{dev}\) for several measurements.}
\label{tab:alpha_fdev}

\end{table}

%Wenn alpha == 0, dann hat man den Standardfall mit 32768 Hz ohne moving Average kann als Vergleich genommen werden

%Mehrere Messungen machen vergleichen. Average error erkl"aren

%Tabelle mit Standarddeviation...

\section{Audio data}
The low-pass filtered audio signal of the microphone is sampled with a period of 1.42 ms.
Every 1.42 ms the timer callback function is invoked for reading the digitized value from the \ac{ADC}.
In the previous implementation of the firmware, not one but several readings are taken at each function call to provide more samples for averaging. This averaging is performed every 50 ms.

\begin{figure}[h]
 \centering
 \includegraphics[height=9cm]{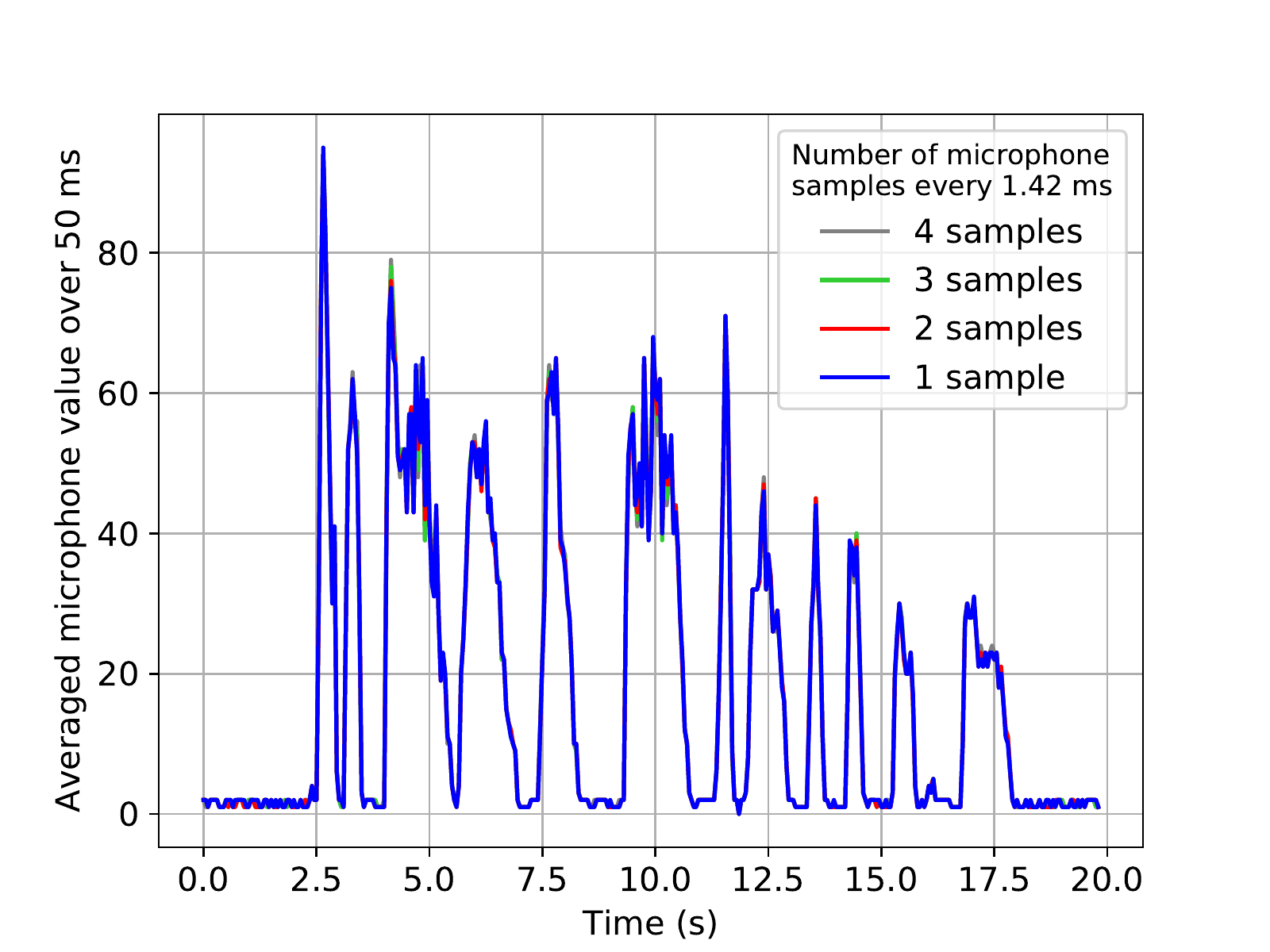} 
 \caption[Audio signal with different sample number]{\textbf{Audio signal with different sample numbers}. The figure shows an exemplarily averaged audio signal with noises for a 20 seconds time interval. The averaged audio signals are computed based on different number of samples every 1.42 ms.}
\label{fig:audio_comparision}
\end{figure}

Since the sampling process of the microphone is rather power-intensive, it was examined in more detail.
The previous implementation spends about 0.1 ms every 1.42 ms to perform microphone readings. During this time interval, two microphone readings can be taken.
To determine the influence of the number of samples, the resulting averaged signal is recorded and compared with different number of samples.
In Figure \ref{fig:audio_comparision} the results are shown.

%\begin{figure}[h]
% \begin{minipage}[c]{0.5\textwidth}
% \centering
% \includegraphics[width=\textwidth, trim=0cm 0cm 1cm %1cm, clip]{img/audio_comparision.pdf}\\
% \vfill 
% (a)
% \end{minipage}
% \begin{minipage}[c]{0.5\textwidth}
% \centering
% \includegraphics[width=\textwidth, trim=0cm 0cm 1cm %1cm, clip]{img/audio_comparision_error.pdf}\\
% \vfill 
% (b)
% \end{minipage}
% \caption[Comparision audio samples quantity]{\textbf{Comparision audio samples quantity}. Jeweils "uber 20 sekunden mehrere Messungen gemacht.}
%\label{fig:}
%\end{figure}

The goal is to minimize the number of samples in order to maximize the time the CPU is in sleep mode, leading to a reduced power consumption of the badge. 
Therefore, the quality of the averaged audio signal as function of the number of samples is investigated.
%Therefore, the difference between the averaged signal when using one sample and the averaged signal when using multiple samples is investigated.
The \ac{MAE} is used as a measure for the difference between two signals. This measure is suitable to quantify the difference between two signals \(x_{i}\) and \(y_{i}\) \cite{Willmott05:AOT}. The \ac{MAE} is computed based on \(N\) observation points with the following equation:

\begin{equation} \label{eq:MAE}
\text{MAE} = \frac{1}{N} \sum_{i=1}^{N}\abs{x_{i}-y_{i}}
\end{equation}

\begin{figure}[h]
 \centering
 \includegraphics[height=9cm]{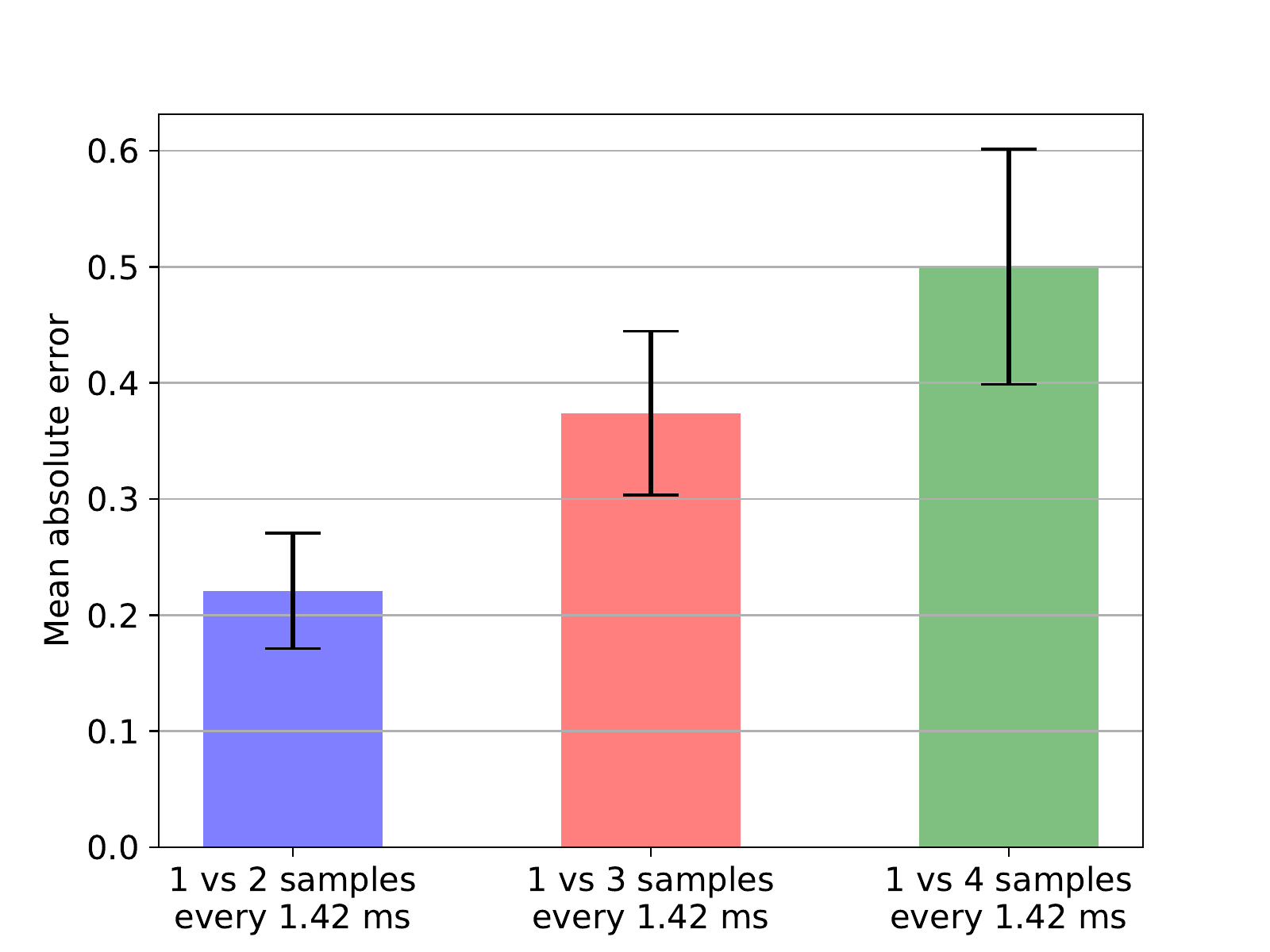} 
 \caption[Audio signal difference]{\textbf{Audio signal difference}. The signal error between different numbers of samples for the calculation of the averaged signal. The error is calculated over several measurements with a time interval of 20 seconds each (due to equation \ref{eq:MAE}).}
% In addition, the standard deviation of the errors is displayed.}
\label{fig:audio_comparision_error}
\end{figure}

\noindent The resulting \ac{MAE} based on multiple measurements is shown in Figure \ref{fig:audio_comparision_error}.
Since the signal difference between using one sample and using multiple samples is negligible, the new implementation reads only one sample every 1.42 ms to reduce power consumption.
% Because of that, one sample is enough to read!

\newpage

\section{Power consumption}
The total power consumption of the badge is composed of the consumption of its various components.
To determine this important quantity, the Power Profiler Kit of Nordic Semiconductor \cite{Nordic16:PPK} is used.
This kit supports a current measurement resolution down to 0.2 \textmu{}A at a sampling rate of 77 kHz.
The measured current of an external \ac{DUT}, such as a badge, is transfered via a nRF51 or nRF52 Development Kit to an application that plots the measured current in real-time.
To obtain a representative value for the power consumption, the measured current is averaged over a configurable time interval.

\begin{figure}[h]
 \centering
 \includegraphics[height=9cm]{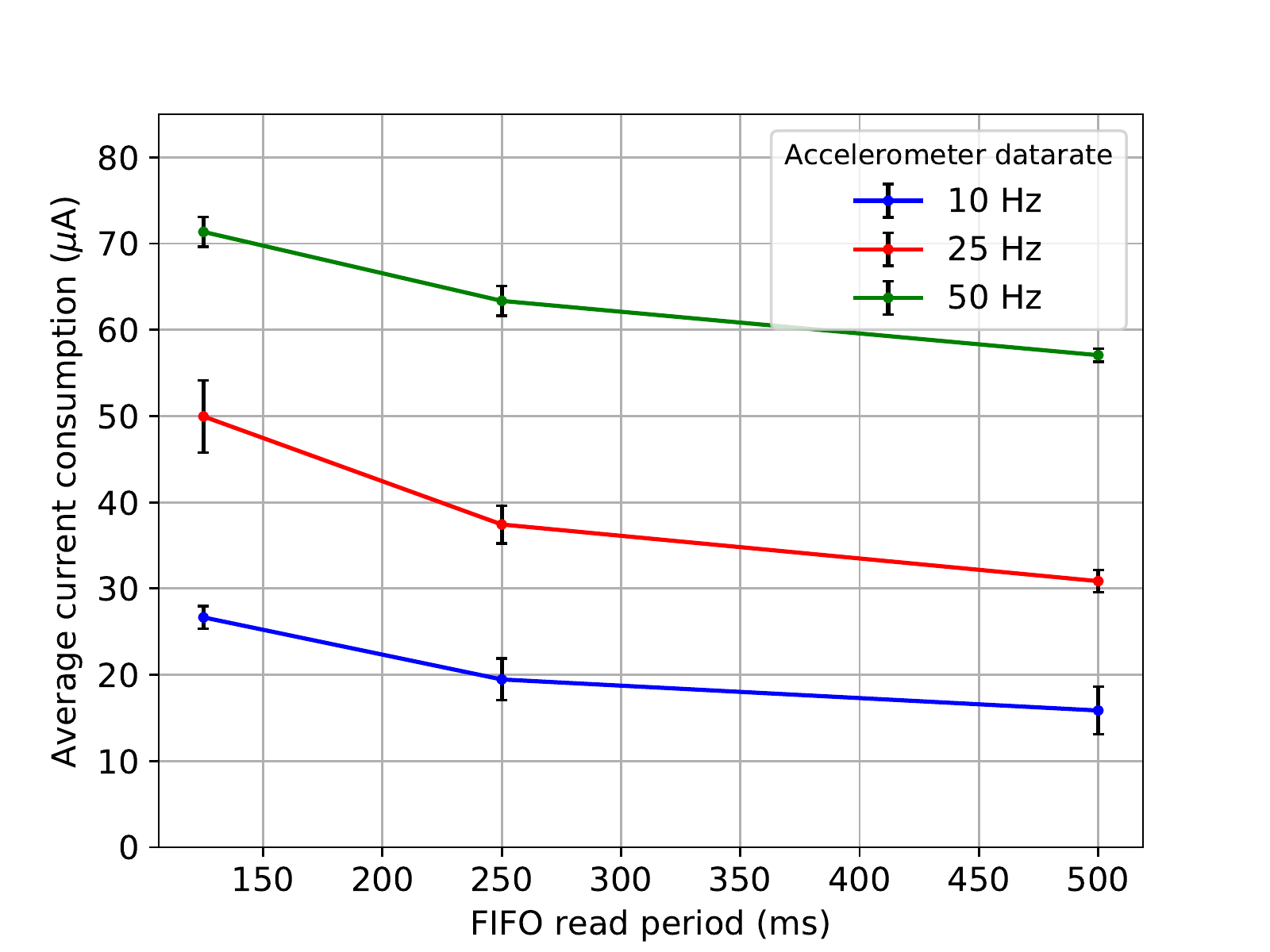} 
 \caption[Accelerometer consumption]{\textbf{Accelerometer consumption}. The figure shows the average power consumption of the badge as function of the \ac{FIFO} read period for different accelerometer datarates.}
\label{fig:current_comsumption_accel}
\end{figure}

The power consumption of the data sources, such as the accelerometer, is heavily dependent on their parameterization.
In the following, the measured value of each data source also includes the consumption of the associated data processing and storage.
Figure \ref{fig:current_comsumption_accel} shows the influence of the accelerometer datarate and the read period of the acceleration \ac{FIFO}.
The more frequently the FIFO is read, the more often the CPU has to leave the sleep mode, which results in increased power consumption.
The datarate of the accelerometer not only influences the power consumption of the accelerometer itself, but also that of the entire system, since the amount of data to be stored is also affected.

\begin{figure}[h]
 \centering
 \includegraphics[trim=1cm 0cm 4cm 2cm, clip, width=\textwidth]{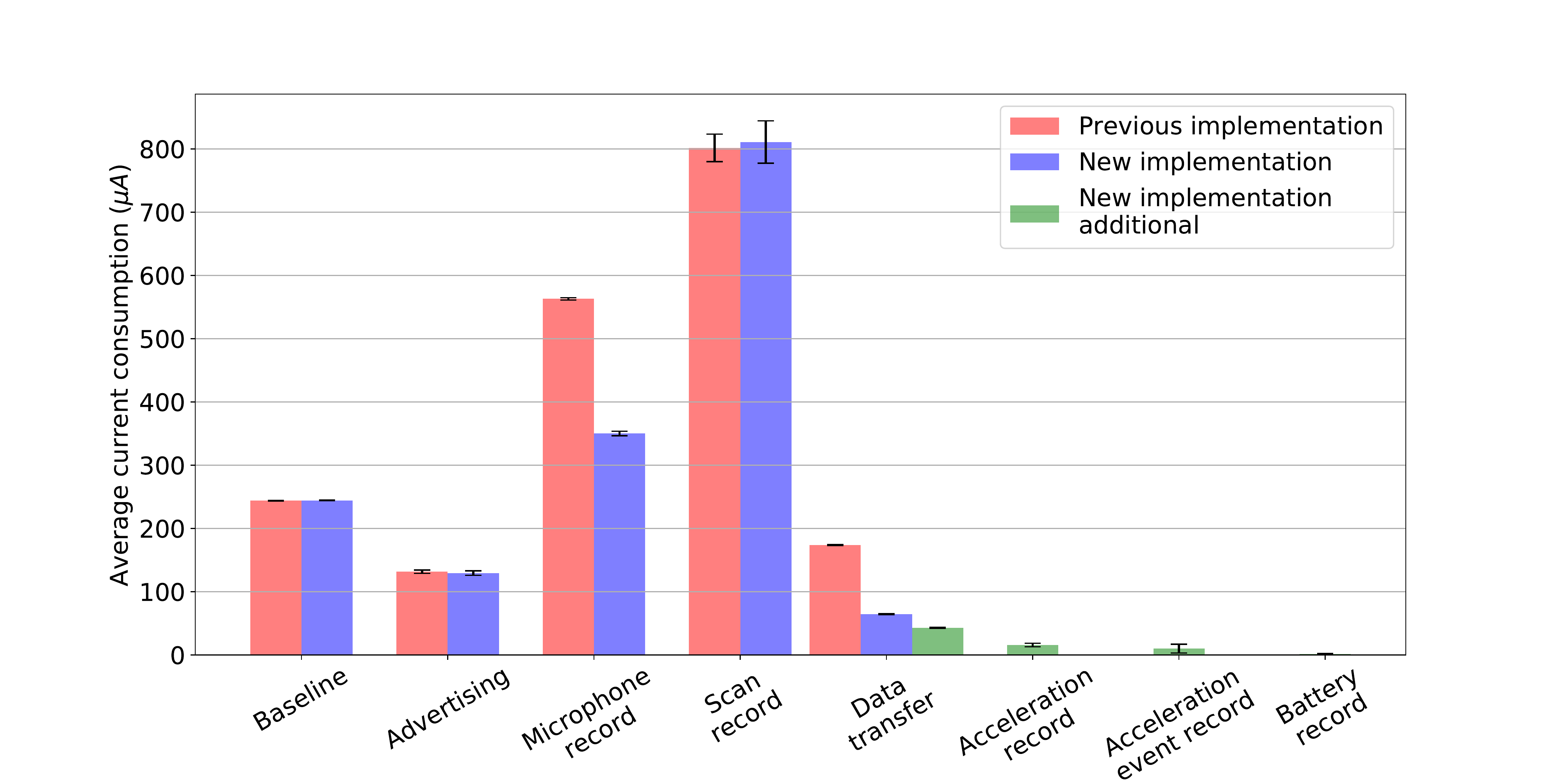} 
 \caption[Average current consumption overview]{\textbf{Average current consumption overview}. The average current consumption of the different components is depicted with bar graphs. For each component, the consumption of the previous implementation is compared to the new implementation. Since the supply voltage and accelerometer are not used as data sources in the previous implementation, the additional consumption of these components in the new implementation is considered separately.}
\label{fig:current_consumption_comparision}
\end{figure}

The averaged current consumption of each component is shown in Figure \ref{fig:current_consumption_comparision}.
The parameters that influence the power consumption of the badges have been selected appropriate to simulate a realistic deployment scenario.
The parameters and a description of the different components are summarized in Table \ref{tab:consumption_parameters}.

%The parameters selected for the data sources are closely to the 
%Since the power consumption depends on the parameterization of the badge, the measured consumptions are based on p

%The power consumption is measured with defined parameters set for the data recording

%The baseline consumption of the badge includes the consumption of the nRF51822 and the analog circuit of the microphone, which is enabled by default.

%The power consumption of the badge can be divided into four categories:
%The baseline consumption of the badge includes the sleep mode of the nRF51822 and the analog circuit of the microphone, which is enabled by default.
%The second category is the consumption of the \ac{BLE} advertising process of the badge.
%The next category is the data transfer to a remote \ac{BLE} device.
%Finally, the last category includes the power consumption of all components related to data recording.

\vfil

\begin{table}[h]
\renewcommand{\arraystretch}{1.5}
\centering
%p{3cm}|l|l|l|l
\begin{tabular}{p{2.5cm}|p{8cm}|p{4cm}}

\textbf{Component} & \textbf{Description} & \textbf{Parameters} \\ \hline
%Baseline & All peripherals enabled by default (nRF51822 sleep, analog microphone circuit) &  - \\ \hline
Baseline & Sleep mode and default enabled peripherals &  - \\ \hline
Advertising & The periodical broadcast of the advertising packet &  200 ms advertising period  \\ \hline
Microphone record & The microphone reading, averaging and chunk storage &   1.42 ms reading period\newline 50 ms average period  \\ \hline
Scan record & The \ac{BLE} device scanning, sorting and chunk storage &  100 ms scan window \newline 300 ms scan interval \newline 3 s scan duration \newline 15 s scan period \\ \hline
Data\newline transfer & Data reading and exchange via the \ac{BLE} connection (assumption: connection established every 5 minutes) & -
%6 ms timer period in new implementation 
\\ \hline
Acceleration\newline record & Acceleration reading, processing and chunk storage & 10 Hz datarate\newline Normal operation mode\newline 500 ms \ac{FIFO} period\\ \hline
Acceleration\newline event record & Acceleration event processing and storage (assumption: event detected every 5 seconds) & 10 Hz datarate\newline Normal operation mode\\ \hline
Battery record & Supply voltage reading and storage & 60 s reading period\\ \hline
\end{tabular}
\caption[Component description power analysis]{\textbf{Component description power analysis}. The table describes various components of the badge that are important for the power consumption analysis. In addition, the parameters used for the analysis are specified.}
\label{tab:consumption_parameters}

\end{table}   % (\chapter{})
\cleardoublepage
\chapter{Discussion} \label{cha:Discussion}
The goal of this work is the development of a new firmware for the badges of the Rhythm project.
%The focus is on the modularity and maintainability of the implementation.
Various methods were developed to provide a modular, testable and maintainable architecture of the application.
These include methods for efficient serialization and storage of structured data, increasing transmission speed, accurate clock synchronization, and optimizing power consumption.
%Furthermore, the new firmware optimizes the energy consumption, the transmission speed, the clock synchronization accuracy and the transmission protocol.
%In this chapter the methods presented in chapter \ref{cha:Methods} and the results of chapter \ref{cha:Results} are discussed in more detail and compared to the previous implementation of the firmware.
This chapter discusses the results in detail and compares them to the previous implementation of the firmware.
%In this chapter the results of chapter \ref{cha:Results} are discussed in more detail and compared to the previous implementation of the firmware.

\section{Serialization performance}
The developed de-/serialization technique Tinybuf offers the possibility to define messages in a flexible way, which can be serialized very efficiently. Tinybuf automatically generates source code for various programming languages.
This leads to faster development processes, since the de-/serialization functionality does not have to be implemented manually.
In the previous implementation of the firmware each message has its own de-/serialization function.
Therefore, in cases where existing messages have to be changed or new messages have to be added, this approch is inflexible.
%Especially if existing messages change or new messages are added, this approach is very inflexible.

To validate the performance of the developed technique, Tinybuf was compared to Google's Protocol Buffers library.
For the evaluation, a simple message was used containing a field of 100 integer elements, each with a size of 2 bytes.
The performance test was splitted into two parts: Evaluation of the resulting encoded length of the binary representation and the evaluation of the execution time for serialization and deserialization.

The results showed, that the encoded length produced by Tinybuf is always 201 bytes long, independent of the chosen integer value of the elements. 
In contrast, the encoded length of the message serialized with Protocol Buffers depends strongly on the individual values of the elements, since varint encoding is applied.
To substantiate the facts, if the integer values are smaller than \(2^{7}\), the encoded length is 102 bytes.
In case, that integer values are in the range from \(2^{7}\) to \(2^{14}\), the binary representation has a size of 203 bytes.
Finally, if all integer values are greater than or equal to \(2^{14}\), the resulting encoded length produced by Protocol Buffers is 303 bytes.
In a realistic scenario, the values are usually distributed randomly over a certain range of values.
In this case, the resulting encoded length of Protocol Buffers is partially smaller than, equal to or greater than the encoded length of Tinybuf, depending on the chosen range and distribution.
%In this case, the resulting encoded length for Tinybuf and Protocol Buffers is more or less comparable, depending on the value range and distribution. 
However, the size of the memory to be reserved in RAM for the encoded message differs significantly. For Protocol Buffers, the worst case must be assumed, which in this case corresponds to an encoded length of 303 bytes.
In contrast, Tinybuf guarantees an encoded length of 201 bytes.
Especially when multiple messages with repeated integer fields are defined, the use of Tinybuf considerably reduces the amount of RAM space to be reserved.

%However, in a realistic scenario the values, for example acceleration data, are not constant, but distributed randomly over a certain value range. Therefore, the encoded length evaluation has been performed again with random values drawn from a uniform distribution.
%In this evaluation, the encoded length of Tinybuf is again constant 201 bytes, but the encoded length of protocol buffers is in general smaller or equal compared to the previous evaluation.
%This can be explained by the fact that the array elements can now contain values smaller than \(2^7\), which can be encoded in only 1 byte.

\begin{figure}[h]
\centering
%\fbox{\includegraphics[page=1, trim=2.5cm 22.6cm 7.5cm 2.5cm]{img/scan_chunk.pdf}}
\includegraphics[page=1, trim=2.5cm 22.6cm 7.5cm 2.5cm, clip]{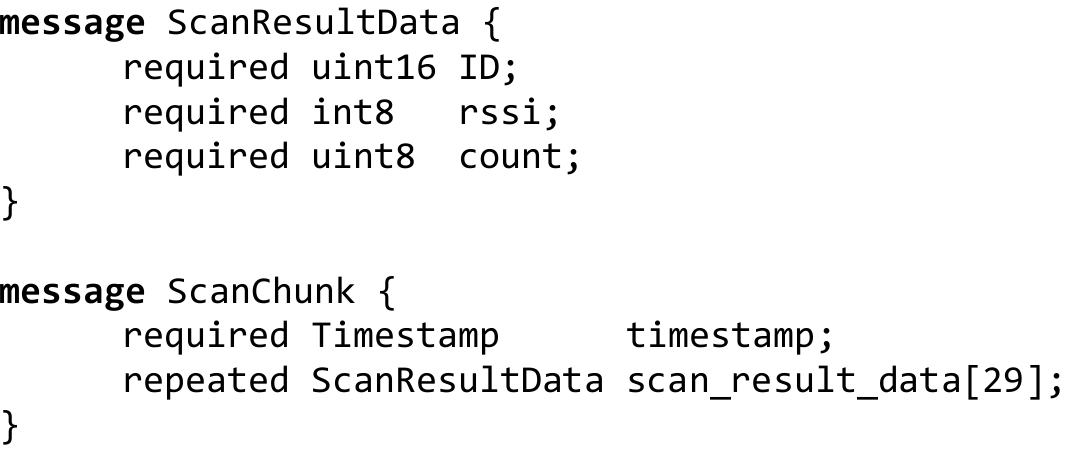}
\caption[Scan Chunk]{\textbf{Scan Chunk}. The message definition of a scan-chunk. It consists of a timestamp, and an array of scan result data (with device \ac{ID}, aggregated \ac{RSSI} value and counter).}
\label{fig:scan_chunk}
\end{figure}

In addition, a significant difference was found regarding the encoded length when using repeated fields with another message as data type.
%between Protocol Buffers and Tinybuf regarding the encoded length can be found when using repeated fields with another message as data type.
This is for example the case in the scan-chunk message. A scan-chunk contains a repeated field of messages, which includes the discovered device, the corresponding \ac{RSSI} value and a counter how often the device has been seen (see Figure \ref{fig:scan_chunk}).
In this case, Protocol Buffers adds a field identifier for each field in the ScanResultData-type, which in turn results in a high overhead in the encoded message. As an example: If 29 devices has been discovered in a scan, the resulting encoded length for Protocol Buffers is approximately 300 bytes. In contrast, Tinybuf encodes the message in only 123 bytes.

%In this case, Protocol Buffers adds a field identifier for each field in this embedded message, which leads to a high overhead.
%In contrast, Tinybuf adds no additional field identifier, but only encodes the necessary information to decode the message again.
%For the encoding of optional fields, Tinybuf inserts an additional byte which specifies whether the optional field is set or not. Protocol Buffers uses the inserted field identifiers to check if an optional field is present in the encoded message.
% Evtl. optional field rauslassen... Viel wichtiger: Warum ist das Protocol Buffer (varint) doof: Es muss immer vom worst case ausgegangen werden, und dadurch ggf. viel Speicher im RAM alloziert werden, um wirklich alles halten zu k"onnen.

%Thus, no additional byte needs to be inserted to indicate the presence of an optional field.

%Bessere Kalkulationen, da Gr"osse genau bekannt istFor data storage this would mean, that each partition has to be dynamic, resulting in a bit more overhead for each stored element, because the length information has to be stored.

The second performance criterion is the execution time of the encoding and decoding operations.
%For this evaluation the message with a repeated field of 100 integer values has been used again.
Both the C and the Python implementations of Tinybuf perform serialization and deserialization faster than Protocol Buffers.
The faster execution time of the encode and decode process can be explained by the fact that Protocol Buffers uses \ac{varint} representation.
This integer representation requires more operations than the simple integer byte representation used by Tinybuf.

%Tinybuf uses efficient memcpy operations in C and struct.pack and struct.unpack functions in Python.

%This can be explained by the usag

%Very efficient struct.pack - .unpack functions to retrieve formatted data from a binary representation.
As mentioned above, Tinybuf enables the efficient encoding and decoding of messages independent of the actual values of the message fields. Especially, the encoding of repeated message fields results in a significantly smaller binary representation compared to Protocol Buffers.
Futhermore, Tinybuf's serialized messages are compatible with the previous protocol implementation and therefore enables the simple integration into the existing system.

\section{Transmission speed}
The Nordic UART Service is used for the bi-directional data exchange between the badges and a remote \ac{BLE} device.
This service allows to transmit a maximum of 20 bytes at once. In order to exchange larger quantities of data, it is necessary to divide the data into 20 byte units and transmit them one after the other.
The time between two consecutive packets is crucial for the resulting transmission speed.
The shorter this time, the higher the resulting data throughput.
To perform the repeated transmission of 20 byte packages until no more bytes have to be transmitted, three different approaches were evaluated.

The timer-based approach repeats the transmission at a configurable period.
The measured effect of the timer period corresponds approximately to the expected theoretical effect.
In fact, the measured data throughput is smaller than the theoretical one.
This can be explained as a result of the measurement technique used to determine the throughput.

During one run, the method sends a microphone data request to the badge whereupon the badge transmits a certain amount of microphone-chunks.
The time recording starts with transmitting the request and stops with the reception of the last chunk.
This implies that not only the actual transmission time is measured, but also the time needed to read the chunks from the \acp{NVM} and to serialize them.
The faster the actual data transmission, the greater the influence of the additional time needed to generate the data packets.

Additionally, the theoretical possible data throughput is limited by the maximum transmission speed of the \ac{BLE} interface.
The maximum possible data throughput of the SoftDevice depends on the connection interval and the number of packets transmitted per connection event. The SoftDevice S130 can transmit up to 6 packets per connection event \cite{SoftDevice16:SS}.
Since the selected connection interval for the evaluation was 50 ms, the resulting maximum possible data throughput is 2400 bytes per second.

As the data throughput cannot be significantly increased with timer periods smaller than 6 ms, this value has been selected as the optimal timer period.
The resulting data throughput of the timer-based approach with a timer period of 6 ms is approximately 2300 bytes per second.
Since this value is by far the highest, this approach has been used in the new firmware.
%Since the data throughput for this timer-based approach is by far the highest, it is used in the new firmware.

In contrast, the scheduler-based approach applied in the previous implementation of the firmware provides a data throughput of approximately 730 bytes per second.
Here, the transmission of 20 bytes is repeated by inserting the function, which is responsible for the transmission, in the scheduler queue until no more bytes are left.
Since other operations are also inserted into this queue, for example the execution of data storage, the repeated transmission process is delayed and consequently the data throughput decreases.

The last approach, which uses the SoftDevice callback to trigger the transmission of the 20 byte packages, has the smallest data throughput with approximately 180 bytes per second.
The SoftDevice invokes the callback function if the packet has been transmitted and acknowledged successfully during a connection event.
Depending on the connection interval settings, the duration until the callback function is called can be rather long.

\section{Clock synchronization}
%For the correlation of the data of several badges, for example the analysis of the speech overlap between participants, it is necessary that the badges have access to an accurate time base.
The incorporated low-power \ac{RTC} oscillator with a nominal frequency of 32768 Hz is used as clock source for the internal time of the badge.
The time is synchronized by messages that are received from a remote \ac{BLE} central device, such as a hub, during a connection. These messages contain the current global timestamp to provide the same time base for all badges.
For an accurate time synchronization, various factors must be considered, such as the internal drift of the oscillator frequency, the accuracy of the external time base, and the duration between the transmission and reception of the synchronization message.

To evaluate the resulting accuracy of the developed technique with a self-adapting oscillator frequency, it is compared to the previous implementation with a constant frequency.
In the latter, long time intervals between the synchronization messages results in a significant time error caused by a deviating oscillator frequency.
The measured time error of the constant frequency approach was up to 40 ms for a synchronization period of 10 minutes.
In contrast, with the self-adapting frequency approach, the error could be reduced to 8 ms for the same synchronization period.
In order to simulate a real deployment scenario, the synchronization messages were transmitted in random time intervals.
To quantify the time error, the \ac{MAE} over a time period of 9.5 hours was computed.
The \ac{MAE} of the constant frequency approach was 19 ms. By using the self-adapting approach with an optimal choice of parameters, the \ac{MAE} was reduced to 5 ms.
The optimal parameter combination was determined over several measurements and multiple badges.
For the \(\alpha\)-coefficient of the \ac{EWMA} slope the value of 0.11 \(\pm\) 0.089 was found. For the frequency deviation \(f_{dev}\) the value of 3.833 \(\pm\) 0.898 Hz was identified.
In general, the smaller the value \(f_{dev}\), the larger \(\alpha\) must be chosen to compensate an oscillator frequency drift.
This can be explained by the fact that with smaller \(f_{dev}\) less influence is exerted on the current frequency at each synchronization point. Therefore, \(\alpha\) must be chosen larger to expedite the correction.
However, a too large \(\alpha\) causes outliers to have a strong influence on the frequency calculation and a too small \(\alpha\) implies a very slow compensation of the oscillator drift.
A similar situation applies for the choice of \(f_{dev}\):
On the one hand, a too small \(f_{dev}\) can prevent a high drift from being completely compensated, because \(f_{dev}\) limits the adjusted oscillator frequency to a predefined range. On the other hand, a too large \(f_{dev}\) allows outliers to have a high impact on the adjusted frequency.
Therefore, a compromise must be found between robustness against outliers, the fast compensation of the oscillator frequency drift and the maximum compensatable drift.
The found optimal parameter combination represents such a compromise.
The presented self-adjusting oscillator frequency technique significantly improves the internal time accuracy compared to the previous implementation.

One limitation of the self-adjusting frequency approach is that it assumes a constant time between the transmission and the reception of the synchronization message.
Since a \ac{BLE} connection is used for the transmission, the synchronization messages are only sent during connection events.
Depending on the connection interval settings, the latency between the generation of the message and its actual transmission can vary between a few milliseconds up to several seconds.
This is the reason, why there is noise in the time synchronization although the drift compensation is applied.
This random time offset can be prevented if the timestamp exchange is implemented on the \ac{MAC} layer of the communication interface as in the \ac{TPSN} \cite{Ganeriwal03:TSP}. Since the used \ac{SDK} provides no interface to control the \ac{BLE} \ac{MAC} layer, this technique cannot be applied.
Another approach to compensate this influence is the use of linear regression over many cycles of synchronization messages \cite{Elson02:FGN}. 
The implementation of the linear regression technique could be investigated in future developments of the system.

\newpage
\section{Audio data}

The recording of the audio signal generated by the microphone has a strong influence on the power consumption of the system.
Therefore, the audio recording process has been specially analyzed in order to find solutions for reducing the power consumption.
The amplified and low-pass filtered audio signal is sampled with a frequency of 700 Hz.
The recorded samples are averaged over a time interval of 50 ms to obtain a representative value for vocal activity within this interval.
The purpose of this averaging is to minimize the amount of data to be stored while still ensuring a good time resolution.

This sampling and averaging process is controlled by two timers: The first timer reads the raw audio signal at a period of 1.42 ms (\(\widehat{=}\) 700 Hz), accumulates the values in a variable and increments the counter in which the number of readings is stored. The second timer calculates the average of the audio signal every 50 ms by dividing the accumulated value by the counter.
In the previous implementation of the firmware, the first timer function reads not only one but two audio samples for the accumulation.
The detailed analysis showed that the quality of the audio signal was not significantly influenced by the number of chosen samples. In other words, the audio signal difference between using one sample and multiple samples is negligible.
To underpin this statement, the \ac{MAE} between the averaged signal over 20 seconds using one sample and two samples is 0.22 \(\pm\) 0.05.

The small difference between using one sample and multiple samples can be explained by the analog low-pass filter.
The low-pass filter limits the maximal frequency in the signal to its cutoff-frequency of approximately 340 Hz.
The time that is required to perform one \ac{ADC} reading, to accumulate the value and to increment the counter is in total approximately 50 \textmu{}s.
When reading multiple samples, this corresponds to a sampling frequency of 20 kHz. 
Since the frequency in the audio signal is limited to a maximum of 340 Hz, the difference between two consecutive samples recorded at a frequency of 20 kHz is negligible.
Based on this result, in the new implementation only one sample is recorded every 1.42 ms, leading to a reduction of power consumption.

For the later communication pattern analysis only the vocal activity, which indicates how often a participant speaks and at what volume, is of relevance.
Therefore, it would be sufficient to record the amplitude envelope of the audio signal.
In future designs an envelope detector \cite{Liao12:ASA} could be integrated to further reduce the sampling rate.

\section{Power consumption}
The power supply of the badge is provided by a 3 Volts coin cell battery. A common value for the capacity of these batteries is 230 mAh \cite{CR2032:CLM}.
As this available amount of energy is rather low, the overall power consumption of the badge must be minimized to ensure an enlarged operating time.

The power consumption of the badge is composed of several components.
In the inactive state the badge consumes about 244 \textmu{}A. 
This baseline consumption includes the sleep state current of the nRF51822, the \ac{EEPROM} and the accelerometer.
The following components, which are turned on by default, have the greatest impact on the consumption in the inactive state: the analog microphone (155 \textmu{}A), the analog amplifier circuit (20 \textmu{}A) and the voltage regulator (65 \textmu{}A).
The periodical broadcast of the advertising packet every 200 ms leads to an average current consumption of 130 \textmu{}A.
The power consumption of the audio signal recording is different for the previous and the new implementation of the firmware.
The first-mentioned one reads two samples every 1.42 ms and consumes about 564 \textmu{}A. 
The new implementation reads only one sample, because the audio quality of the averaged signal is comparable.
As a result, the average current consumption can be reduced to 350 \textmu{}A.

The scan process for surrounding \ac{BLE} devices has the highest consumption of all components. 
This can be explained by the high consumption of the active receive operation of the nRF51822, which draws up to 13 mA \cite{NRF51822:PS}. The averaged power consumption of the scan process is approximately 800 \textmu{}A.
Although the current consumption during the \ac{BLE} data transfer is slightly higher in the new implementation, the average power consumption (64 \textmu{}A) was considerably decreased compared to the previous one (173 \textmu{}A).
This can be attributed to the significantly higher data throughput.
Since the new implementation includes additional data sources, such as the accelerometer, an additional consumption (42 \textmu{}A) is generated for their data transmission.
The recording of raw acceleration data consumes about 15 \textmu{}A on average, the acceleration event recording about 10 \textmu{}A and the supply voltage recording about 2 \textmu{}A.
As already mentioned in the previous chapter, the consumption of the components is highly dependent on the parameter settings.

Finally, for the developed firmware, the total average power consumption of the badge including the additional data sources is 1.66 mA.
This results in a theoretically possible lifetime of approximately 5.7 days (using a 3 V, 230 mAh coin cell battery).
This theoretical lifetime cannot be achieved as the available capacity of coin cells decreases due to high current peaks that are typical for BLE applications \cite{Furset11:HPD}.

%Improvement:Current consumption: For beacon it would make sense to disable the microphone circuit (immerhin x uA) baseline current

%Es konnte keine Differenz zwischen Anzahl der Reads bei Flash und RAM bei Microphone festgestellt werden. Deshalb wird das noch Verglichen (eingehen auf Peaks, Problematisch f"ur CoinCells (Diskussion)).

\section{Filesystem}
The implemented filesystem and virtual memory abstraction enables the developer to easily access the \acp{NVM} for sequential storing and reading.
The filesystem divides the memory into partitions with a configurable size.
A partition can be initialized as static or dynamic. In a static partition all elements, such as data chunks, must have the same size. In contrast, a dynamic partition allows a variable length of elements.
Within a dynamic partition an \ac{XOR} linked list is used to efficiently manage sequential store and read operations.

In the previous implementation, the two physical memories, flash-memory and \ac{EEPROM}, are considered separately. In the flash-memory the recorded microphone-chunks are stored. In the \ac{EEPROM} the assigned ID and group number as well as the scan-chunks are stored.
Due to this strict separation it is not possible to divide the available storage space evenly between the data sources.
In addition, the data chunks must have a certain size so that they can be stored word- and page-aligned in memory.
To meet these alignment requirements, the data chunks must be partially padded.
The use of padding might waste usable memory space.
Since in the previous implementation scan-chunks have a fixed size, memory is wasted if only few devices have been discovered during the scan process.

The developed filesystem based on the virtual memory with byte level access overcomes these limitations.
Since there is only one large virtual memory, the available memory space can be divided arbitrarily between the data sources.
In addition, padding is not required anymore due to the byte level access.
The storage of variable size scan-chunks can be realized in a dynamic partition without wasting memory space.
Another advantage of a dynamic partition is that it enables data compression, which generally results in a variable size of data \cite{Szpankowski11:MEL}.

The implemented filesystem has also some restrictions: For instance, the filesystem does not provide a mechanism for deleting data elements, but only for sequentially overwriting old elements. Therefore, the filesystem is not suitable for applications which are not based on sequential data generation and retrieval.
Furthermore, the filesystem requires access to the memory at byte level. This implies that an abstraction layer must be implemented if the underlying memory does not allow byte level access.

%Of course some additional information has to be incorporated (such as the record-id and 

%Old implementation adds a 4 byte check, and uses padding just to enable word alignment (not good) (Scanner)

%But through dynamical approach, scanning can be optimizd, den sinnvollen einsatz compression can be done...
%Numbers for devices available for storage is limited to a constant number (and cannot be changed??)(Theoretical compare: Scanning space)

%Filesystem: less space in flash, because of bigger program (care has to be taken, to set flash boundaries correctly)

%Disadvantage: Can not delete elements (only optimized to read and store elements sequentially)

%Search key independent of data?? E.g. timestamp?

%\section{Testing framework}
   % (\chapter{})
\cleardoublepage
\chapter{Conclusion} \label{cha:Conclusion}

Teamwork is becoming increasingly important these days.
Collaboration in groups is often used to solve complex problems or to develop new solution strategies.
In order to improve the collaboration and productivity of a group, the behavior within and between groups must be understood.
For the analysis of group dynamics, special techniques must be applied to quantify the social interaction and communication behavior of members.
The open-source Rhythm project \cite{Lederman18:RAU} provides a platform with various tools and devices for analyzing face-to-face interactions, such as turn-taking and speech overlap.
These include electronic badges, which each participant wears in front of his chest.
The badges are able to record vocal activity, movement, proximity to others and absolute location.
The recorded data can be retrieved during the deployment of the badges to perform analysis and to optionally provide real-time feedback to the participants.

As part of this work, the firmware of the badges was reimplemented and optimized for modularity, maintainability and extensibility.
For this purpose, a strict separation between the different functional units and a high degree of abstraction was introduced.
Furthermore, a framework was designed that enables the simple creation, execution and assessment of tests. By applying thoroughly testing, implementation errors can be detected and corrected quickly.

Besides the data sources already supported by the previous firmware, such as audio and proximity, the new firmware provides raw acceleration data, the detection of certain acceleration events and the supply voltage as new data sources.
Since the recording of audio data is power-intensive, it has been examined more closely.
The evaluation results clearly showed that the number of samples read every 1.42 ms has negligible influence on the averaged audio signal.
Therefore, the new firmware reads only one sample every 1.42 ms, which significantly reduces power consumption compared to the previous implementation.

A generic serialization library called Tinybuf was developed for the efficient and platform-independent de-/serialization of structured data. It processes predefined messages to automatically create optimized source code for encoding and decoding. The new firmware uses Tinybuf for the serialization of structured data prior to storage. Furthermore, it is used to provide a flexible and efficient protocol for the communication with remote \ac{BLE} devices.
In a benchmark it was compared to Google's Protocol Buffers de-/serialization library. Tinybuf executes the encoding and decoding faster than Protocol Buffers, has a smaller binary representation in most cases, and is compatible with the previous protocol implementation.

In addition, a filesystem was developed in this work, which enables an efficient storage of sequentially generated data in the \acp{NVM}. It is based on a virtual memory representation with byte level access, which abstracts the different memory types.
The filesystem is designed to withstand power loss during memory accesses.
It physically separates the available memory into partitions for the different types of data.
The partitions can be configured to support either fixed or variable size elements.
In addition to data storage, the new firmware also supports data streaming.

For the communication between the badges and remote \ac{BLE} devices, a flexible communication protocol on the basis of the Tinybuf serialization library was implemented.
The protocol provides various commands to control, monitor and retrieve data from the badge.
Furthermore, a technique was applied to significantly increase the data throughput via the \ac{BLE} connection compared to the previous implementation.
This reduces the time required to transmit data and thus reducing the power consumption of the badge.
Additionally, a clock sychronization method was developed that enables the effective compensation of oscillator frequency drifts.
It leads to a preciser internal time computation and is compatible with the previous protocol implementation.

Due to the modularization and strict separation of functionalities in the new firmware, the developed methods can also be applied in related projects and on other platforms.
In addition, the project can serve as a reference for modular embedded software development that allows simple verification of functionalities.

% Data exchange:
% Faster transmission (leading to less power consumption)
% To compensate the frequency offsets and drifts (clock synchronization)
% 

% Due to the high degree of abstraction the developed techniques can be applied in different other projects.
% Open-source project, educational platform, basis for modular development
%Filesystem that is independent (can be eingesetzt unter verschiedenen Platformern..)

%Outlook: Accelerometer evaluation, Nice to have (compression, encryption (Daten sind persönlich)), Tinybuf...

In future work various further developments of the system could be realized.
One important part is the evaluation of the accelerometer to determine reasonable parameters for datarate, operating mode and event detection threshold.
Furthermore, the accelerometer could be used to automatically deactivate data recording if no movement of the badge/user has been detected within a predefined time interval.
For even further reduction of power consumption, the individual parameters of the other data sources could be investigated in more detail.
%A possible future application of the accelerometer is the automatic deactivation of data recording if no movement has occurred within a certain time interval.
%Additionally, the parameters of the other data sources can be intensively investigated in order to identify further potentials for power reduction.

A useful technique for increasing the amount of data that can be stored in the \acp{NVM} is data compression.
Since the developed filesystem already supports the storage of elements with variable length, the integration of data compression should be realizable.
Depending on the type of data, different data compression techniques have to be examined.
%However, different techniques for data compression have to be examined. As there are different types of data, it may be useful to apply different compression methods depending on the type of data.

Since the recorded data are personal, measures for data protection, such as encryption, should be implemented. 
For this purpose, an appropriate encryption technique (symmetric, asymmetric, hybrid) must be chosen.
Additionally, a special protocol might be required to enable secure key exchange.
%Additionally, a more complex protocol is required which, for example, enables secure key exchange.

Besides further software developments, hardware improvements should be considered as well.
This includes, for example, chip migration to the nRF52 from Nordic Semiconductor. Compared to the nRF51822, the nRF52 has a significantly reduced power consumption as well as a larger \ac{RAM} and flash-memory \cite{NRF52840:PS}.
For further increase of the available memory, a larger external memory chip (flash-memory or \ac{EEPROM}) might be inserted.
%To increase the available memory for data, a larger external memory chip (flash-memory or \ac{EEPROM}) can be inserted. 
Finally, a digital microphone could be integrated in order to simplify the circuit of the badge.
%, because it would replace the analog microphone and amplification circuit.
% Hardware side: Switching to nRF52, größerer externer Speicherchip, DCDC Wandler, digital microphone, Envelope detector
   % (\chapter{})
\cleardoublepage

   % (\chapter{})
\cleardoublepage

   % Ausblick (\chapter{Ausblick} TEXT)
\cleardoublepage

   % Zusammenfassung (\chapter{Zusammenfassung}  TEXT)
\cleardoublepage

\appendix
\cleardoublepage
\chapter{Glossar}

\begin{acronym}[EEPROM]
 \acro{IR}{infrared}\\
 \acro{RFID}{radio-frequency identification}
 \acro{BLE}{Bluetooth Low Energy}\\
 \acro{ISM}{Industrial Scientific Medical}\\
 \acro{GAP}{Generic Access Profile}\\
 \acro{ATT}{Attribute Protocol}\\
 \acro{GATT}{Generic Attribute Profile}\\
 \acro{NUS}{Nordic UART Service}\\
 \acro{PCB}{printed circuit board}\\
 \acro{SoM}{system on module}\\
 \acro{SoC}{system on chip}\\
 \acro{RAM}{random access memory}\\
 \acro{ADC}{analog-to-digital converter}\\
 \acro{SPI}{serial peripheral interface}\\
 \acro{UART}{universal asynchronous receiver-transmitter}\\
 \acro{I2C}{inter-integrated circuit}\\
 \acro{I/O}{input/output}\\
 \acro{RTC}{real-time clock}\\
 \acro{NVM}{non-volatile memory}\\
 \acro{EEPROM}{electrical erasable programmable read-only memory}\\
 \acro{SMD}{surface-mounted device}\\
 \acro{ID}{identification number}\\
 \acro{MAC}{medium access control}\\
 \acro{RSSI}{received signal strength indicator}\\
 \acro{MEMS}{microelectromechanical system}\\
 \acro{FIFO}{first-in first-out}\\
 \acro{LED}{light-emitting diode}\\
 \acro{NTP}{Network Time Protocol}\\
 \acro{API}{application programming interface}\\
 \acro{XML}{Extensible Markup Language}\\
 \acro{JSON}{JavaScript Object Notation}\\
 \acro{HW}{hardware}\\
 \acro{SDK}{software development kit}\\
 \acro{CPU}{central processing unit}\\
 \acro{SS}{slave select}\\
 \acro{GCC}{GNU Compiler Collection}\\
 \acro{HTML}{Hypertext Markup Language}\\
 \acro{g}{gravitational force}\\
 \acro{mg}{milli gravitational force}\\
 \acro{EWMA}{exponentially weighted moving average}\\
 \acro{varint}{variable-length integer}\\
 \acro{CRC}{cyclic redundancy check}\\
 \acro{XOR}{exclusive or}\\
 \acro{TPSN}{Timing-sync Protocol for Sensor Networks}\\
 \acro{MAE}{mean absolute error}\\
 \acro{DUT}{device under test}\\
 
\end{acronym}
   % Glossar (\chapter{Glossar}  TEXT)io
\cleardoublepage
\chapter{Patents}

\newpage

\section{US7216088 (B1)} \label{sec:US7216088}

%https://patents.google.com/patent/US7216088

\begin{table}[h]
\renewcommand{\arraystretch}{2.5}
\centering

\begin{tabular}{p{5cm}p{11cm}} 
\textbf{Title} & \textit{System and method for managing a project based on team member interdependency and impact relationships} \\
\textbf{Publication Number} & US7216088 (B1) \\
\textbf{Publication Date} & May 8, 2007 \\
\textbf{Inventor(s)} & Oscar A. Chappel, \newline Christopher T. Creel  \\
\textbf{Assignee(s)} & NTT Data Services Corp \\
\textbf{Abstract} & A system and method for determining interdependencies between project team members working on a development project. The method includes receiving data indicative of a temporal relationship between a first and a second project team member having modified at least one artifact of the development project. The data indicative of the temporal relationship between the project team members may be statistically analyzed. At least one metric representative of an interdependency relationship between the first and second project team members may be formed. The metric(s) representative of the interdependency relationship may be stored. \\

\end{tabular}
\end{table}

\newpage

\section{WO2010099488 (A1)} \label{sec:WO2010099488}
%https://patents.google.com/patent/WO2010099488A1/en

\begin{table}[h]
\renewcommand{\arraystretch}{2.5}
\centering

\begin{tabular}{p{5cm}p{11cm}} 
\textbf{Title} & \textit{Contact tracking using wireless badges} \\
\textbf{Publication Number} & WO2010099488 (A1) \\
\textbf{Publication Date} & September 9, 2010 \\
\textbf{Inventor(s)} & Theodore Herman, \newline Philip Polgreen   \\
\textbf{Assignee(s)} & University Of Iowa Research Foundation \\
\textbf{Abstract} & This application discusses apparatus and methods for tracking contacts between various entities in a region. An apparatus includes portable transceiver device having a radio, memory and a power supply. The portable transceiver device can detect and record a history of proximity to other transceiver devices. The portable transceiver device can receive information related to disease exposure of the transceiver device and can include a processing component to compute an exposure potential to a contagious disease using the history of proximity and the disease exposure information. \\

\end{tabular}
\end{table}

\newpage

\section{US10049336 (B2)} \label{sec:US10049336}
%https://patents.google.com/patent/US10049336B2/en

\begin{table}[h]
\renewcommand{\arraystretch}{2.5}
\centering

\begin{tabular}{p{5cm}p{11cm}} 
\textbf{Title} & \textit{Social sensing and behavioral analysis system} \\
\textbf{Publication Number} & US10049336 (B2) \\
\textbf{Publication Date} & August 14, 2018 \\
\textbf{Inventor(s)} & Daniel Olguin Olguin, \newline Tuomas Jaanu, \newline Derek Heyman, \newline Benjamin Waber   \\
\textbf{Assignee(s)} & Sociometric Solutions Inc  \\
\textbf{Abstract} & A method and system for capturing and analyzing human behavior data is disclosed. The present disclosure describes a method and system for a plurality of people, wherein each person wears a badge. The badge transmits data collected from a plurality of sensors from the badge to a base station. The data is sent from the base station to a server, which aggregates the data from a plurality of base stations, and then analyzes and processes the data to create raw human behavior data. From the raw human behavior data and plurality of metrics is calculated, which can be displayed on a computer screen according to whichever metrics a user wishes to view. \\

\end{tabular}
\end{table}

\cleardoublepage

\cleardoublepage

%%%%%%%%%%%%%%%%%%%%%%%%%%%%%%%%%%%%%%%%%%%%%%%%%%%%%%%%%%%%%%%%%%%%%%%%%%
% Diese Datei nicht veraendern!
%%%%%%%%%%%%%%%%%%%%%%%%%%%%%%%%%%%%%%%%%%%%%%%%%%%%%%%%%%%%%%%%%%%%%%%%%%
\addcontentsline{toc}{chapter}{\listfigurename}
\listoffigures
 % Bilderverzeichnis
\cleardoublepage
%%%%%%%%%%%%%%%%%%%%%%%%%%%%%%%%%%%%%%%%%%%%%%%%%%%%%%%%%%%%%%%%%%%%%%%%%%
% Diese Datei nicht veraendern!
%%%%%%%%%%%%%%%%%%%%%%%%%%%%%%%%%%%%%%%%%%%%%%%%%%%%%%%%%%%%%%%%%%%%%%%%%%
\addcontentsline{toc}{chapter}{\listtablename}
\listoftables
 % Tabellenverzeichnis
\cleardoublepage
%%%%%%%%%%%%%%%%%%%%%%%%%%%%%%%%%%%%%%%%%%%%%%%%%%%%%%%%%%%%%%%%%%%%%%%%%%
% Diese Datei nicht veraendern!
%%%%%%%%%%%%%%%%%%%%%%%%%%%%%%%%%%%%%%%%%%%%%%%%%%%%%%%%%%%%%%%%%%%%%%%%%%
\addcontentsline{toc}{chapter}{\bibname}
\bibliography{mad}

\begin{thebibliography}{{VAR}16}

\bibitem[Abr12]{RTC:ABS07}
Abracon LLC:
\newblock {\em ABS07-120-32.768kHz-T; TUNING FORK CRYSTAL}, 10 2012.

\bibitem[Aim12]{Aimonen12:NPB}
P.~Aimonen:
\newblock {\em Nanopb - Protocol Buffers for Embedded Systems}, 2012,
\newblock \url{https://github.com/nanopb/nanopb}. Accessed 2018-11-05.

\bibitem[Bab02]{Babcock02:MAI}
B.~Babcock, S.~Babu, M.~Datar, R.~Motwani, J.~Widom:
\newblock {\em Models and issues in data stream systems},
\newblock in {\em Proceedings of the twenty-first ACM SIGMOD-SIGACT-SIGART
  symposium on Principles of database systems}, ACM, 2002, pp. 1--16.

\bibitem[Bar06]{Barr06:PES}
M.~Barr, A.~Massa:
\newblock {\em Programming embedded systems: with C and GNU development tools},
\newblock O'Reilly Media, Inc., 2006.

\bibitem[Ber05]{Berdine05:SMA}
J.~Berdine, C.~Calcagno, P.~W. O'hearn:
\newblock {\em Smallfoot: Modular automatic assertion checking with separation
  logic},
\newblock in {\em International Symposium on Formal Methods for Components and
  Objects}, Springer, 2005, pp. 115--137.

\bibitem[Bla03]{Blanchet03:ASA}
B.~Blanchet, P.~Cousot, R.~Cousot, J.~Feret, L.~Mauborgne, A.~Min{\'e},
  D.~Monniaux, X.~Rival:
\newblock {\em A static analyzer for large safety-critical software},
\newblock in {\em ACM SIGPLAN Notices}, Vol.~38, ACM, 2003, pp. 196--207.

\bibitem[Bou97]{Bouten97:TAA}
C.~V. Bouten, K.~T. Koekkoek, M.~Verduin, R.~Kodde, J.~D. Janssen:
\newblock {\em A triaxial accelerometer and portable data processing unit for
  the assessment of daily physical activity},
\newblock {\em IEEE transactions on biomedical engineering}, Vol.~44, No.~3,
  1997, pp. 136--147.

\bibitem[Bro14]{Brown14:TAO}
C.~Brown, C.~Efstratiou, I.~Leontiadis, D.~Quercia, C.~Mascolo, J.~Scott,
  P.~Key:
\newblock {\em The architecture of innovation: Tracking face-to-face
  interactions with ubicomp technologies},
\newblock in {\em Proceedings of the 2014 ACM International Joint Conference on
  Pervasive and Ubiquitous Computing}, ACM, 2014, pp. 811--822.

\bibitem[Cal16]{Calacci16:BAO}
D.~Calacci, O.~Lederman, D.~Shrier, A.~Pentland:
\newblock {\em Breakout: An open measurement and intervention tool for
  distributed peer learning groups},
\newblock {\em Proc. 2016 Int'l Conf. Social Computing Behavioral-Cultural
  Modeling \& Prediction and Behavior Representation in Modeling and Simulation
  (SBP-BRIMS 16) Social Cultural and Behavioral Modeling}, 2016.

\bibitem[Car02]{Carley02:COS}
K.~M. Carley:
\newblock {\em Computational organizational science and organizational
  engineering},
\newblock {\em Simulation Modelling Practice and Theory}, Vol.~10, No.~5-7,
  2002, pp. 253--269.

\bibitem[Cat10]{Cattuto10:DOP}
C.~Cattuto, W.~Van~den Broeck, A.~Barrat, V.~Colizza, J.-F. Pinton,
  A.~Vespignani:
\newblock {\em Dynamics of person-to-person interactions from distributed RFID
  sensor networks},
\newblock {\em PloS one}, Vol.~5, No.~7, 2010, p. e11596.

\bibitem[Cha14]{Chang14:BAV}
K.-H. Chang:
\newblock {\em Bluetooth: a viable solution for IoT?[Industry Perspectives]},
\newblock {\em IEEE Wireless Communications}, Vol.~21, No.~6, 2014, pp. 6--7.

\bibitem[Che04]{Chess04:SAF}
B.~Chess, G.~McGraw:
\newblock {\em Static analysis for security},
\newblock {\em IEEE Security \& Privacy}, Vol.~2, No.~6, 2004, pp. 76--79.

\bibitem[Cho02]{Choudhury02:TSA}
T.~Choudhury, A.~Pentland:
\newblock {\em The sociometer: A wearable device for understanding human
  networks},
\newblock in {\em CSCW'02 Workshop: Ad hoc Communications and Collaboration in
  Ubiquitous Computing Environments}, 2002.

\bibitem[Dab98]{Daberko98:OSI}
N.~Daberko:
\newblock {\em Operating system including improved file management for use in
  devices utilizing flash memory as main memory}, July 28 1998,
\newblock US Patent 5,787,445.

\bibitem[Dab02]{Dabek02:EDP}
F.~Dabek, N.~Zeldovich, F.~Kaashoek, D.~Mazi{\`e}res, R.~Morris:
\newblock {\em Event-driven programming for robust software},
\newblock in {\em Proceedings of the 10th workshop on ACM SIGOPS European
  workshop}, ACM, 2002, pp. 186--189.

\bibitem[Des06]{Desikan06:STP}
S.~Desikan, G.~Ramesh:
\newblock {\em Software testing: principles and practice},
\newblock Pearson Education India, 2006.

\bibitem[Dia17]{Dian17:ASI}
F.~J. Dian, A.~Yousefi, K.~Somaratne:
\newblock {\em A study in accuracy of time synchronization of BLE devices using
  connection-based event},
\newblock in {\em Information Technology, Electronics and Mobile Communication
  Conference (IEMCON), 2017 8th IEEE Annual}, IEEE, 2017, pp. 595--601.

\bibitem[DiM04]{Dimicco04:IGP}
J.~M. DiMicco, A.~Pandolfo, W.~Bender:
\newblock {\em Influencing group participation with a shared display},
\newblock in {\em Proceedings of the 2004 ACM conference on Computer supported
  cooperative work}, ACM, 2004, pp. 614--623.

\bibitem[Don10]{Dong10:QGP}
W.~Dong, A.~S. Pentland:
\newblock {\em Quantifying group problem solving with stochastic analysis},
\newblock in {\em International Conference on Multimodal Interfaces and the
  Workshop on Machine Learning for Multimodal Interaction}, ACM, 2010, p.~40.

\bibitem[Doy93]{Doyle93:HTM}
M.~Doyle, D.~Straus:
\newblock {\em How to make meetings work},
\newblock The Berkley Publishing Group, 1993.

\bibitem[Dre96]{Drew96:TTT}
S.~Drew, C.~Coulson-Thomas:
\newblock {\em Transformation through teamwork: the path to the new
  organization?},
\newblock {\em Management Decision}, Vol.~34, No.~1, 1996, pp. 7--17.

\bibitem[Els02]{Elson02:FGN}
J.~Elson, L.~Girod, D.~Estrin:
\newblock {\em Fine-grained network time synchronization using reference
  broadcasts},
\newblock {\em ACM SIGOPS Operating Systems Review}, Vol.~36, No.~SI, 2002, pp.
  147--163.

\bibitem[Fav07]{Favalli07:WC}
L.~Favalli, M.~Lanati, P.~Savazzi:
\newblock {\em Wireless communications 2007 CNIT thyrrenian symposium}, 2007.

\bibitem[For18]{Forsyth18:GD}
D.~R. Forsyth:
\newblock {\em Group dynamics},
\newblock Cengage Learning, 2018.

\bibitem[Fur11]{Furset11:HPD}
K.~Furset, P.~Hoffman:
\newblock {\em High pulse drain impact on CR2032 coin cell battery capacity},
\newblock {\em Nordic Semiconductor and Energizer}, 2011.

\bibitem[Gal05]{Gal05:ATF}
E.~Gal, S.~Toledo:
\newblock {\em A transactional flash file system for microcontrollers.},
\newblock in {\em USENIX Annual Technical Conference, General Track}, 2005, pp.
  89--104.

\bibitem[Gan03]{Ganeriwal03:TSP}
S.~Ganeriwal, R.~Kumar, M.~B. Srivastava:
\newblock {\em Timing-sync protocol for sensor networks},
\newblock in {\em Proceedings of the 1st international conference on Embedded
  networked sensor systems}, ACM, 2003, pp. 138--149.

\bibitem[GCC05]{Gcov05:GAT}
GCC:
\newblock {\em gcov - a Test Coverage Program}, 2005,
\newblock \url{https://gcc.gnu.org/onlinedocs/gcc/Gcov.html}. Accessed
  2018-10-23.

\bibitem[Gom12]{Gomez12:OAE}
C.~Gomez, J.~Oller, J.~Paradells:
\newblock {\em Overview and evaluation of bluetooth low energy: An emerging
  low-power wireless technology},
\newblock {\em Sensors}, Vol.~12, No.~9, 2012, pp. 11734--11753.

\bibitem[Goo08a]{Google08:GT}
Google:
\newblock {\em Google Test}, 2008,
\newblock \url{https://github.com/google/googletest}. Accessed 2018-10-15.

\bibitem[Goo08b]{Google08:PB}
Google:
\newblock {\em Protocol Buffers}, 2008,
\newblock \url{https://developers.google.com/protocol-buffers/}. Accessed
  2018-11-05.

\bibitem[Gre03]{Greenberg03:BIO}
J.~Greenberg, R.~A. Baron:
\newblock {\em Behavior in Organizations: Understanding and Managing the Human
  Side of Work},
\newblock Pearson Prentice Hall Upper Saddle River, NJ, 2003.

\bibitem[Ham04]{Hamill04:UTF}
P.~Hamill:
\newblock {\em Unit test frameworks: tools for high-quality software
  development},
\newblock O'Reilly Media, Inc., 2004.

\bibitem[Hol04]{Holt04:FSA}
C.~C. Holt:
\newblock {\em Forecasting seasonals and trends by exponentially weighted
  moving averages},
\newblock {\em International journal of forecasting}, Vol.~20, No.~1, 2004, pp.
  5--10.

\bibitem[Hwa04]{Hwang04:ESD}
S.-Y. Hwang, D.-H. Yu, K.-J. Li:
\newblock {\em Embedded system design for network time synchronization},
\newblock in {\em International Conference on Embedded and Ubiquitous
  Computing}, Springer, 2004, pp. 96--106.

\bibitem[Jeh01]{Jehn01:TDN}
K.~A. Jehn, E.~A. Mannix:
\newblock {\em The dynamic nature of conflict: A longitudinal study of
  intragroup conflict and group performance},
\newblock {\em Academy of management journal}, Vol.~44, No.~2, 2001, pp.
  238--251.

\bibitem[Jia09]{Jianwu09:ROD}
Z.~Jianwu, Z.~Lu:
\newblock {\em Research on distance measurement based on RSSI of ZigBee},
\newblock in {\em Computing, Communication, Control, and Management, 2009. CCCM
  2009. ISECS International Colloquium on}, Vol.~3, IEEE, 2009, pp. 210--212.

\bibitem[Jia14]{Jianyong14:RBB}
Z.~Jianyong, L.~Haiyong, C.~Zili, L.~Zhaohui:
\newblock {\em RSSI based Bluetooth low energy indoor positioning},
\newblock in {\em Indoor Positioning and Indoor Navigation (IPIN), 2014
  International Conference on}, IEEE, 2014, pp. 526--533.

\bibitem[Kim08]{Kim08:MME}
T.~Kim, A.~Chang, L.~Holland, A.~S. Pentland:
\newblock {\em Meeting mediator: enhancing group collaborationusing sociometric
  feedback},
\newblock in {\em Proceedings of the 2008 ACM conference on Computer supported
  cooperative work}, ACM, 2008, pp. 457--466.

\bibitem[Kin18]{Kindt18:NDL}
P.~H. Kindt, M.~Saur, M.~Balszun, S.~Chakraborty:
\newblock {\em Neighbor discovery latency in BLE-like protocols},
\newblock {\em IEEE Transactions on Mobile Computing}, Vol.~17, No.~3, 2018,
  pp. 617--631.

\bibitem[Kno12]{SPU0414HR5H-SB:PDS}
Knowles Corporation:
\newblock {\em SPU0414HR5H-SB Product Datasheet}, 12 2012,
\newblock Rev. E.

\bibitem[Led16a]{Lederman16:OB}
O.~Lederman, D.~Calacci, A.~MacMullen, D.~C. Fehder, F.~E. Murray, A.~Pentland:
\newblock {\em Open badges: A low-cost toolkit for measuring team communication
  and dynamics},
\newblock in {\em 2016 International Conference on Social Computing,
  Behavioral-Cultural Modeling, \& Prediction and Behavior Representation in
  Modeling and Simulation , Washington DC, USA, June 28-July 1}, 2016.

\bibitem[Led16b]{Lederman16:OHP}
O.~Lederman et~al.:
\newblock {\em Openbadge Hub Python}, 2016,
\newblock \url{https://github.com/HumanDynamics/openbadge-hub-py}. Accessed
  2018-10-12.

\bibitem[Led16c]{Lederman16:OBP}
O.~Lederman et~al.:
\newblock {\em OpenBadge project}, 2016,
\newblock \url{https://github.com/HumanDynamics/openbadge}. Accessed
  2018-10-12.

\bibitem[Led18]{Lederman18:RAU}
O.~Lederman, A.~Mohan, D.~Calacci, A.~S. Pentland:
\newblock {\em Rhythm: A Unified Measurement Platform for Human Organizations},
\newblock {\em IEEE MultiMedia}, Vol.~25, No.~1, 2018, pp. 26--38.

\bibitem[Les09]{Leshed09:VRT}
G.~Leshed, D.~Perez, J.~T. Hancock, D.~Cosley, J.~Birnholtz, S.~Lee, P.~L.
  McLeod, G.~Gay:
\newblock {\em Visualizing real-time language-based feedback on teamwork
  behavior in computer-mediated groups},
\newblock in {\em Proceedings of the SIGCHI Conference on Human Factors in
  Computing Systems}, ACM, 2009, pp. 537--546.

\bibitem[Lev15]{Levi15:GDF}
D.~Levi:
\newblock {\em Group dynamics for teams},
\newblock Sage Publications, 2015.

\bibitem[Lia12]{Liao12:ASA}
T.-T. Liao, L.-S. Lo:
\newblock {\em Audio system and related method integrated with ultrasound
  communication functionality}, April 17 2012,
\newblock US Patent 8,160,276.

\bibitem[Mae12]{Maeda12:PEO}
K.~Maeda:
\newblock {\em Performance evaluation of object serialization libraries in XML,
  JSON and binary formats},
\newblock in {\em Digital Information and Communication Technology and it's
  Applications (DICTAP), 2012 Second International Conference on}, IEEE, 2012,
  pp. 177--182.

\bibitem[Mar12]{Marks12:ITS}
R.~J.~I. Marks:
\newblock {\em Introduction to Shannon sampling and interpolation theory},
\newblock Springer Science \& Business Media, 2012.

\bibitem[Mas17]{Masumoto17:MAV}
K.~Masumoto, T.~Yaguchi, H.~Matsuda, H.~Tani, K.~Tozuka, N.~Kondo, S.~Okada:
\newblock {\em Measurement and visualization of face-to-face interaction among
  community-dwelling older adults using wearable sensors},
\newblock {\em Geriatrics \& gerontology international}, Vol.~17, No.~10, 2017,
  pp. 1752--1758.

\bibitem[Mer14]{Merkel14:DLL}
D.~Merkel:
\newblock {\em Docker: lightweight linux containers for consistent development
  and deployment},
\newblock {\em Linux Journal}, Vol.~2014, No.~239, 2014, p.~2.

\bibitem[Mil91]{Mills91:ITS}
D.~L. Mills:
\newblock {\em Internet time synchronization: the network time protocol},
\newblock {\em IEEE Transactions on communications}, Vol.~39, No.~10, 1991, pp.
  1482--1493.

\bibitem[Nad79]{Nadler79:TEO}
D.~A. Nadler:
\newblock {\em The effects of feedback on task group behavior: A review of the
  experimental research},
\newblock {\em Organizational Behavior and Human Performance}, Vol.~23, No.~3,
  1979, pp. 309--338.

\bibitem[Nah94]{Nahavandi94:RTF}
A.~Nahavandi, E.~Aranda:
\newblock {\em Restructuring teams for the re-engineered organization},
\newblock {\em Academy of Management Perspectives}, Vol.~8, No.~4, 1994, pp.
  58--68.

\bibitem[Nor16a]{NRF51822:PS}
Nordic Semiconductor:
\newblock {\em nRF51822 Product Specification}, 7 2016,
\newblock Rev. 3.3.

\bibitem[{Nor}16b]{Nordic16:PPK}
{Nordic Semiconductor}:
\newblock {\em Power Profiler Kit}, 2016,
\newblock \url{https://www.nordicsemi.com/eng/Products/Power-Profiler-Kit}.
  Accessed 2018-11-15.

\bibitem[Nor16c]{SoftDevice16:SS}
Nordic Semiconductor:
\newblock {\em S130 SoftDevice Specification}, 4 2016,
\newblock Rev. 2.0.

\bibitem[{Nor}17]{Nordic17:NT5}
{Nordic Semiconductor}:
\newblock {\em Nordic Thingy:52}, 2017,
\newblock \url{https://www.nordicsemi.com/eng/Products/Nordic-Thingy-52}.
  Accessed 2018-11-02.

\bibitem[{Nor}18a]{NordicSemiconductor18:INF}
{Nordic Semiconductor}:
\newblock {\em Infocenter}, 2018,
\newblock \url{http://infocenter.nordicsemi.com/}. Accessed 2018-10-16.

\bibitem[Nor18b]{NRF52840:PS}
Nordic Semiconductor:
\newblock {\em nRF52840 Product Specification}, 3 2018,
\newblock Rev. 1.0.

\bibitem[Obe08]{Lcov08:LAG}
P.~Oberparleiter:
\newblock {\em lcov - a graphical GCOV front-end}, 2008,
\newblock \url{https://linux.die.net/man/1/lcov}. Accessed 2018-10-23.

\bibitem[Olg06]{Olguin06:WCB}
D.~O. Olgu{\'\i}n, J.~A. Paradiso, A.~Pentland:
\newblock {\em Wearable communicator badge: Designing a new platform for
  revealing organizational dynamics},
\newblock in {\em Proceedings of the 10th international symposium on wearable
  computers (student colloquium)}, 2006, pp. 4--6.

\bibitem[Olg09]{Olguin09:SOT}
D.~O. Olgu{\'\i}n, B.~N. Waber, T.~Kim, A.~Mohan, K.~Ara, A.~Pentland:
\newblock {\em Sensible organizations: Technology and methodology for
  automatically measuring organizational behavior},
\newblock {\em IEEE Transactions on Systems, Man, and Cybernetics, Part B
  (Cybernetics)}, Vol.~39, No.~1, 2009, pp. 43--55.

\bibitem[Olg10]{Olguin10:SBO}
D.~O. Olgu{\'\i}n, A.~Pentland:
\newblock {\em Sensor-based organisational design and engineering},
\newblock Vol. 1(1-2), 2010, pp. 69--97.

\bibitem[Osh15]{Osherove15:TAO}
R.~Osherove:
\newblock {\em The art of unit testing},
\newblock MITP-Verlags GmbH \& Co. KG, 2015.

\bibitem[Pen10]{Pentland10:DYR}
A.~Pentland:
\newblock {\em Defend your research: We can measure the power of
  charisma-Harvard Business Review},
\newblock {\em Harvard Business Review Case Studies, Articles, Books,
  Pamphlets-Harvard Business Review. Retrieved from http://hbr.
  org/2010/01/defend-your-research-we-can-measure-the-power-of-charisma/ar/1},
  2010.

\bibitem[Pen12]{Pentland12:TNS}
A.~Pentland:
\newblock {\em The new science of building great teams},
\newblock {\em Harvard Business Review}, Vol.~90, No.~4, 2012, pp. 60--69.

\bibitem[Raz15]{Raza15:BS}
S.~Raza, P.~Misra, Z.~He, T.~Voigt:
\newblock {\em Bluetooth smart: An enabling technology for the Internet of
  Things},
\newblock in {\em Wireless and Mobile Computing, Networking and Communications
  (WiMob), 2015 IEEE 11th International Conference on}, IEEE, 2015, pp.
  155--162.

\bibitem[Rig17]{Rigado:BMD200}
Rigado Inc.:
\newblock {\em BMD-200 Module for Bluetooth 4.2 LE}, 8 2017,
\newblock Rev. 1.7.

\bibitem[{Ruu}18]{Ruuvi18:WIR}
{Ruuvi Innovations Oy}:
\newblock {\em What is RuuviTag?}, 2018,
\newblock \url{https://ruuvi.com/ruuvitag-specs/}. Accessed 2018-11-02.

\bibitem[Sam18]{Samrose18:CCC}
S.~Samrose, R.~Zhao, J.~White, V.~Li, L.~Nova, Y.~Lu, M.~R. Ali, M.~E. Hoque:
\newblock {\em CoCo: Collaboration Coach for Understanding Team Dynamics during
  Video Conferencing},
\newblock {\em Proceedings of the ACM on Interactive, Mobile, Wearable and
  Ubiquitous Technologies}, Vol.~1, No.~4, 2018, p. 160.

\bibitem[Shc14]{Shchekotov14:ILM}
M.~Shchekotov:
\newblock {\em Indoor localization method based on Wi-Fi trilateration
  technique},
\newblock in {\em Proceeding of the 16th conference of fruct association},
  2014, pp. 177--179.

\bibitem[STM17]{LIS2DH12:DS}
STMicroelectronics:
\newblock {\em LIS2DH12 Datasheet: MEMS digital output motion sensor:
  ultra-low-power high-performance 3-axis 'femto' accelerometer}, 6 2017,
\newblock Rev. 6.0.

\bibitem[STM18]{EEPROM:M95M02}
STMicroelectronics:
\newblock {\em M95M02-DR}, 9 2018,
\newblock Rev. 11.0.

\bibitem[Str12]{Stryker12:FFT}
J.~B. Stryker, M.~D. Santoro:
\newblock {\em Facilitating face-to-face communication in high-tech teams},
\newblock {\em Research-Technology Management}, Vol.~55, No.~1, 2012, pp.
  51--56.

\bibitem[Szp11]{Szpankowski11:MEL}
W.~Szpankowski, S.~Verd{\'u}:
\newblock {\em Minimum expected length of fixed-to-variable lossless
  compression without prefix constraints},
\newblock {\em IEEE Transactions on Information Theory}, Vol.~57, No.~7, 2011,
  pp. 4017--4025.

\bibitem[Tal02]{Tal02:TFT}
A.~Tal:
\newblock {\em Two flash technologies compared: NOR vs NAND},
\newblock {\em White Paper of M-Systems}, 2002.

\bibitem[Tan95]{Tannen95:TPO}
D.~Tannen:
\newblock {\em The power of talk: Who gets heard and why},
\newblock {\em Harvard Business Review}, Vol.~73, No.~5, 1995, pp. 138--148.

\bibitem[{VAR}16]{CR2032:CLM}
{VARTA Microbattery GmbH}:
\newblock {\em CR2032 Lithium Manganese Dioxide Data Sheet}, 9 2016.

\bibitem[Wak09]{Wakisaka09:BSS}
Y.~Wakisaka, N.~Ohkubo, K.~Ara, N.~Sato, M.~Hayakawa, S.~Tsuji, Y.~Horry,
  K.~Yano, N.~Moriwaki:
\newblock {\em Beam-scan sensor node: Reliable sensing of human interactions in
  organization},
\newblock in {\em Networked Sensing Systems (INSS), 2009 Sixth International
  Conference on}, IEEE, 2009, pp. 1--4.

\bibitem[Wat14]{Watanabe14:ERB}
J.-i. Watanabe, N.~Ishibashi, K.~Yano:
\newblock {\em Exploring relationship between face-to-face interaction and team
  performance using wearable sensor badges},
\newblock {\em PloS one}, Vol.~9, No.~12, 2014, p. e114681.

\bibitem[Why91]{Whyte91:DFW}
G.~Whyte:
\newblock {\em Decision failures: Why they occur and how to prevent them},
\newblock {\em Academy of Management Perspectives}, Vol.~5, No.~3, 1991, pp.
  23--31.

\bibitem[Wil05]{Willmott05:AOT}
C.~J. Willmott, K.~Matsuura:
\newblock {\em Advantages of the mean absolute error (MAE) over the root mean
  square error (RMSE) in assessing average model performance},
\newblock {\em Climate research}, Vol.~30, No.~1, 2005, pp. 79--82.

\bibitem[Wuc07]{wuchty07:TID}
S.~Wuchty, B.~F. Jones, B.~Uzzi:
\newblock {\em The increasing dominance of teams in production of knowledge},
\newblock {\em Science}, Vol.~316, No.~5827, 2007, pp. 1036--1039.

\bibitem[Xu10]{Xu10:DMM}
J.~Xu, W.~Liu, F.~Lang, Y.~Zhang, C.~Wang:
\newblock {\em Distance measurement model based on RSSI in WSN},
\newblock {\em Wireless Sensor Network}, Vol.~2, No.~08, 2010, p. 606.

\bibitem[Yan12]{Yang12:LIF}
Z.~Yang, C.~Wu, Y.~Liu:
\newblock {\em Locating in fingerprint space: wireless indoor localization with
  little human intervention},
\newblock in {\em Proceedings of the 18th annual international conference on
  Mobile computing and networking}, ACM, 2012, pp. 269--280.

\end{thebibliography}
 % Literaturverzeichnis

\end{document}